\documentclass{nature}

\usepackage{amssymb}
\usepackage{amsmath}
\usepackage{graphicx}
\usepackage{url}
\usepackage{xcolor}
\usepackage{subcaption}
\usepackage{lineno}
\usepackage{orcidlink}

\title{Formation of Fast-spinning Neutron Stars in Close Binaries and Magnetar-driven Stripped-envelope Supernovae}

\author{Rui-Chong Hu$^{\orcidlink{0000-0002-6442-7850}1,2\ast}$,
Jin-Ping Zhu$^{\orcidlink{0000-0002-9195-4904}3,4,5\ast}$,
Ying Qin$^{\orcidlink{0000-0002-2956-8367}6,1\ast}$,
Yong Shao$^{\orcidlink{0000-0003-2506-6906}7\ast}$,
Bing Zhang$^{\orcidlink{0000-0002-9725-2524}2,8}$,
Yun-Wei Yu$^{\orcidlink{0000-0002-1067-1911}9,10}$,
En-Wei Liang$^{\orcidlink{0000-0002-7044-733X}1}$,
Liang-Duan Liu$^{\orcidlink{0000-0002-8708-0597}9,10}$,
Bo Wang$^{\orcidlink{0000-0002-3231-1167}11}$,
Xin-Wen Shu$\orcidlink{0000-0002-7020-4290}^6$,
Jian-Feng Liu$^{9,10}$}

\begin{document}

\maketitle

\begin{affiliations}
 \item Guangxi Key Laboratory for Relativistic Astrophysics, School of Physical Science and Technology, Guangxi University, Nanning 530004, China
  \item Department of Physics and Astronomy, University of Nevada, Las Vegas, NV 89154, USA
 \item Department of Astronomy, School of Physics, Peking University, Beijing 100871, China
 \item School of Physics and Astronomy, Monash University, Clayton Victoria 3800, Australia
 \item OzGrav: The ARC Centre of Excellence for Gravitational Wave Discovery, Clayton Victoria 3800, Australia
 \item Department of Physics, Anhui Normal University, Wuhu, Anhui, 241002, China
 \item Department of Astronomy, Nanjing University, Nanjing 210023, China
 \item Nevada Center for Astrophysics, University of Nevada, Las Vegas, NV 89154, USA
 \item Institute of Astrophysics, Central China Normal University, Wuhan 430079, China
 \item Key Laboratory of Quark and Lepton Physics (Central China Normal University), Ministry of Education, Wuhan 430079, China
 \item Yunnan Observatories, Chinese Academy of Sciences, Kunming 650216, China
\end{affiliations}
\noindent{$^\ast$Co-First Authors. These authors contributed equally: Rui-Chong Hu, Jin-Ping Zhu, Ying Qin, and Yong Shao}
\bigskip

\begin{abstract}

Extreme stripped-envelope supernovae (SESNe), including Type Ic superluminous supernovae (SLSNe-I), broad-line Type Ic SNe (SNe Ic-BL), and fast blue optical transients (FBOTs), are widely believed to harbor a newborn fast-spinning highly-magnetized neutron star (``magnetar''), which can lose its rotational energy via spin-down processes to accelerate and heat the ejecta. The progenitor(s) of these magnetar-driven SESNe, and the origin of considerable angular momentum (AM) in the cores of massive stars to finally produce such fast-spinning magnetars upon core-collapse are still under debate. Popular proposed scenarios in the literature cannot simultaneously explain their event rate density, SN and magnetar parameters, and the observed metallicity. Here, we perform a detailed binary evolution simulation that demonstrates that tidal spin-up helium stars with efficient AM transport mechanism in close binaries can form fast-spinning magnetars at the end of stars' life to naturally reproduce the universal energy-mass correlation of these magnetar-driven SESNe. Our models are consistent with the event rate densities, host environments, ejecta masses, and energetics of these different kinds of magnetar-driven SESNe, supporting that the isolated common-envelope formation channel could be a major common origin of magnetar-driven SESNe. The remnant compact binary systems of magnetar-driven SESNe are progenitors of some gravitational-wave transients and galactic systems.

\end{abstract}

SESNe display little to no hydrogen and/or helium in their spectra\cite{filippenko1997}, whose ejecta kinetic energy is generally provided by neutrinos and whose radiation energy is mostly powered by the radioactive decay of $^{56}$Ni and $^{56}$Co. One of the exotic subgroups is the particularly luminous SLSNe-I, defined to be brighter at peak than absolute magnitude $-21\,{\rm{mag}}$ which are about $\sim10-100$ times brighter than ordinary SNe\cite{galyam2019}. The lightcurve, the unusually bright peak, and the {extreme explosion energy} of SLSNe-I cannot be well explained by the {classical core-collapse SN} model, with the most popular alternative hypothesis {being that a millisecond magnetar engine injects energy into the remnant, which is mostly converted into the ejecta kinetic energy and in the meantime heats the ejecta}\cite{kasen2010}. Another particular subgroup of SESNe is represented by SNe Ic-BL, in which the ``BL'' is typically attached if the observed spectra show much larger ejecta velocities than that of traditional neutrino-powered SESNe. Virtually all the discovered SNe accompanied with long-duration gamma-ray bursts (lGRBs) were identified as SNe Ic-BL\cite{zhang2018}, {but a large fraction of SNe Ic-BL were found without the generation of lGRB jets.}. In the literature, there are two types of post-collapse central engine models proposed for lGRBs and SNe Ic-BL, i.e., extra energy injection from an accreting stellar-mass black hole\cite{woosley1993,popham1999} (BH) or a spinning-down magnetar\cite{usov1992,dai1998a,wheeler2000,zhang2001}. It has been widely suggested that a newborn magnetar should play a crucial role in making a plateau and/or afterglow flares that appeared in a good fraction of lGRB X-ray afterglows\cite{metzger2011,lv2014}, as well as the magnetar-driven shock breakout emissions in some SNe Ic-BL\cite{zhang2022}. FBOTs are fast-evolving, very blue, and luminous transients\cite{drout2014,pursiainen2018}, whose progenitor and power source are still mysterious. The observations of FBOT hosts revealed that these transients explode exclusively in star-forming host galaxies, which could favor a core-collapse origin in connection with massive stars\cite{drout2014,pursiainen2018}. However, the peaks of most FBOTs are much brighter than the maximum possible luminosity from a standard $^{56}$Ni-powered SN, and hence an additional power source is required\cite{drout2014,pursiainen2018,liu2022,sawada2022}. Since the star formation rates and metallicities of the FBOT hosts are similar to those of SLSN-I and lGRB host galaxies\cite{wiseman2020}, it is reasonable to suspect that SLSNe-I and lGRBs might be related to FBOTs in terms of the progenitor and power source (i.e., magnetar engine). Indeed, the existence of magnetar central engine can provide a better explanation than the model with pure $^{56}$Ni-decay heating for the lightcurves of some FBOTs\cite{fang2019,liu2022,sawada2022}. Thus, the magnetar engine can be ubiquitous in the SLSN-I, lGRB, SN Ic-BL, and FBOT phenomena.

In the literature, there are widely divergent views on the nature of the progenitors and the evolutionary channel of extreme SESNe. A crucial question is how to attain and sustain a rapidly rotating core of a massive star to finally form a fast-spinning magnetar until the time of the explosion. {A key constraint is that the magnetic torque {(Tayler-Spruit dynamo)} can significantly alter the AM transport processes via wind mass loss in the star to slow down their cores\cite{Spruit2002}, in which case models invoking single stars cannot simultaneously meet the requirements on maintaining enough AM over the stars' life and losing their outer envelope through winds to form stripped-envelope stars\cite{woosley2006GRB}.}
One of the most popular models is that single stars with rapid rotation at birth\cite{yoon2005,woosley2006GRB,aguileradena2018,Song2023} or stars in binaries spun-up via mass transfer, tidal effects, or stellar mergers\cite{cantiello2007,eldridge2011,demink2013,mandel2016,eldridge2019,ghodla2023} can experience the so-called quasi-chemically homogeneous evolution (CHE) to directly evolve into fast-spinning Wolf-Rayet stars (i.e., helium stars). However, the CHE channel is generally predicted to occur at very low metallicity (e.g., $Z\lesssim0.3\,Z_\odot$)\cite{woosley2006GRB,chrimes2020}. In observations, except SLSNe-I whose host galaxies have a metallicity threshold of $\sim0.5\,Z_\odot$\cite{lunnan2014,chen2017}, a lot of other SESNe, including lGRBs, SNe Ic-BL and FBOTs, can be found to occur in super-solar host galaxies, even though they are heavily suppressed in more metal-rich environments with metallicity $Z\gtrsim2-3\,Z_\odot$\cite{japelj2018,wiseman2020,modjaz2020}. Because a large fraction of these extreme SESNe were discovered in the environments exceeding the CHE metallicity threshold and their rate densities are much larger than the CHE theoretical predictions\cite{chrimes2020}, this formation channel cannot be the sole and primary contributing channel. Another promising scenario is that helium stars in close binaries originate through the classical channels of common-envelope evolution (CEE) and stable mass transfer (SMT), which can tidally spin up the helium stars through interactions with their companions\cite{izzard2004,detmers2008,Bogomazov2009,qin2018,chrimes2020,bavera2022,fuller2022}. {Compared to the CHE channel, these channels} may allow envelope-stripping and formation of fast-spinning magnetars at random metallicity environments.  {However, previous theoretical studies exploring the tidal scenario as the origin of these extreme SESNe often lack a connection with observations or are inconsistent with the observed explosion results, making it difficult to provide strong evidence for this hypothesis. Whether this tidal scenario can resolve the origin issues of these extreme SESNe remains an open question in the literature.}

{By considering the magnetar engines to power these extreme SESNe, Figure \ref{fig:M_ej_E_rot_MainText} shows that the observationally inferred distributions between the initial magnetar rotational energies $E_{\rm{rot,i}}$ and ejecta masses $M_{\rm{ej}}$ of SLSNe-I, FBOTs, SNe Ic-BL with and without lGRBs (respectively abbreviated as GRB-SNe and SNe Ic-BL, hereafter) have a very strong positive universal correlation (see Methods for the data collection). The $E_{\rm{rot,i}}-M_{\rm{ej}}$ distributions between SLSNe-I, GRB-SNe, and SNe Ic-BL overlap substantially, while a clear criterion to define FBOTs is their small ejecta masses with an upper limit of $\sim1\,M_\odot$ around the lower limit of the masses of other explosion phenomena. This universal correlation, indicating that these extreme SESNe could share a common origin, provides a powerful link between the SN parameters, magnetar parameters, and the nature of the progenitor population. Thus, modeling this observed correlation allows us to infer the origin of these extreme SESNe.} In this paper, we explore the origin of these magnetar-driven SESNe by simultaneously modeling their universal $E_{\rm{rot,i}}-M_{\rm{ej}}$ correlation and rate densities with observed metallicities {through the binary tidal scenario formed} via the isolated CEE and SMT channels. 

\section*{Results}

Here, we perform detailed binary evolution modeling of stars in close orbits (see Methods for details) from the moment when the hydrogen envelope is completely stripped off via mass loss and/or Roche lobe overflow until close to the time of the explosions, during which the helium stars {would further lose their helium envelopes} by undergoing wind mass loss and/or mass transfer (see Supplementary Information Section \ref{sup_sec:mass_transfer_case} for the mass transfer case) and can be tidally spun-up by main-sequence (MS) or compact companions continuously. {In our simulations, the initial helium star mass $M_{\rm{He,i}}$ covers a range of $2.4-40\,M_\odot$ corresponding to an inferred zero-age main-sequence (ZAMS) mass $M_{\rm{ZAMS}}$ of $\sim10-90\,M_\odot$, while we consider the initial orbital period $P_{\rm{orb,i}}$ within $0.1-10\,{\rm{d}}$.} A fast-spinning magnetar can be formed upon core-collapse of the star to power a magnetar-driven SESN. Figure \ref{fig:M_ej_E_rot_MainText} compares our simulated initial magnetar rotational energy $E_{\rm{rot,i}}$ to the observationally inferred $E_{\rm{rot,i}}$ as a function of ejecta mass $M_{\rm{ej}}$ with metallicity $Z = 0.3\,Z_\odot$ by considering an efficient AM transport mechanism in the stellar interior\cite{Spruit2002}. {The reason why these relationships increase linearly with initial helium star mass is mainly due to more massive helium stars having larger-sized cores with higher moments of inertia, allowing more AM to be stored in the part of the core assumed to collapse and to form a magnetar.} {The metallicity of the SN host galaxy is typically believed to be a great estimator of the SN progenitor metallicity (e.g., ref\cite{bravo2011}). Because most of magnetar-driven SESNe were discovered in metallicity environments between  $\sim0.1\,Z_\odot$ and $\sim2-3\,Z_\odot$ (e.g., ref.\cite{japelj2018,lunnan2014,chen2017,wiseman2020,modjaz2020}), we thus select four different metallicities of $Z=\{0.1,0.3,1,3\}\,Z_\odot$} to simulate the $E_{\rm{rot,i}}-M_{\rm{ej}}$ relationship displayed in Figure \ref{fig:Dependence_Metallicity_MainText}. {Because the events derived from lower-metallicity environments mainly occur at high redshifts, we briefly show our simulated the $E_{\rm{rot,i}}-M_{\rm{ej}}$  relationship at $Z=\{0.01,0.03\}\,Z_\odot$ in \ref{fig:Dependence_Metallicity_Supplement}, which is nearly consistent with those at $0.1\,Z_\odot$.} We note that unphysical results with $E_{\rm{rot,i}}$ much larger than the maximum available energy of a millisecond magnetar are excluded in our simulations. {We conservatively set this maximum available energy as $\sim2-3\times10^{52}\,{\rm{erg}}$, although some works, e.g., ref.\cite{metzger2015}, also suggested that it could be even as high as $1-2\times10^{53}\,{\rm{erg}}$.} More details on the metallicity dependence and model dependence of AM transport mechanism are presented in Supplementary Section \ref{sup_sec:parameter_dependence}.

In very low-metallicity to solar-metallicity environments, Figures \ref{fig:M_ej_E_rot_MainText} and \ref{fig:Dependence_Metallicity_MainText} reveal that the $E_{\rm{rot,i}}-M_{\rm{ej}}$ distributions of most SLSNe-I, GRB-SNe, and SNe Ic-BL, can be achievable for helium stars with initial masses of ${M}_{\rm{He,i}}\sim5-40\,M_\odot$ in close binaries (initial orbital periods of $P_{\rm{orb,i}}\lesssim1-2\,{\rm{d}}$). Since tides can be significant at $P_{\rm{orb,i}}\lesssim1-2\,{\rm{d}}$ (e.g., refs.\cite{qin2018,fuller2022}), it is expected that tidal effects in close binaries should play a crucial role in forming fast-spinning magnetars to account for the explosions of these magnetar-driven SESNe. The energy release of radioactive decay of $^{56}$Ni and $^{56}$Co, as the source of ordinary SESN emissions, is derived from the energy provided by neutrinos which is typically\cite{sukhbold2016} $\lesssim2\times10^{51}\,{\rm{erg}}$. In principle, only if the initial magnetar rotational energy is much higher than neutrino-powered energy would magnetar spin-down energy injection dominate SN explosion. Energy injection from magnetars born in slightly wider binaries (i.e., $P_{\rm{orb,i}}\gtrsim1-2\,{\rm{d}}$) might be close to or even much lower than the energy provided by neutrinos, in which case the SN emission would be ordinary $^{56}$Ni-powered SESNe or SNe partially powered by a magnetar engine (e.g., ref.\cite{gomez2022}). At higher-metallicity environments (i.e., $Z>Z_\odot$), mainly due to stronger wind mass loss during the helium-burning phase to remove AM within stars, helium stars in close binaries can hardly generate fast-spinning magnetars that have an initial rotational energy much larger than the neutrino-powered energy. Thus, the explosions of SLSNe-I, GRB-SNe, and SNe Ic-BL would be heavily suppressed in super-solar metallicity environments, so that SNe in close binaries from high-mass helium stars would always be ordinary SNe Ic. Indeed, recent observations on the late-time nebular spectra and environments supported that the SN Ic progenitors could be massive helium stars in close binaries\cite{fang2019,sun2022}, while the hosts of observed SN Ic, many of which have super-solar environments, are systematically more metal-rich than those of these magnetar-driven SESNe\cite{japelj2018,modjaz2020}.

As shown in Figures \ref{fig:M_ej_E_rot_MainText} and \ref{fig:Dependence_Metallicity_MainText}, in sub-solar environments (i.e., $Z\lesssim{Z}_\odot$), FBOTs likely originate from low-mass helium stars with initial masses of $M_{\rm{He,i}}\lesssim5\,M_\odot$. The core-collapse explosions of these low-mass helium stars in binary systems are also called as ultra-stripped SNe\cite{tauris2015,suwa2015,sawada2022}, whose explosion energy derived from neutrinos is typically from $\sim5\times10^{49}\,{\rm{erg}}$ to $\sim1.7\times10^{50}\,{\rm{erg}}$ based on simulations\cite{suwa2015}, much lower than that of typical high-mass SESNe. We find that the $E_{\rm{rot,i}}$ and $M_{\rm{ej}}$ distributions of $\gtrsim80\%$ FBOTs can result from helium stars with initial masses from $M_{\rm{He,i}}\sim2.4-2.5\,M_\odot$ to $\sim5\,M_\odot$. Thus, the $M_{\rm{ej}}$ distributions of most of FBOTs can be well explained by the core-collapse explosions of low-mass helium stars in binary systems. Different from SLSNe-I, GRB-SNe, and SNe Ic-BL that need strong tides in close binaries to significantly spin-up the progenitors, FBOTs can even occur in relatively wide binaries with $P_{\rm{orb,i}}\lesssim10\,{\rm{d}}$, whose magnetar remnants still have enough rotational energy larger than neutrino-powered energy to likely dominate SN explosions. In super-solar environments, due to a higher mass loss rate, more lower-mass helium stars would have a final core with a mass $M_{\rm{CO}}\lesssim1.37\,M_\odot$, a floor below which the final fate of the helium stars can generate white dwarfs (WDs; e.g., ref\cite{tauris2015}) instead of forming fast-spinning magnetars to power FBOTs. Furthermore, we find that the initial rotational energy of magnetars from more helium stars with larger orbital periods would be close to or lower than the neutrino-powered energy so that the SN emissions could be $^{56}$Ni-powered. It is thus expected that the explosions of FBOTs could be partially suppressed in high-metallicity environments. 

In order to test whether the deaths of helium stars formed via the CEE and SMT channels can achieve the observed event rate density of magnetar-driven SESNe, we perform a binary population synthesis simulation to generate a series of binary populations (see Methods). {Within the range of the observed metallicities of different types of magnetar-driven SESNe, we select binary systems composed of a helium star with an MS or a compact companion, whose initial orbital period and initial helium star mass fall into their constrained $P_{\rm{orb,i}}-M_{\rm{He,i}}$ parameter space following our detailed binary simulations shown in Figures \ref{fig:M_ej_E_rot_MainText} and \ref{fig:Dependence_Metallicity_MainText}, to calculate their redshift-dependent event rate densities (see Figure \ref{fig:EventRate_MainText})} {by adopting a reasonable CE efficiency of $\alpha_{\rm CE}=5$ and the default mass-transfer model (see details of our model conditions in Methods) as an example.} We simulate the redshift-dependent rate densities of SLSNe-I by selecting close binary helium stars {from our population} under the conditions of $M_{\rm{He,i}}>5\,M_\odot$ and $P_{\rm{orb,i}}<2\,{\rm{d}}$ {based on our detailed binary simulation results} with the observed metallicity in a range of $Z<0.5\,Z_\odot$ \cite{chen2017}. SLSNe-I could have event rate densities of $\mathcal{R}_{\rm{SLSN-I}}=35^{+25}_{-13}$, $91^{+76}_{-36}$, $\sim400$, and $\sim400\,{\rm{Gpc}}^{-3}\,{\rm{yr}}^{-1}$ at $z\sim0.1$, $1.1$, $2$ and $4$ in observations\cite{galyam2019,frohmaier2021,prajs2017,cooke2012}, which can be essentially reproduced by our simulated SLSN-I population {under the assumptions of our population synthesis model.} Since a large fraction of super-solar metallicity hosts are found among SNe Ic-BL, we extend the selection range of metallicity to $Z<2\,Z_\odot$ following the observations\cite{japelj2018} to model their rate densities. The predicted SN Ic-BL rate density throughout the redshift range of $z\lesssim2$ might be $\sim2-4$ times larger than those of SLSNe-I, while SLSNe-I and SNe Ic-BL may have similar evolutionary histories at $z\gtrsim2$. {Present observed constraints on the SN Ic-BL rate density in the literature, e.g., $3.7^{+2.9}_{-3.7}\%$ of SESNe\cite{shivvers2017} which was obtained based on only one discovered event in the Lick Observatory Supernova Search volume-limited sample, are relatively ambiguous with large errors. Thus, we cannot conclude whether our population synthesis model can explain the observed rate density of SNe Ic-BL.} By selecting FBOTs in the conditions of $2.4\leq{M}_{\rm{He,i}}<5\,M_\odot$ and $P_{\rm{orb,i}}<10\,{\rm{d}}$ with observed threshold metallicity\cite{wiseman2020} of $Z<Z_\odot$ from our simulated binary population, the local FBOT rate density is $\sim1000\,{\rm{Gpc}}^{-3}\,{\rm{yr}}^{-1}$. {The latest constrained FBOT rate density is $\sim1\%$ of the core-collapse SN rate density (i.e., $\mathcal{R}_{\rm{FBOT}}\gtrsim1000\,{\rm{Gpc}}^{-3}\,{\rm{yr}}^{-1}$) in the local universe, estimated based on a sample of 37 events at a redshift of $0.05<z<1.56$ in the Dark Energy Survey\cite{pursiainen2018,wiseman2020}. Therefore, based on our assumptions, our population synthesis simulation for the FBOT rate density are basically consistent with the observed value. Due to the observed uncertainties in the beaming factors of lGRBs and the ratio of lGRBs to SNe Ic-BL, we do not explore whether our population synthesis simulations can reproduce the observed lGRB rate density. In Supplementary Information Section \ref{sup_sec:population_synthesis}, we explore different CE efficiencies and mass-transfer models that affect the event rate densities, with results largely falling within the range of observations.} 

We further explore the possible companions at the moment of magnetar-driven SESNe based on our simulated populations. The filter criterion of metallicity is roughly set as $Z<Z_\odot$. As shown in Figure \ref{fig:Companions_MainText}, $\sim65-80\%$ magnetar-driven SESNe would occur in helium star binaries hosting an MS companion, while a compact companion (mostly a helium star, NS or BH) presents in other magnetar-driven SESNe. Among binaries with an MS companion, we find that $\sim80-85\%$ of their progenitors are expected to experience a CEE. Furthermore, the formation of binaries containing two helium stars involves a double-core CE between two post-MS stars. Thus, the CEE channel should be the major scenario for the formation of magnetar-driven SESNe. In contrast with binaries for a higher CE efficiency adopted in the population synthesis simulation, compact-object companions are easier to merge with stars during the CE stage, which can result in a lower ratio to expected companions of magnetar-driven SESNe. 

{In summary, by adopting reasonable parameter assumptions, our detailed binary evolution simulations naturally reproduce the $E_{\rm{rot,i}}-M_{\rm{ej}}$ correlation, providing strong evidence that SLSNe-I, GRB-SNe, SNe Ic-BL, and FBOTs can share a unified origin through the tidal interactions in close binaries. This coherent finding also self-consistently supports that the magnetar engine--a long-held hypothesis in the literature--can commonly exist in the explosions of these extreme SESNe. Based on different model conditions, our population synthesis results with observed metallicities can basically match the observed rate densities of these magnetar-driven SESNe, suggesting the formation of these helium-star close binaries through isolated CEE and SMT channels and, hence, a potential connection between magnetar-driven SESNe and galactic neutron star (NS) binaries as well as gravitational-wave events due to NS mergers.  }

\section*{Discussion}

\begin{figure*}
\centering
\includegraphics[width=\linewidth, trim = 30 0 40 0, clip]{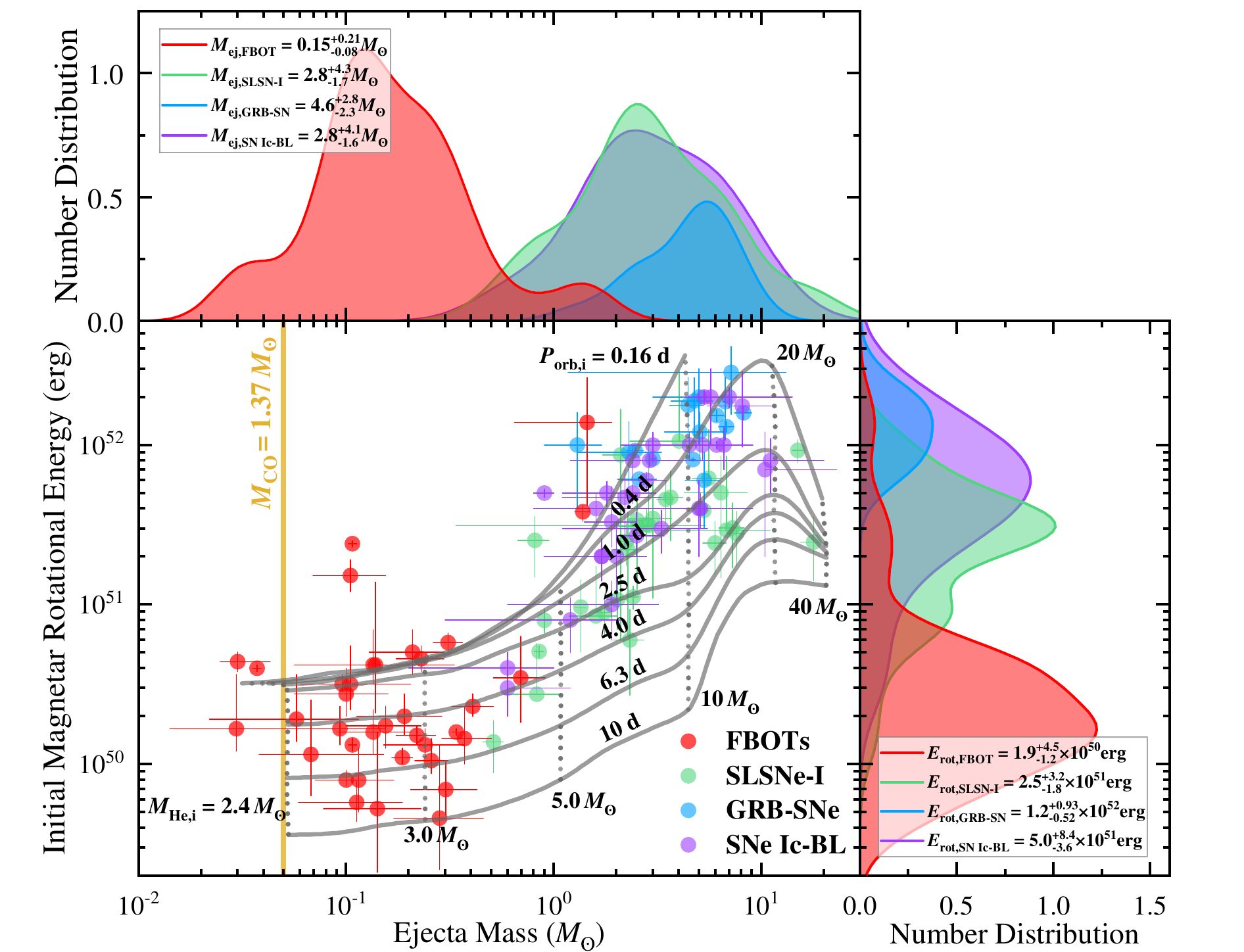}
\caption{{\bf Observationally inferred $E_{\rm{rot,i}}-M_{\rm{ej}}$ distributions of magnetar-driven SESNe and our simulated relationships by evolving helium stars in close binaries as a function of $P_{\rm{orb,i}}$ and $M_{\rm{He,i}}$ with metallicity $Z=0.3\,Z_\odot$.} The red, green, blue, and violet points represent the observations 
for FBOTs (see Supplementary Information Table \ref{table:FBOT}), SLSNe-I\cite{yu2017}, GRB-SNe\cite{lv2018}, and SNe Ic-BL\cite{lyman2016,taddia2019}. The solid and dashed lines correspond to different initial orbital periods $P_{\rm{orb,i}}$ and initial helium star masses $M_{\rm{He,i}}$ of the progenitor systems as labeled. The orange line shows the estimated ejecta mass for a final metal core with a mass of $M_{\rm{CO}}=1.37\,M_\odot$ (see Methods). The top and right panels display the number distributions of $M_{\rm{ej}}$ and $E_{\rm{rot,i}}$, derived by the method of kernel density estimation, for these magnetar-driven SESNe. }
\label{fig:M_ej_E_rot_MainText}
\end{figure*}

Firstly, as shown in Figure \ref{fig:M_ej_E_rot_MainText},of the $\sim60\%-70\%$ observed SLSNe-I, GRB-SNe, and SNe Ic-BL have ejecta masses of $1\lesssim{M}_{\rm{ej}}\lesssim4\,M_\odot$, corresponding to progenitors with mass $5\lesssim{M}_{\rm{He,i}}\lesssim10\,M_\odot$. These initial helium star masses imply ZAMS masses $18\lesssim{M}_{\rm{ZAMS}}\lesssim30\,M_\odot$, for which stars can almost leave behind NSs and, hence, have successful explosions (e.g., ref.\cite{sukhbold2016}). Secondly, other observed SLSNe-I, GRB-SNe, and SNe Ic-BL with $M_{\rm{ej}}\gtrsim4\,M_\odot$ come from helium stars with $10\lesssim{M}_{\rm{He,i}}\lesssim20\,M_\odot$, i.e., ZAMS stars with $30\lesssim{M}_{\rm{ZAMS}}\lesssim50\,M_\odot$. The most massive discovered SLSNe-I, with $M_{\rm{He,i}}\sim40\,M_\odot$, likely require a progenitor with $M_{\rm{ZAMS}}\sim90\,M_\odot$. {The nature of the final remnant, whether it is an NS or a BH, formed after the core collapse of a helium star with initial mass ${M}_{\rm{He,i}}\gtrsim10\,M_\odot$, is still under debate. For example, the numerical simulation results by ref.\cite{AguileraDena2023} suggest that most $\lesssim40\,M_\odot$ helium stars may form NSs, supporting the observed mass range of magnetar-driven SNe. Meanwhile, some numerical simulations\cite{ertl2020,schneider2021} suggest that a fraction of $\gtrsim10\,M_\odot$ helium stars can either undergo failed explosions resulting in direct BH formation or have successful SN explosions with BH formation due to significant fallback, although the specific mass range for BH formation varies depending on the model assumption.}

{Our models suggest that the explosions of helium stars in close binaries primarily formed through the CEE channel are capable to reproduce not only the universal $E_{\rm{rot,i}}-M_{\rm{ej}}$ correlation, but also event rate densities of magnetar-driven SESNe with observed metallicities. In Figure \ref{fig:illustration}, we show an illustration of {the} formation channel of magnetar-driven SESNe and subsequent close-orbit NS binaries. Based on our binary population synthesis simulations, most magnetar-driven SESNe are produced by the first-formed tidally spun-up helium star in a close binary with an MS secondary. In order to form such a binary system, the two progenitor stars might initially be in a binary system close enough to experience a CE phase when the primary MS evolves into a giant star. Furthermore, a small fraction of magnetar-driven SESNe could originate from explosions of the second-formed fast-spinning helium star in a binary system with a compact-object companion formed via CE evolution. In the system, the first SN could be a weak SN, an ordinary SN, or a magnetar-driven SESN. Therefore, one to two magnetar-driven SESNe could occur in a binary system during its evolution. }

A fraction of magnetar-driven SESNe are considered to finally generate post-SN close-orbit NS binaries, including an NS with an MS, a helium star or a compact-object companion, if the systems can survive the SN kicks. An NS with a compact-object companion, as well as the binary NS (BNS) systems formed after the second SN of the NS--star systems, are potential progenitors of gravitational-wave sources in the LIGO and/or LISA bands. We show the fractions of NSs with compact-object companions that can survive and merge within Hubble time in Supplement Section \ref{sup_sec:kick} by considering the effects of natal kicks on binary systems. Since FBOT magnetars usually receive small kicks due to their extremely small amount of ejecta, most binary systems would survive and usually have low eccentricities\cite{tauris2015,tauris2017,ViganaGomez2018}. Only those binary systems with close orbits of $P_{\rm{orb,i}}\lesssim0.7\,{\rm{d}}$ can more easily merge within Hubble time. If the initial helium star mass is $5\lesssim{M}_{\rm{He,i}}\lesssim10\,M_\odot$, there are only a fraction of binary systems, whose initial orbital periods are mostly $P_{\rm{orb,i}}\lesssim1\,{\rm{d}}$, would not be disrupted even for large SN kicks and would finally merge within Hubble time. {Moreover, the second SNe of some galactic BNS systems could originate from these magnetar-driven SESNe. For those BNS systems with low eccentricities and pre-SN orbital periods shorter than $\sim10\,{\rm{d}}$ (e.g., ref.\cite{tauris2017}), their second SNe are likely magnetar-driven ultra-stripped SNe (i.e., FBOTs). If the second SN of a BNS system is SLSN-I, lGRB, or SN Ic-BL,} even though binary systems still exist after SNe, they can have some extreme characteristics, e.g., large proper motion velocities and large orbital eccentricities. The well-known Hulse–Taylor pulsar\cite{hulse1975}, {whose second SN explosion was thought to be a normal SN Ib or Ic\cite{tauris2017}}, has a binary parameter with the orbital period of $P_{\rm{orb}}=0.323\,{\rm{d}}$ and the eccentricity of $e=0.612$. A kick of at least $200\,{\rm{km}}\,{\rm{s}}^{-1}$ during the second SN explosion is needed to explain the observed properties of the binary system of the Hulse–Taylor pulsar\cite{tauris2017}, indicating the young NS orbiting with the Hulse–Taylor pulsar might have a helium star progenitor with $M_{\rm{He,i}}\gtrsim6.3\,M_\odot$ (see Supplement Section \ref{sup_sec:hulse_taylor_pulsar} for details) and could be a millisecond magnetar when it was born. However, due to the long inspiral timescale of this system, the second-born magnetar could have a present-value rotation period of $\sim700\,{\rm{s}}$, which suggests that it is difficult to observe as a pulsar (see Supplement Section \ref{sup_sec:hulse_taylor_pulsar} for detailed calculations). Therefore, the Hulse-Taylor pulsar BNS system could be the aftermath of a historical SLSN-I, lGRB, or SN Ic-BL in our galaxy, {rather than a normal SN Ib or Ic\cite{tauris2017}.}  

\begin{addendum}

 \item We thank Ilya Mandel, {Philipp Podsiadlowski,} and Jim Fuller for valuable discussions, {Ping Chen,} He Gao, { Ryosuke Hirai,} Yacheng Kang, Wenbin Lu, {Sergei B. Popov,} Kai Wang, and Yuan-Pei Yang for helpful comments. Y.Q. acknowledges support from the Anhui Provincial Natural Science Foundation (Grant No. 2308085MA29) and the National Natural Science Foundation of China (Grant No. 12473036). E.W.L. and R.C.H. were supported the support by the National Natural Science Foundation of China (Grant No. 12133003). J.P.Z. was partially supported by the National Basic Research Program of China (Grant No. 2014CB845800). Y.S. was supported by the Natural Science Foundation of China (Grant No. 11973026). Y.W.Y., and L.D.L were supported by the National Key R\&D Program of China (2021YFA0718500), the China Manned Spaced Project (CMS CSST-2021-A12), and the National Natural Science Foundation of China (Grant No. 11833003). B.W. was supported by the National Key R\&D Program of China (No. 2021YFA1600404), the National Natural Science Foundation of China (No. 12225304), the Western Light Project of CAS (No. XBZG-ZDSYS-202117), and the Yunnan Fundamental Research Project (No. 202001AS070029). X.W.S. was partially supported by the National Natural Science Foundation of China (Grant Nos. 12192220, 12192221).

 \item[Competing Interests] The authors declare that they have no competing interests.
 \item[Contributions] Y.Q., J.P.Z., R.C.H., B.Z., and E.W.L. initiated the work. R.C.H., Y.Q., and J.P.Z. performed the detailed binary evolution simulations. Y.S. and R.C.H. performed the population synthesis simulations. J.P.Z. collected the SN parameters and rate densities from literature. J.P.Z. fitted the FBOT lightcurves to obtain SN and magnetar parameters. J.P.Z. and R.C.H. created the figures, illustration and tables. J.P.Z. wrote the initial draft with important contributions from Y.Q., Y.S., R.C.H., and B.Z., under the supervision by B.Z., E.W.L., and Y.W.Y. All authors have reviewed, discussed and commented on the modelling, data analysis and manuscript.
 \item[Corresponding Author] Correspondence should be addressed to JPZ (email: zhujp@pku.edu.cn), YQ (email: yingqin2013@hotmail.com) and YS (email: shaoyong@nju.edu.cn).
\end{addendum}

\clearpage

\begin{figure*}
\centering
\includegraphics[width=0.77\linewidth, trim = 6 0 8 10, clip]{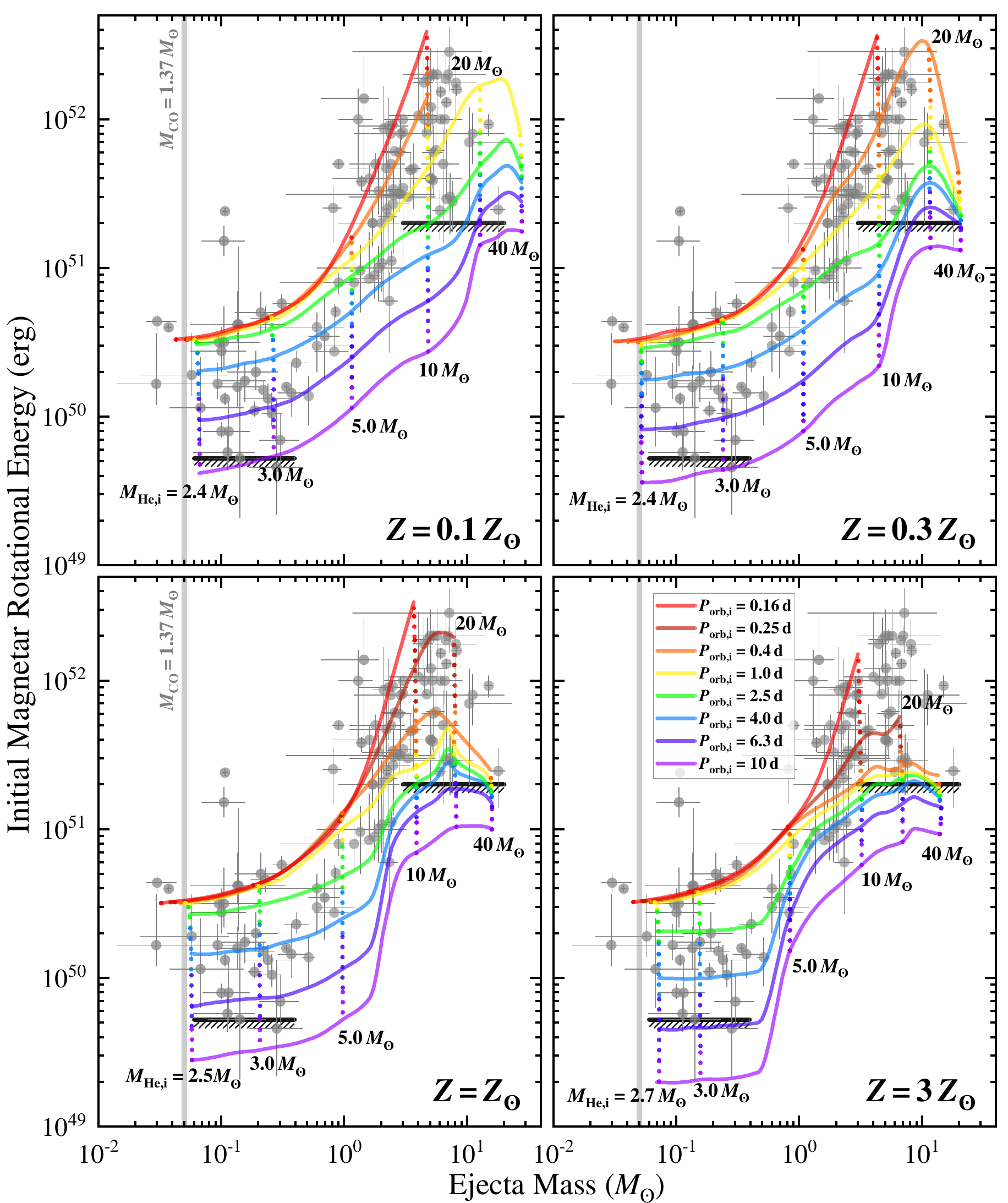}
\caption{{\bf Dependence of simulated $E_{\rm{rot,i}}-M_{\rm{ej}}$ relationships on metallicity.} The colored solid lines (as marked in the color scheme in the inset) correspond to different $P_{\rm{orb,i}}$, and the vertical dashed lines correspond to different $M_{\rm{He,i}}$ values as labeled. We consider four different metallicity environments to evolve helium stars in close binaries, including $0.1\,Z_\odot$ (top left panel), $0.3\,Z_\odot$ (top right panel), $Z_\odot$ (bottom left panel), and $3\,Z_\odot$ (bottom right panel). The gray points represent the observations between $M_{\rm{ej}}$ and $E_{\rm{rot,i}}$ for four types of magnetar-driven SESNe. The light gray line shows the estimated ejecta mass for a CO core with $M_{\rm{CO}}=1.37\,M_\odot$. Black solid lines at $\sim5\times10^{49}\,{\rm{erg}}$ are the lower limit of neutrino-powered energy from core-collapse of low-mass helium stars based on numerical simulations\cite{suwa2015}, while black solid lines at $\sim2\times10^{51}\,{\rm{erg}}$ are the maximum neutrino-powered energy of the original core-collapse SNe\cite{sukhbold2016}.}
\label{fig:Dependence_Metallicity_MainText}
\end{figure*}

\clearpage

\begin{figure*}
\centering
\includegraphics[width=0.75\linewidth, trim = 70 15 100 20, clip]{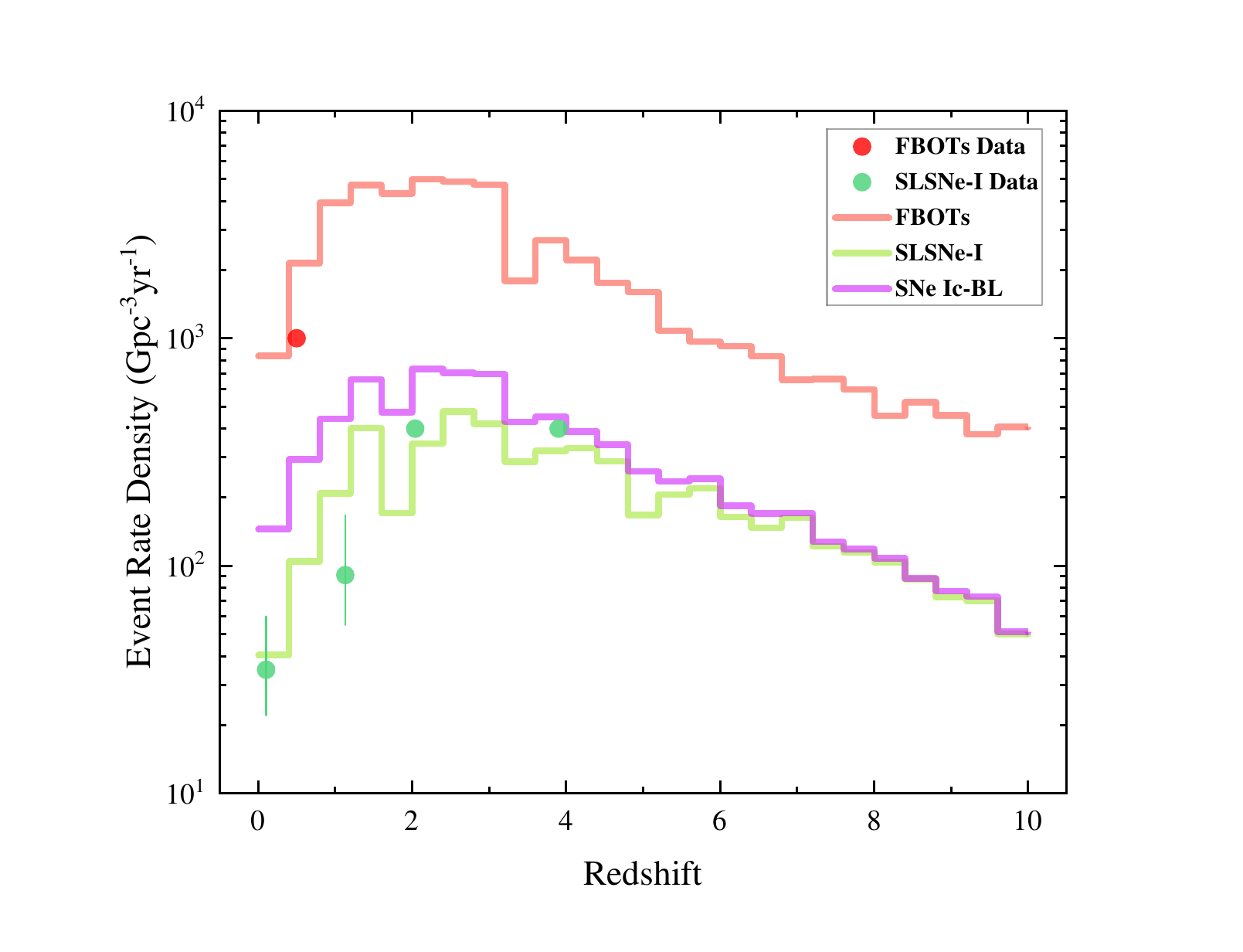}
\caption{{\bf Comparisons between observed and simulated redshift-dependent rate densities of magnetar-driven SESNe.} The red and green points represent the observed event rate densities of FBOTs\cite{wiseman2020} and SLSNe-I\cite{galyam2019} respectively. The predicted rate densities of FBOTs, SLSNe-I, and SNe Ic-BL are marked as light red, light green, and purple lines, respectively. Here, the CE efficiency is set as $\alpha_{\rm{CE}}=5$.}
\label{fig:EventRate_MainText}
\end{figure*}

\clearpage

\begin{figure*}
\centering
\includegraphics[width=0.75\linewidth, trim = 85 90 110 120, clip]{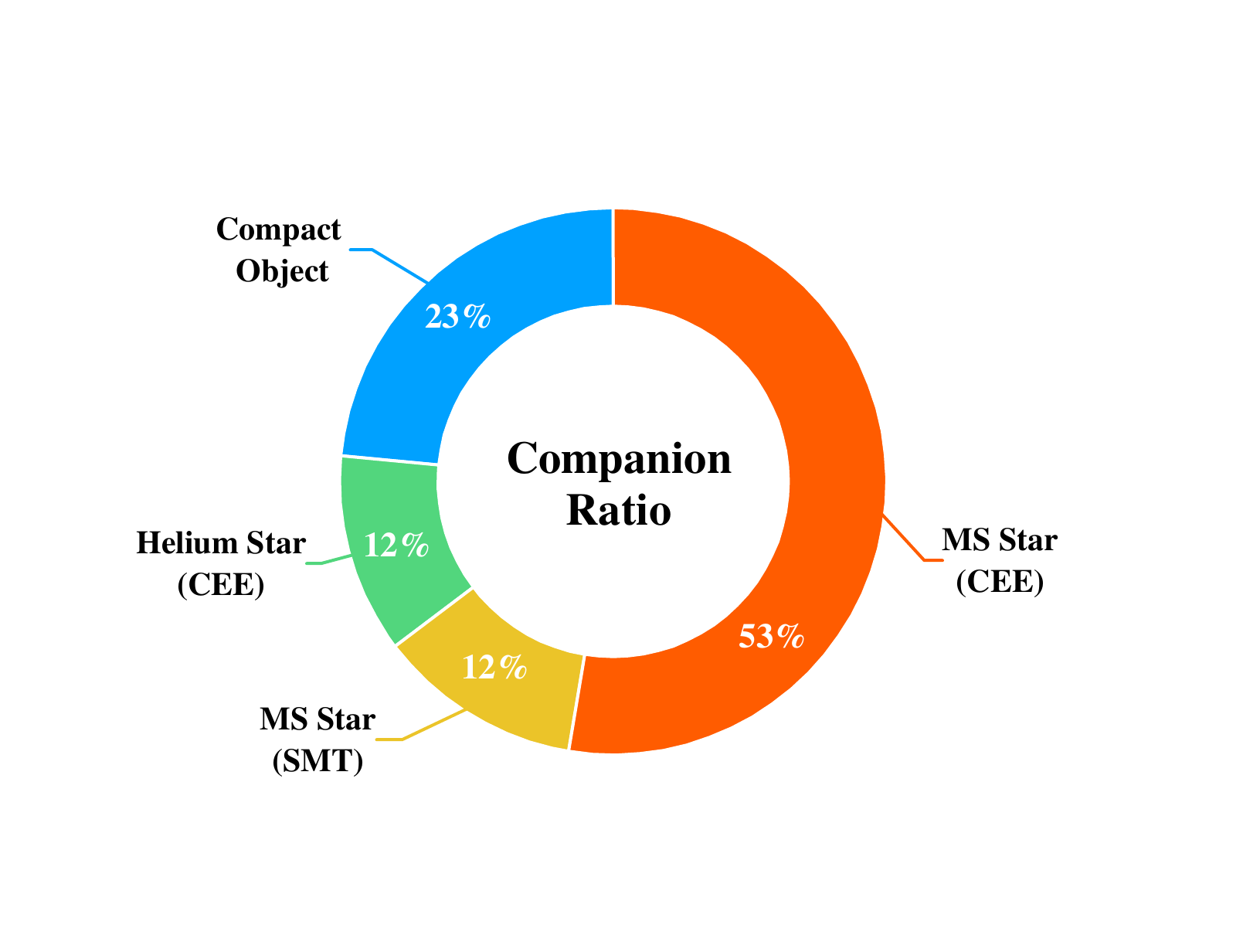}
\caption{{\bf Possible companions at the moment of magnetar-driven SESNe.} CEE and SMT channels with MS companions at the time of magnetar-driven SESNe are shown in orange and yellow segments, respectively. Helium star companion is marked as green segment, while blue segment indicate progenitors with compact-object companion. {Here, the CE efficiency of $\alpha_{\rm{CE}}=5$ is adopted.}}
\label{fig:Companions_MainText}
\end{figure*}

\begin{figure*}
\centering
\includegraphics[width=0.75\linewidth, trim = 0 0 0 0, clip]{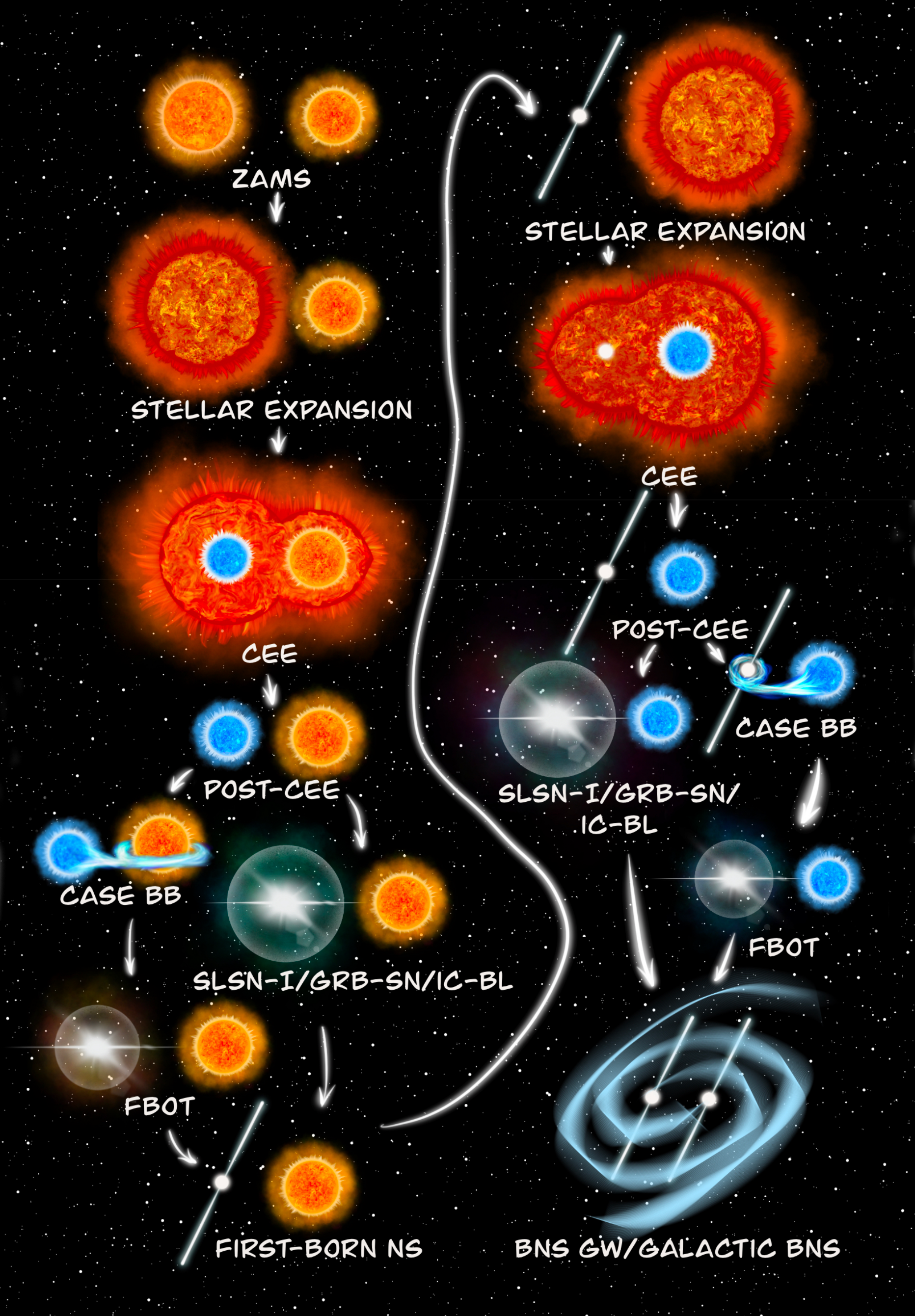}
\caption{{\bf Illustration of the formation channel of magnetar-driven SESNe \textcolor{cyan}{and binary NS systems}.} Magnetar-driven SESNe could be the first and/or second SNe in binary systems. These systems could finally leave behind close-orbit NS binaries if the systems are not disintegrated by the SN kicks.   }
\label{fig:illustration}
\end{figure*}

\clearpage

\begin{methods}

\subsection{Data collection of magnetar-driven SESNe.}

{The newborn magnetars can transfer rotational energy into heating and accelerating the ejecta, producing powerful and bright SN emissions if the spin-down timescale is comparable to the effective diffusion time through the SN ejecta\cite{kasen2010}.} The initial magnetar rotational period $P_{\rm{rot,i}}$ and ejecta mass $M_{\rm{ej}}$ of each SLSN-I and FBOT can be estimated by modeling their lightcurves as well as temperature and velocity evolutions assuming SN emission is powered by dipole spin-down of a rapidly rotating magnetar\cite{kasen2010}. We collect the $P_{\rm{rot,i}}$ and $M_{\rm{ej}}$ distributions of SLSNe-I from ref.\cite{yu2017}{, which are also consistent with those of ref.\cite{nicholl2017,blanchard2020,chen2023}}. The initial rotational energy of magnetar is calculated by $E_{\rm{rot,i}}=0.5I_{\rm{mag}}(2\pi/P_{\rm{rot,i}})^2$, where the magnetar moment of inertia is set as $I_{\rm{mag}}=10^{45}\,{\rm{g}}\,{\rm{cm}}^2$. 

The $E_{\rm{rot,i}}$ and $M_{\rm{ej}}$ distributions of FBOTs have been calculated by ref.\cite{liu2022} recently. However, the ejecta mass $M_{\rm{ej}}$ of FBOTs might be underestimated, because neutrino-powered energy $E_\nu$ is ignored when calculating FBOT ejecta masses\cite{liu2022}. By considering an initial kinetic energy $E_\nu\sim10^{50}\,{\rm{erg}}$ derived from neutrinos\cite{suwa2015} and setting a constant opacity $\kappa=0.07\,{\rm{cm}}^2{\rm{g}}^{-1}$, we use {the software \texttt{emcee}\cite{foremanmackey2013}} to refit the lightcurve data of FBOTs collected in ref.\cite{liu2022} based on the magnetar-driven model\cite{kasen2010}. We list three main fitting parameters of 36 FBOTs, including the ejecta mass $M_{\rm{ej}}$, initial magnetar rotational energy $E_{\rm{rot,i}}$ and polar magnetic field strength $B_{\rm{p}}$, in Supplementary Information Table \ref{table:FBOT}, in which the results of $E_{\rm{rot,i}}$ {and $B_{\rm{p}}$} are nearly consistent with those shown in ref.\cite{liu2022}. One can also refer to Figures 4 and 5 in ref.\cite{liu2022} for our multi-band fitting lightcurves and posteriors of these 36 collected FBOTs. Four events, i.e., DES14S2anq, DES14S2plb, DES15C3mgq, and PS1-12bb are ignored, since their lightcurves cannot be explained by the magnetar engine model.

{Different from FBOTs and SLSNe-I, the energy released from the magnetars of GRB-SNe and SNe Ic-BL is mainly used to accelerate the ejecta rather than to partially power the SN emission due to more rapid energy injection of the mangetars (e.g., ref.\cite{piro2011,yu2017}). Thus, GRB-SN and SN Ic-BL emissions might be predominantly $^{56}$Ni-powered.} The explosion energy and ejecta mass for each GRB-SN and SN Ic-BL can be estimated by fitting the lightcurves and photospheric velocity profiles near peak with the semianalytic Arnett model\cite{arnett1982}. The $E_{\rm{rot,i}}$ values for GRB-SNe and SNe Ic-BL are taken from ref.\cite{lv2018} and refs.\cite{lyman2016,taddia2019}, respectively, by assuming that the SN explosion energy $E_{\rm{SN}}$ is derived from the rotational energy of the magnetar. This assumption leads to a slight overestimation for $E_{\rm{rot,i}}$, since the explosion energy also include neutrino-powered energy which is secondary compared with the magnetar's energy injection. Our collected data of $E_{\rm{rot,i}}$ and $M_{\rm{ej}}$ are summarized in Supplementary Information Table \ref{table:SLSN}. {We note that the data uncertainties of GRB-SNe and SNe Ic-BL are mainly derived from the observational measurements, while the data uncertainties of SLSNe-I and FBOTs also include the effect of model fitting.}

The combined distributions between $E_{\rm{rot,i}}$ and $M_{\rm{ej}}$ of these magnetar-driven SESNe, displayed in Figure \ref{fig:M_ej_E_rot_MainText}, have a Pearson correlation coefficient of $\rho=0.834$, indicating a very strong universal correlation.

\subsection{{Data collection of event rate densities.}}

{The observed event rate densities of SLSNe-I and FBOTs we use are directly collected from the literature.}

{For low-redshift SLSNe-I, ref.\cite{quimby2013} derived a rate density of $\mathcal{R}_{\rm{SLSN-I}}\sim35^{+77}_{-26}\,{\rm{Gpc}^{-3}\,{yr}^{-1}}$ at a redshift of $z\approx0.17$ based on one event discovered with the Robotic Optical Transient Search Experiment-IIIb telescope between 2004 and 2009, while ref.\cite{frohmaier2021} derived a rate of $\mathcal{R}_{\rm{SLSN-I}}\sim35^{+25}_{-15}\,{\rm{Gpc}^{-3}\,{yr}^{-1}}$ out to $z\approx0.2$ by using the detection of 8 events discovered by the Palomar Transient Factory from 2010 to 2012. We use the later one because of its lower errors. Ref.\cite{prajs2017} reported that three identified SLSNe-I over the duration of the Supernova Legacy Survey at $0.2<z<1.6$ could give a rate density of $\mathcal{R}_{\rm{SLSN-I}}\sim91^{+76}_{-36}\,{\rm{Gpc}^{-3}\,{yr}^{-1}}$ at a volume-weighted redshift of $z=1.13$. Further, ref.\cite{cooke2012} reported two SLSNe without spectral confirmation, i.e., SN\,2213-1745 at $z=2.05$ and SN\,1000+0216 at $z=3.90$, to estimate the high-redshift rate density of $\mathcal{R}_{\rm{SLSN-I}}\sim400\,{\rm{Gpc}^{-3}\,{yr}^{-1}}$ with a large error. These two SLSNe could still be hydrogen-poor, since the lack of Ly$\alpha$ emission of the former provided the spectroscopic evidence against a hydrogen-rich SLSN classification while the evolution of the later event was similar to that of hydrogen-poor SLSN\,2007bi\cite{cooke2012}.}

{The local FBOT rate density of $\mathcal{R}_{\rm{FBOT}}\sim4800-8000\,{\rm{Gpc}^{-3}\,{yr}^{-1}}$ ($\sim4\%-7\%$ of the core-collapse SN rate density) at $z\approx0.2$ was first measured by ref.\cite{drout2014} for the Pan-STARRS1 Medium Deep Survey sample of 12 FBOTs. Ref.\cite{pursiainen2018,wiseman2020} constrained a local FBOT rate density to be $\mathcal{R}_{\rm{FBOT}}\gtrsim1000\,{\rm{Gpc}^{-3}\,{yr}^{-1}}$ (i.e., $\sim1\%$ of the core-collapse SN rate density), estimated based on a sample of 37 events at a redshift of $0.05<z<1.56$ in the Dark Energy Survey. Since the Dark Energy Survey considered more FBOT events, we used their constrained value as the local FBOT rate density.}

{In the literature, present observed constraints on the SN Ic-BL rate density are relatively ambiguous with a large error, e.g., $3.7^{+2.9}_{-3.7}\%$ of SESNe\cite{shivvers2017} which was obtained based on only one discovered event in the Lick Observatory Supernova Search volume-limited sample. Thus, we do not think present SN Ic-BL rate density is representative to compare with our simulated rate density. }

\subsection{Detailed binary evolution modeling.}

We perform detailed binary evolution modeling using {the released version \texttt{mesa-r15140} of} the Modules for Experiments in Stellar Astrophysics (\texttt{MESA}) stellar evolution code\cite{Paxton2011,Paxton2013,Paxton2015}. We first create pure single helium stars at the zero-age helium MS following the same method of ref.\cite{qin2018,hu2022} and then relax the created helium stars to reach the thermal equilibrium, where the helium-burning luminosity just exceeds $99\%$ of the total luminosity. {The initial helium stars are assumed to be tidally locked.} We model convection using the mixing-length theory\cite{MLT1958} with a mixing-length $\alpha_{\rm mlt}=1.93$. We adopt the Ledoux criterion to treat the boundaries of the convective zone and consider the step overshooting as an extension given by $\alpha_p=0.1H_p$, where $H_p$ is the pressure scale height at the Ledoux boundary limit. Semiconvection\cite{Langer1983} with an efficiency parameter $\alpha_{\sc}=1.0$ is included in our modeling. The network of {\texttt{approx21.net}} is chosen for nucleosynthesis. {We adopt the ``\texttt{Dutch}'' scheme for mass loss\cite{Nugis2000,Glebbeek2009}, calibrated by multiplying with a \texttt{Dutch$_{-}$scaling$_{-}$factor} = 0.667 to match the recently updated modeling of helium stars' winds\cite{Higgins2021}.} We treat rotational mixing and AM transport as diffusive processes\cite{Heger2000}, including the effects of the Goldreich–Schubert–Fricke instability, Eddington–Sweet circulations, as well as secular and dynamical shear mixing. We adopt diffusive element mixing from these processes with an efficiency parameter of $f_c=1/30$\cite{Heger2000}.  We consider the Eddington-limited accretion rate of mass transfer from a helium star onto its companion. {Beyond this Eddington limit, the mass lost via Roche lobe overflow is 100\% ejected.} {Tides play a critical role in the evolution of the orbit and the internal AM of the helium stars in close binary systems. In this study, we adopt the dynamical tides for helium stars and we kindly refer readers of interest to more details in Ref. \cite{qin2018}.}

{For AM transport within stars, the Tayler-Spruit dynamo\cite{Spruit2002} (TS dynamo), produced by differential rotation in the radiative layers, has been considered as the main candidate for efficient coupling between the stellar core and its outer layers. Assume that the TS dynamo occurred in the stars. It is expected that most of the internal AM would be lost via wind mass loss and/or mass transfer onto its companion. However, the stellar core that could potentially form into an NS or a BH can have AM replenished through tidal interaction with its companion in a close orbit. Therefore, the AM of the progenitor of compact objects is determined by the interplay between the tides and the efficiency of the internal AM transport. We note that a revised version of the TS dynamo, proposed recently by ref.\cite{fuller2019}, would predict much more efficient AM transport, resulting in even lower rotation rates of resultant compact objects. As a comparison, the revised version of TS dynamo is included in our modeling.}

Following ref.\cite{qin2018} which presented that the companion mass has little effect on the final mass of the helium star and the rotational energy of the remnant, we thus only employ \texttt{MESA} to perform detailed binary evolution of helium stars and $1.4\,M_{\odot}$ NSs in close orbits. {In addition to an NS as the companion star, we also take into account other types of companions, mostly including MS star, helium star and BH, to compute the event rates using a rapid population synthesis (see more details in the next section). The radii of MS and helium star companions remain almost the same during the late evolution of primary helium stars, so the mass transfer is avoided. Therefore, MS and helium star companions are expected to have a similar effect on the tidal interaction of the the primary helium stars, which results in quantitatively comparable properties (ejecta mass and final rotational energy).} For the parameter space considered in this study (see \ref{fig:MassTransferCase_Supplement}), we cover the helium star mass in a range of $2.3-40\,M_{\odot}$ (linear variation from 2.3 to $3.1\,M_{\odot}$, linear variation in logarithmic space from 3.2 to $40\,M_{\odot}$) and the orbital period from $0.04 -10\,{\rm{d}}$ (linear variation in logarithmic space). In \ref{fig:MassTransferCase_Supplement}, we present various interactions of our modeling for different initial values of helium star mass and orbital period (see detailed descriptions in Supplements). 

{The numerical simulation results by ref.\cite{AguileraDena2023} suggest that most $\lesssim40\,M_\odot$ helium stars may form NSs, supporting the observed mass range of magnetar-driven SNe. However, there are also some numerical simulations in the literature exploring the final remnant after the core-collapse of helium stars suggest that a certain mass range of high-mass helium stars, which we are interested in, may ultimately allow for the formation of BH rather than NS (e.g., Ref.\cite{ertl2020,schneider2021,Heger2023}). The results of these numerical simulations, revealing the mass range for BH formation, may not be applicable to magnetar-driven SNe. This is because the magnetar central engine in our studied magnetar-driven SN explosions, which is not considered in the previous numerical simulations, could have an injection energy higher than the neutrino-powered energy and, hence, may suppress BH formation. Thus, we simply assume that all helium stars can form an NS upon core collapse in our study to align with observations of high-mass magnetar-driven SNe and then develop an empirical method below to estimate the NS baryonic mass. } After the carbon depletion in the center of the helium star, we terminate the evolution of the helium star to estimate the mass of the final CO core $M_{\rm{CO}}$. For a final CO core with a mass of $M_{\rm{CO}}\lesssim2\,M_\odot$, the NS baryonic mass is approximately linear with the core mass, which is $M_{\rm{NS,b}}\approx0.45M_{\rm{CO}}+0.69$, based on numerical simulations\cite{suwa2015}, {where $M_{\rm CO}$ and $M_{\rm NS,b}$ are in units of $M_\odot$}. While for a helium star with a much larger final core mass of $M_{\rm{CO}}\gg2\,M_\odot$, we define a maximum NS baryonic mass as $M_{\rm{NS,b}}^{\rm{max}}=2.36\,M_\odot$ corresponding to a maximum NS gravitational mass of $M_{\rm{NS,g}}^{\rm{max}}=2\,M_\odot$ if considering an APR NS equation of state\cite{worley2008}. We thus construct an empirical equation between $M_{\rm{CO}}$ and $M_{\rm{NS,b}}$, i.e., $M_{\rm{NS,b}} = \left[(0.45M_{\rm{CO}}+0.69)^{-\eta}+(M_{\rm{NS,b}}^{\rm{max}})^{-\eta}\right]^{-1/\eta},$ where $\eta=5$. Assuming AM conservation during the core collapse, we calculate the initial magnetar rotational energy by $E_{\rm{rot,i}}=0.5I_{\rm{mag}}(2\pi/P_{\rm{rot,i}})^2$, where the magnetar rotation period is $P_{\mathrm{rot,i}}=2\pi{I}_{\mathrm{mag}}/J$ with $I_{\mathrm{mag}}$ taken from ref.\cite{worley2008} by applying an APR NS equation of state and AM $J$ contained within the inner mass $M_{\rm{NS,b}}$ given by our modeling in \texttt{MESA}. Here, $I_{\mathrm{mag}}= (M_{\mathrm{NS,g}}-0.1)\times 10^{45}\,{\rm g}\,{\rm cm}^2$ where $M_{\mathrm{NS,g}}$ is the NS gravitational mass. {We convert the baryonic mass of the NS to its gravitational mass using the empirical formula of $M_{\rm NS,b} = M_{\rm NS,g} + A\times M_{\rm NS,g}^2$, where $A=0.08$ is adopted based on the latest fitting results from ref.\cite{gao2020}. Thus, the formula can be further expressed as $M_{\mathrm{NS,g}}=[(1+0.32M_{\mathrm{NS,b}})^{1/2}-1]/0.16$.} 

Due to the absence of hydrogen and most of helium envelopes when the explosions of SLSNe-I, GRB-SNe, SNe Ic-BL, and partial FBOTs, their ejecta mass is estimated as the difference value between the final mass of CO core and NS baryonic mass, i.e., $M_{\rm{ej}}=M_{\mathrm{CO}}-M_{\mathrm{NS,b}}$. We note that the final CO core for low-mass helium stars with a large initial orbital period of $P_{\rm{orb,i}}\gtrsim2\,{\rm{d}}$ might be surrounded by a non-negligible helium envelope\cite{tauris2015}. {However, most of our FBOT sample\cite{pursiainen2018} has peak temperatures mainly distributed in the range of $\sim10,000-15,000\,{\rm K}$, corresponding to the measured temperature at the photosphere and, hence, the outer helium  would have a lower temperature. Given that the helium cannot be ionized and becomes effectively transparent if the temperature drops below $\sim13,000\,{\rm K}$\cite{piro2014}, it is expected that the helium does not contribute to the shaping of the main FBOT lightcurve and the ejecta mass obtained from our FBOT lightcurve modeling mostly reflects the difference in mass between the CO core and the NS. Thus, we still use $M_{\rm{ej}}=M_{\mathrm{CO}}-M_{\mathrm{NS,b}}$ to estimate the ejecta mass of FBOTs. }

\subsection{Binary Population Synthesis Simulation.}
We use the \texttt{BSE} code\cite{Hurley2002} to follow binary evolutionary tracks until the formation of the binary systems containing a helium star, and estimate the event {rate density} of magnetar-driven SESNe according to our obtained parameter distributions of orbital periods and helium star masses. This code has been updated in many aspects including the treatments of stellar winds, mass-transfer processes, CEE, and SN-explosion mechanisms\cite{sl2014,sl2021}. {In the community of binary evolution, it is known that the treatments of mass-transfer processes and CEE are still subject to big uncertainties. For the mass transfer from the more-massive primary to the less-massive secondary during the evolution of primordial binaries initially containing two zero-age MS stars, we consider three models with different mass-transfer  efficiencies $\beta$ (i.e., the fractions of the accreted matter by the secondary stars among all the transferred matter) \cite{sl2014}. In Model I, which we treat as the default mass-transfer model, the mass-transfer efficiency is assumed to be dependent on the rotational velocity of the secondary star with $\beta = 1-\Omega/\Omega_{\rm cr}$, where $\Omega$ is the angular velocity of the secondary star and $\Omega_{\rm cr}$ is
its critical value. In Model II, it is
assumed that the secondary star accretes half of the transferred
matter with $\beta=0.5$. In Model
III, the mass-transfer efficiency is set to be $\beta = \min [10(\tau_{\dot{M}}/\tau_{\rm KH}),1]$, where $\tau_{\dot{M}}$ is the mass-transfer timescale and $\tau_{\rm KH}$ the thermal timescale of the secondary star \cite{Hurley2002}. We assume that the escaped matter from the binary system carries away the specific orbital AM of the secondary star. Under different mass-transfer models, the criteria of mass-transfer stability, that determining whether the binaries undergo SMT or CEE, have been obtained and incorporated into the \texttt{BSE} code\cite{sl2014}.} During the CEE we adopt the $\alpha_{\rm CE}\lambda$ prescription\cite{Webbink1984} to deal with binary orbital decay, taking two different CE ejection efficiencies of $\alpha_{\rm CE}=1$ and $\alpha_{\rm CE}=5$\cite{Fragos2019}. 

Each simulation contains $N = 2\times10^7$ binary systems that begin the evolution from the primordial binaries with two zero-age main sequence stars. For the parameter configurations of primordial binaries, we assume that the mass $M_{1}$ of the primary stars obeys the Kroupa's initial mass function (i.e., $\xi(M_{1})\propto M_{1}^{-2.7}$)\cite{Kroupa1993}, the mass $M_{2}$ of the secondary stars follows a flat distribution between 0 and $M_{1}$\cite{Kobulnicky2007}, and the orbital separation $a$ is uniformly distributed in the logarithmic space\cite{Abt1983}. In our simulations, we take $M_{1}$ in the range of $5-100M_{\odot}$, $M_{2}$ in the range of $1-100M_{\odot}$, and $a$ in the range of $3-10000R_{\odot}$. All these parameters are randomly chosen in a 3-dimensional logarithmic space, so $\Phi(\ln M_{1}) = M_{1}\xi(M_{1})$, $\varphi(\ln M_{2}) = M_{2}/M_{1}$ and $\Psi(\ln a) = \rm const$ (see also ref.\cite{Hurley2002}). If a specific primordial binary $j$ evolves to become a helium-star binary, it contributes the helium-star binary population with a rate density of 
\begin{equation}
    \mathcal{R}_{\rm j} = \left( \frac{f_{\rm bin}}{2}\right) \left( \frac{\mathcal{SFR}}{M_{\rm av}}\right)\Phi(\ln M_{1})\varphi(\ln M_{2})\Psi(\ln a)\delta \ln M_{1}\delta \ln M_{2}\delta \ln a,
\end{equation}
where $f_{\rm bin}$ is the binary fraction among all stars, $\mathcal{SFR}$ is the cosmic star-formation rate density, $M_{\rm av}\sim 0.5M_{\odot}$ is the average mass of all stars, and $\delta (\ln \chi) = (\ln \chi_{\rm max}-\ln \chi_{\rm min})/N^{1/3}$ for each binary parameter $\chi$ (see ref.\cite{sl2021} for more details).
We simply assume that all stars are formed in binaries, i.e., $f_{\rm bin} = 1$. Since $\gtrsim 60\%$ of massive stars as magnetar's progenitors are observed as members of binary systems\cite{Sana2012,Moe2017}, this assumption may lead to an overestimation of the calculated event rate by a factor of less than 2.  For stars formed at different redshifts $z$, we adopt an updated determination of the cosmic star formation history and metallicity evolution\cite{Madau2017}. The star-formation rate density is given by 
\begin{equation}
    \mathcal{SFR}({\rm z}) = 0.01\frac{(1+z)^{2.6}}{1+[(1+z)/3.2]^{6.2}}M_{\odot}\rm\,yr^{-1}\,Mpc^{-3},
\end{equation}
which improves the previous formula of ref.\cite{Madau2014} by better reproducing the recent results at $4 < z < 10$. In addition, the mean metallicity $\langle Z\rangle$ has the form of
\begin{equation}
    \log \langle Z/Z_{\odot}\rangle = 0.153-0.074z^{1.34}.
\end{equation}
Following ref.\cite{Belczynski2020}, it is  assumed that $\log (Z/Z_{\odot})$ obeys a Gaussian distribution centered at the mean metallicity with a dispersion $\sigma = 0.5\rm\,dex$. The metallicity of each primordial binary is randomly taken in the logarithmic space.

As shown in Figure \ref{fig:Dependence_Metallicity_MainText}, the close-orbit helium stars formed at super-solar metallicity environments can hardly generate fast-spinning magnetars that have an initial rotational energy much larger than the neutrino-powered energy. Magnetar-driven SESNe could be rare at $Z\gtrsim2-3\,Z_\odot$ both in observations and simulations and, hence, we set the upper limit of metallicity for our population synthesis simulations as $\sim2\,Z_\odot$. Most of the magnetar-driven SESNe were observed at environments of $\sim0.1\,Z_\odot$ (e.g., ref.\cite{japelj2018,lunnan2014,chen2017,wiseman2020,modjaz2020}). Helium stars in close binaries with lower metallicity environments can result in the similar $E_{\rm rot}-M_{\rm ej}$ relationship with those formed at $0.1\,Z_\odot$ (see \ref{fig:Dependence_Metallicity_Supplement}) and, hence, the core-collapses of these helium stars can still lead to the formation of high-spin magnetars and magnetar-driven SESNe in the distant high-redshift universe. For our population synthesis simulations, in order to cover the stars formed in high-redshift distance range, we use $0.005\,Z_\odot$ as the simulated lower metallicity. After the simulations, we record relevant information for the binary systems with a helium star, including the companion type, the component masses, the orbital period, and the metallicity. To calculate a specific event rate {density} of magnetar-driven SESNe in each redshift bin, the parameter distribution of the primordial binaries {with different metallicities} has been normalized to unity. Our {simulations reveal} that the vast majority of the (close) helium-star binaries host an MS companion.

\end{methods}

\begin{addendum}

 \item [Data Availability] The data from our detailed binary evolution simulations and population synthesis simulations stored in \texttt{Zenodo} (\url{https://zenodo.org/records/13984623}). The data of magnetar-driven SESNe are listed in Supplementary Information Tables \ref{table:SLSN} and \ref{table:FBOT}.
 
 \item [Code Availability] The inlists and input files to reproduce our detailed binary evolution simulations are available at \url{https://zenodo.org/records/13984623}. The specific codes are available upon request to the corresponding author.

\end{addendum}

\clearpage

\begin{supplinfo}

\renewcommand{\thefigure}{Supplementary Information Figure \arabic{figure}}
\renewcommand{\figurename}{}
\setcounter{figure}{0}

We note that if not mentioned otherwise, all simulations in Supplementary Information are computed under the metallicity environment of $Z=0.3\,Z_\odot$.

\section{{Detailed stellar and binary evolution}}\label{sup_sec:mass_transfer_case}

{In \ref{fig:RadiusAge_Supplement}, we show the evolution of helium star radius with different initial masses. For less massive ($\lesssim 8\,M_{\odot}$) helium stars, they will become giant stars due to their envelope inflation after the central helium is exhausted. This implies that less massive helium stars in binary systems will undergo mass exchanges in initially wide orbits. We first present in \ref{fig:MassTransferRate_Supplement} three binary sequences of a 3 $M_{\odot}$ helium star with an NS star as its companion in initially different orbital periods. First, starting with an orbital period of 0.063\,${\rm{d}}$, the mass transfer initiates during core helium burning (case BA mass transfer). {Due to the steadily increasing mass transfer rate, we terminate the evolution when this system triggers dynamical delayed instability\cite{Ivanova2003}, as it is more likely to experience a common envelope or merge.} Second, in a slightly wider orbit ($P_{\rm orb,i}$ = 0.16\,{\rm{d}}), the system will delay the mass transfer phase due to later helium-star's radius expansion during the shell helium burning phase (case BB mass transfer). {Mass transfer during the late evolutionary stage (i.e., Case BB) can significantly impact the AM of the helium star. This is because a helium star expands as it starts to burn helium in the shell (see the third panel in Figure 2 in Supplementary). In particular, a lower-mass helium star expands more significantly due to weaker wind mass-loss rates. Therefore, mass lost via mass transfer removes more AM of helium stars' core through strong coupling with the envelope (i.e., efficient AM transport).} Third, for $P_{\rm orb,i}=1.0\,{\rm{d}}$, mass transfer of the system starts to occur after the shell helium burning phase (case BC mass transfer).}

Various interactions of a helium star with its companion NS for different initial-parameter configurations are presented in \ref{fig:MassTransferCase_Supplement}. For binaries in very tight orbits (e.g., $P_{\rm{orb,i}}\lesssim0.06\,{\rm{d}}$), the helium star overflows its Roche lobe through the first Lagrangian point ($L_1$). In a slightly wider orbit ($0.06\,{\rm{d}}\lesssim P_{\rm{orb,i}}\lesssim0.2\,{\rm{d}}$), {the binary systems are expected to undergo Case BA mass transfer.} More massive helium stars tend to overflow their second Lagrangian points ($L_2$), while for low-mass helium stars the mass transfer is dynamically unstable, leading to binary merge. For systems with lower-mass ($\lesssim10.0\,M_{\odot}$) helium stars and initially short orbits, binary systems will go through Case BB (green circles) or Case BC (yellow circles) mass transfer phase, while no mass transfer phase is expected for more massive helium stars in much wider orbits. 

{In this work, we explore the detailed evolution of naked helium stars, assuming no residual hydrogen remains from the CEE or SMT phases. However, some numerical simulations suggest that the hydrogen-rich envelope of expanding stars may not be fully removed during the CEE and SMT phases, especially for stars at low metallicity\cite{Ivanova2011,Gotberg2020,Ivanova2020,Klencki2022,VignaGomez2022}. Compared to naked helium star models, the residual hydrogen-envelope may enhance the parameter space of Case BB/BC\cite{laplace2020} or even lead to a prolonged case BA in wider binaries with periods $>1\,{\rm d}$\cite{Klencki2022}. The newborn magnetar formed from helium stars with a hydrogen envelope can have a lower initial rotational energy relative to the one formed from a naked helium star, as the hydrogen envelope can carry a certain amount of AM within the star when it is lost. }

\section{Parameter Dependence} \label{sup_sec:parameter_dependence}

\subsection{AM Transport Mechanism.}

In \ref{fig:Dependence_Model_Supplement}, we show a comparison of the initial magnetar rotational energies by considering two kinds of AM transport mechanisms during the magnetar formation: an efficient transport by the TS dynamo\cite{Spruit2002} we adopt in this work and a very efficient transport by the revised TS dynamo\cite{fuller2022}. {In \ref{fig:Dependence_Jin_Model_Supplement}, we present a comparison of the AM distribution at different phases of the helium star under two AM transport mechanisms, with all other assumptions kept the same. We find that the magnetars commonly have much slower initial magnetar rotational energies if a very efficient AM transport mechanism\cite{fuller2019} is adopted. After helium shell burning, the later helium-star's envelope expansion and the mass transfer of the low-mass helium stars can extract AM, leading to a nearly non-rotating helium envelope and a rotating core. However, by considering a very efficient AM transport mechanism, due to the tight coupling between the core and the envelope for the very efficient AM transport mechanism, the core AM can be quickly extracted and tends to be similar when the helium stars explode.} A very efficient AM transport mechanism predicts much lower initial rotational energies of magnetars formed after the core-collapses of low-mass helium stars than those of observationally inferred FBOT magnetars. However, $\gtrsim6-7\,M_\odot$ helium stars with a very efficient AM transport mechanism in very close binaries with $P_{\rm orb,i}\lesssim1\,{\rm d}$ can still generate fast-spinning magnetars to explain a fraction of SLSNe-I, GRB-SNe, and SNe Ic-BL. {However, due to the very efficient AM transport, which leads to faster AM loss, tidal forces are insufficient to maintain the rotation of the helium stars in wider orbits. As a result, the final core AM is lower compared to the scenarios invoking efficient AM transport. }

\subsection{Metallicity.} As shown in Figure \ref{fig:Dependence_Metallicity_MainText}, with consideration of four different metallicity environments, i.e., $Z=\{0.1,0.3,1,3\}\,Z_\odot$, the $E_{\rm{rot,i}}-M_{\rm{ej}}$ relationships are simulated by evolving helium stars in binary systems. Due to the higher mass loss rate in a higher-metallicity environment, the pre-SN helium stars can generally eject {more} materials and lose more AM inside the cores to generate NSs with slower spins. 

If $M_{\rm{He,i}}\gtrsim5\,M_\odot$, the parameter space of the initial orbital period that allows the formation of fast-spinning magnetars to explain the explosions of SLSNe-I, GRB-SNe, and SNe Ic-BL would reduce significantly for $Z\gtrsim{Z}_\odot$. In an environment with $Z=3\,Z_\odot$, only helium stars in very close binaries with $P_{\rm{orb,i}}\lesssim{0}.25\,{\rm{d}}$ can finally generate fast-spinning magnetars that have an initial rotational energy much larger than the neutrino-powered energy. Thus, in super-solar metallicity environments, the explosions of magnetar-driven SESN would be heavily suppressed. {It is worth noting that the magnetars formed from helium stars at higher metallicities (i.e., $Z\gtrsim{Z}_\odot$) with initial masses of $5\,M_\odot\lesssim{M}_{\rm{He,i}}\lesssim10\,M_\odot$ would have faster spins with increasing metallicity. This is because high-metallicity helium stars in this mass range are less compact compared to the ones with initial lower metallicities, which induces less efficient coupling and thus allows them to retain more AM inside their cores.} 

We now discuss the effect of metallicity on helium stars with initial masses of $M_{\rm{He,i}}\lesssim5\,M_\odot$. For a final carbon-oxygen (CO) core with $M_{\rm{CO}}\lesssim1.37\,M_\odot$, the star would leave behind an oxygen–neon-magnesium WD or a CO WD (e.g., ref.\cite{tauris2015}) instead of forming fast-spinning magnetars to power FBOTs. Higher metallicity usually results in final CO cores with lower mass, so that the final fate of more low-mass helium stars would not be NSs rather than WDs. For example, as shown in Figure \ref{fig:Dependence_Metallicity_MainText}, helium stars with $M_{\rm{He,i}}\lesssim2.4\,M_\odot$ are more likely to form WDs if $Z\lesssim0.3\,Z_\odot$, while the floor of $M_{\rm{He,i}}$ that lead to the formation of NS increase to $\sim2.7\,M_\odot$ in a metallicity environment of $Z=3\,Z_\odot$. We find that the $E_{\rm{rot,i}}-M_{\rm{ej}}$ relationships show slight changes for helium stars in close binaries with $P_{\rm{orb,i}}\lesssim1\,{\rm{d}}$ as metallicity increases. In low-metallicity environments with $Z<0.3\,Z_\odot$, FBOTs can be achievable in binary systems with $P_{\rm{orb,i}}\lesssim10\,{\rm{d}}$. However, for those systems with $P_{\rm{orb,i}}\gtrsim6-10\,{\rm{d}}$, we find that magnetars from solar and super-solar metallicity systems can hardly have enough rotational energies that significantly larger than neutrino-powered energy. Thus, the explosions of FBOTs are expected to be partially suppressed in high-metallicity environments.

{A few magnetar-driven SESNe were detected in environments of $Z\lesssim0.1\,Z_\odot$ at high redshifts. As shown in \ref{fig:Dependence_Metallicity_Supplement}, we display our simulated $E_{\rm{rot,i}}-M_{\rm{ej}}$ relationships with the consideration of three lower metallicities, including $Z=\{0.01,0.03,0.1\}\,Z_\odot$. At such environments, metallicity mainly affects the initial mangetar rotational energy of $M_{\rm{He,i}}\gtrsim20\,M_\odot$ helium stars. Otherwise, the simulated $E_{\rm{rot,i}}-M_{\rm{ej}}$ relationships mostly have slight changes with different metallicities.}

\subsection{Companion Mass and Initial Rotation of Helium Stars.}  {In our study, the helium stars are always evolved with a companion mass of $M_{\rm com}=1.4,M_\odot$. Here, we investigate how different companion masses affect the $E_{\rm rot,i}-M_{\rm ej}$ relationships. By introducing two additional companion masses, $M_{\rm com}=8\,M_\odot$ and $30\,M_\odot$, the comparisons of the $E_{\rm rot,i}-M_{\rm ej}$ relationships are presented in  \ref{fig:Dependence_Companion_Mass_Supplement}. Under our assumption that the helium stars are initially tidally locked by their companions, it is clear that the effect of companion mass on the initial magnetar rotational energy is limited. This is due to the following facts. On one hand, for low-mass helium stars, mass transfer can always quickly extract AM of the envelope, leading to differential rotation between the rapidly rotating core and envelope. After the shell helium burning, the helium star can expand and initiate mass transfer to the companion once it reaches the Roche lobe radius. The core can continue to lose AM via the coupling with the slowly rotating envelope through the TS dynamo. Although the Roche lobe radius of helium stars for a given mass and a given orbital period can be smaller with a more massive companion,  the duration of mass transfer usually does not vary significantly. Therefore, the final rotation of the core in the helium star is not greatly affected. On the other hand, for high-mass helium stars that do not undergo mass transfer, their rotation is determined by both the orbital period and the size of the interior core, which are typically unrelated to the companion mass. }

{The initial rotation of the helium stars emerging from the CEE or SMT stages is highly uncertain, as it has not yet been, and is difficult to be, constrained by either observations or numerical simulations. For simplicity, the helium stars in our \texttt{MESA} computations are assumed to be initially synchronized with the orbits, i.e., the initial surface angular velocities $\omega_{\rm He,i}$ are equal to the initial orbital angular velocities $\omega_{\rm orb,i}=2\pi/P_{\rm orb,i}$, which are commonly used in the literature (e.g., \cite{Marchant2016,qin2018}). Here, we show the ratio between the synchronization timescale $T_{\rm syn}$ and lifetime of the helium star $T_{\rm He}$, obtained from our \texttt{MESA} simulations, in \ref{fig:Dependence_Lifetime_Supplement}. Synchronization of the helium stars with masses of $M_{\rm He,i}\lesssim5\,M_\odot$ before core collapse can always be achieved if the initial orbital periods are $P_{\rm orb,i}\lesssim0.5-0.6\,{\rm d}$. For helium stars with masses of $M_{\rm He,i}\gtrsim5\,M_\odot$, tidal synchronization can be influenced by the companion mass. If the companion is an NS, synchronization can always happen if the initial orbital periods are $P_{\rm orb,i}\lesssim0.6-0.7\,{\rm d}$; if the companion is an MS star or a massive BH with masses of $M_{\rm com}\gtrsim8\,M_\odot$, the allowed initial orbital periods for synchronization can be extended to $P_{\rm orb,i}\lesssim1-2\,{\rm d}$, which align well with the parameter spaces for SLSNe, GRB-SNe, and SNe Ic-BL. Beyond these parameter spaces, the final rotation of the helium stars might be influenced by their initial rotation. }

{Then, with the companion mass set at $1.4\,M_\odot$, we explore the effect of the initial rotation of the helium stars on the initial rotational energy of magnetars, by considering three other initial surface angular velocities $\omega_{\rm orb,i}=0$, $0.004\,\omega_{\rm crit}$, and $0.04\,\omega_{\rm crit}$, where $\omega_{\rm crit} = \sqrt{(1-L_{\rm He,i}/L_{\rm Edd})GM_{\rm He,i}/R^3_{\rm He,i}}$ is the critical surface angular velocity with $L_{\rm He,i}$ representing the luminosity, $L_{\rm Edd}$ the Eddington luminosity, and $R_{\rm He,i}$ the radius. As a reference, the surface angular velocity of the Sun is $\sim0.004-0.005\,\omega_{\rm crit}$. The results of the $E_{\rm rot,i}-M_{\rm ej}$ relationships are shown in \ref{fig:Dependence_Initial_Rotation_Supplement}. If the helium stars are initially non-rotating, only those with the orbits that can achieve tidal locking have explosions within the parameter space of magnetar-driven SESNe, while the magnetar rotation in other orbits is very low. It is expected that a larger parameter space can achieve tidal locking as the companion mass ($M_{\rm com}$) increases which has been presented in \ref{fig:Dependence_Lifetime_Supplement}. When we increase $\omega_{\rm He,i}$ to the level of the Sun ($0.004\,\omega_{\rm crit}$), we find that helium stars that cannot be tidally locked before core collapse can lead to a lower limit for the magnetar rotation of a given helium star mass. However, the lower limit for $\lesssim5\,M_\odot$ helium stars falls within the region of FBOTs, indicating that their SN explosions should always be classified as FBOTs. In contrast, the lower limit for helium stars with masses of $\gtrsim5\,M_\odot$ remains significantly lower than the neutrino-driven energy. If $\omega_{\rm He,i}\sim0.04\,\omega_{\rm crit}$, the initial rotation energy of magnetars from $\lesssim5\,M_\odot$ helium stars  is concentrated around $\sim4\times10^{50}\,{\rm erg}$, which is the observed upper limit of FBOTs. For higher-mass helium stars, we find their SN explosions should always be magnetar-driven SESNe. Considering the uncertainty regarding the initial rotation of helium stars, we cannot determine whether our assumption of initial tidal locking for these stars is optimistic or pessimistic; rather, it serves as a great and pragmatic approximation.}

\subsection{Conservation of AM.}  {In our study, we calculate the initial NS rotational energy by assuming that AM is always conserved during the core collapse of the star, a commonly used assumption in the literature (e.g., ref\cite{aguileradena2018,fuller2022}). If the progenitor's magnetic field strength is much higher than the typical value, recent numerical simulations of SN explosions suggested that the proto-NS can be significantly spun down due to magnetic torques, with extracted rotational energy enhancing GW emission. (e.g., ref\cite{Powell2023}). This causes that the expected NS rotational energy may be much lower than our simulations. Here, we present a comparison of the scenarios with conserved AM and those without it. In \ref{fig:Dependence_Conservation_Supplement}, we show the $E_{\rm rot,i}-M_{\rm ej}$ relationships considering 50\% and 100\% AM conservation during the formation of the NS from the stellar core. One can see that the initial magnetar rotational energies in the 100\% conserved AM scenario can be approximately four times higher than those in the 50\% conserved AM scenario. In the 50\% conserved AM scenario, we find that most magnetar-driven SESNe observations cannot be explained by our models. Only very tight orbits with $P_{\rm orb,i} \gtrsim 0.4\,{\rm d}$ can reproduce the $E_{\rm rot,i}-M_{\rm ej}$ data with a relatively low initial magnetar rotational energy. It is therefore expected that the progenitor's magnetic field strength cannot be much higher than the typical value, as this would result in excessive AM and NS rotational energy being transferred to GW emission. }

\subsection{Different maximum NS gravitational masses.}  {In \ref{fig:Dependence_NS_Maximum_Mass_Supplement}, we show the $E_{\rm rot,i}-M_{\rm ej}$ relationships for three different maximum NS gravitational masses, including $M_{\rm NS,g}^{\rm max}=1.4\,M_\odot$, $1.7\,M_\odot$, and $2.0\,M_\odot$, which correspond to the maximum NS baryonic masses of $M_{\rm NS,b}^{\rm max}=1.56\,M_\odot$, $1.93\,M_\odot$, and $2.36\,M_\odot$, respectively. The difference in the initial magnetar rotational energy is roughly proportional to the selected maximum NS baryonic mass. For instance, if the maximum NS baryonic mass differs by a factor of 1.5, the rotational energy also varies by approximately the same factor. A larger maximum NS mass, particularly when $M_{\rm NS,g}^{\rm max}\gtrsim1.7\,M_\odot$ ($M_{\rm NS,b}^{\rm max}\gtrsim1.93\,M_\odot$),  enables broader and greater coverage of the observational data.   }

\section{{Event Rate Densities and Detailed Formation Channels}} \label{sup_sec:population_synthesis}

{In \ref{fig:EventRate_Comp_Supplement}, we present the local event rate densities of different types of magnetar-driven SESNe, obtained from the different population synthesis models described in Methods. When considering a model where the mass-transfer efficiency ($\beta$) depends on the accretor's spin (i.e., Model I), the local event rate densities are not sensitive to the CE ejection efficiency ($\alpha_{\mathrm{CE}}$). However, in other mass-transfer models, such as fixed $\beta$ (Model II) or where $\beta$ depends on the timescale ratio (Model III), an increase in $\alpha_{\mathrm{CE}}$ leads to higher local event rate densities. This is because more progenitors of magnetar-driven SESNe undergo a CEE phase in Models II and III, as shown in  \ref{fig:Channel_Ratio_Supplement}, making the rate densities sensitive to $\alpha_{\mathrm{CE}}$. Expect for the models with $\alpha_{\rm CE}=5$ and the mass-transfer models of Model II and III, we find the magnetar-driven SESN rate densities of our population synthesis simulations basically fall within the range of the observed values.}

{Assuming initially non-rotating helium stars, we estimate the lower limit of the simulated event rate densities for magnetar-driven SESNe. \ref{fig:Period_distribution_Supplement} shows the orbital period distributions of helium star binaries with main-sequence and compact-object companions, based on our various population synthesis models. For main-sequence companions, the orbital period distribution of systems at the time of core collapse peaks around $P_{\rm orb,i}\sim0.5\,{\rm d}$. In contrast, for compact-object companions, the orbital period distribution peaks at $P_{\rm orb,i}\lesssim0.25,{\rm d}$. Considering that only systems with $P_{\rm orb,i}\lesssim0.5-0.6\,{\rm d}$ can achieve tidal synchronization for helium stars with masses of $M_{\rm He,i}\lesssim5\,M_\odot$, we expect that the lower limit of the simulated FBOT event rate density can be $\sim10\%-20\%$ of the simulated rate density based on the simulated orbital period distribution shown in Figure \ref{fig:Period_distribution_Supplement}. This corresponds to a local simulated FBOT rate density of $\sim100-200\,{\rm Gpc}^{-3}{\rm yr}^{-1}$, and up to $\sim500-1000\,{\rm Gpc}^{-3}{\rm yr}^{-1}$ within a redshift range of $0.05<z<1.56$, if we include the cosmic star formation history. The latest constrained FBOT rate density is $\mathcal{R}_{\rm{FBOT}}\gtrsim1000\,{\rm{Gpc}}^{-3}\,{\rm{yr}}^{-1}$ at a redshift of $0.05<z<1.56$ in the Dark Energy Survey\cite{pursiainen2018,wiseman2020}. Thus, due to the large uncertainty, the lower limit of FBOT event rate densities are still within the observed range. Since the orbital period ranges for SLSNe, GRB-SNe, and SNe Ic-BL are approximately consistent with those required for synchronization, we expect that the initial non-rotation conditions do not significantly affect the rate density estimations of these extreme transients.}

{If a portion of $\lesssim5\,M_\odot$ helium stars have initial surface angular velocities less than $0.004\,\omega_{\rm crit}$, the magnetar central engine would not contribute the emissions of their SN explosions. Thus, these SN explosions can be normal $^{56}$Ni-powered USSNe. In observations, there are indeed USSNe that show no evidence of magnetar engine contribution (e.g., ref\cite{de2018,Yao2020}); however, some bright USSNe require magnetar contributions (e.g., ref\cite{sawada2022,Moore2024}). Therefore, it is possible that a portion of helium stars have initial surface angular velocities less than $0.004\,\omega_{\rm crit}$, which means that FBOTs could be part of the SN explosions from ultra-stripped helium stars. We note that mass transfer of our close-orbit helium star progenitors involves Case A (occurs during the slow growth) and Case B (during the first rapid expansion) mass transfer. Additionally, a small portion of close-orbit helium star progenitors can be formed through Case C mass transfer. Case C mass transfer occurs during the helium shell burning phase of the donor star, i.e., the final expansion phase. Helium stars formed through Case C mass transfer typically have a non-convective CO or ONeMg core, which cannot be influenced by the tidal forces of their companions. Moreover, since these helium stars would quickly explode, the pre-collapse core rotation of these helium stars and the rotation periods of magnetars are primarily determined by their rotation after Case C mass transfer, rather than by the tidal scenario explored and discussed in our study. }

{Based on our population synthesis simulations, we can provide a more detailed understanding of the formation channels of magnetar-driven SESNe and close-orbit NS binaries, as illustrated in \ref{fig:illustration_Supplement}. The illustration does not consider stellar mergers and disruptions of binary systems due to SN kicks. (1) The majority of close-orbit helium star–MS binaries can form through the CEE channel, while a smaller fraction may form via the SMT channel. If the primary helium star is massive enough, it can lead to the formation of a BH. Then, the secondary MS star can undergo the CEE or SMT channel, if the binary system does not experience a stellar merger, eventually becoming a helium star that could explode as a magnetar-driven SESN. The final binary system may either be an NSBH binary with an eccentric orbit or be disintegrated by the SN kick. (2) For the popular CEE and SMT scenario, the first SN explosion could be a magnetar-driven SESN or a normal Ib/Ic/II SN if the primary is less massive, while the secondary SN explosion usually occurs in a close-orbit binary that is typically a magnetar-driven SESN. Two MS stars can experience double-core CEE evolution to form helium star-helium star binary, which may also make two magnetar-driven SESN explosions duration the evolution. Finally, these two scenarios can allow the formation of BNS systems. (3) After the first magnetar-driven SESN explosion in either the CEE or SMT scenario, the secondary may involve a less massive star, ultimately leading to the formation of a WD without a SN explosion. The resulting binary system would then be a NSWD binary. }

\section{{NS or BH?}}\label{sup_sec:ns_or_bh}

{Despite the uncertainty regarding the final remnant of these massive helium stars with initial masses ${M}_{\rm{He,i}}\gtrsim10\,M_\odot$, we want to emphasis that a magnetar central engine should always play a crucial role in the explosions of SLSNe, GRB-SNe, and SNe Ic-BL regardless of whether the remnant is a long-lived NS or a BH. On one hand, the emission of SLSNe-I requires a long-lasting central engine, which can be naturally accounted for by a long-lived, spinning-down magnetar. In contrast, the BH accretion model always requires unrealistic accretion masses to explain the SLSN-I lightcurves. Therefore, the final remnant of SLSNe-I is expected to be a long-lived magnetar. On the other hand, since the lightcurves of GRB-SNe and SNe Ic-BL are usually $^{56}$Ni-powered, rapid energy injection is required after the explosions. For the magnetar central engine model, this can be easily achieved when the magnetar spin-down timescale is much shorter than the diffusion timescale. Alternatively, the death of rapidly rotating massive stars can form a BH fed by infalling stellar material at a rate of $\gtrsim 1\,M_\odot\,{\rm yr}^{-1}$, in which the accretion process can drive a relativistic jet and a powerful disk wind as the agent of potential energy injection, i.e., the so-called ``collapsar'' model\cite{woosley1993,macfadyen1999}. However, the simulations presented by ref.\cite{zenati2020} showed that the total kinetic energy of the collapsar disk wind could be $\sim1\times10^{49}-6\times10^{50}\,{\rm erg}$, much smaller than the kinetic energy of GRB-SNe and SNe Ic-BL. Thus, in the BH collapsar scenario, to explain the usually high explosion energy of GRB-SNe and SNe Ic-BL, the formation of a short-lived proto-magnetar after the core collapse of a helium star may still be necessary. Before collapsing into a BH, the proto-magnetar can lose most of its rotational energy into the ejecta via magnetic spin-down\cite{Gottlieb2024} or accretion-propeller spin-down\cite{piro2011}, although its survival time may only be $\sim1-2\,{\rm s}$. }

\section{Effects of SN Kicks} \label{sup_sec:kick}

\subsection{Modelling.} Following a Gaussian distribution of SN kicks, the mean natal kick for NSs is\cite{mandel2020}
\begin{equation}
\label{equ:mu_kick}
    \mu_{\mathrm{kick}}=v_{\mathrm{NS}}\frac{M_{\rm{CO}}-M_{\rm{NS,b}}}{M_{\rm{NS}}},
\end{equation}
where $v_{\mathrm{NS}}$ is the NS scaling prefactor. We adopt $v_{\mathrm{NS}}=265\,\mathrm{km\,s^{-1}}$ and add an additional component drawn from a Gaussian distribution with the standard deviation of $0.3\mu_{\mathrm{kick}}$. 

We update the semi-major axis and the eccentricity of the binary after the SN following ref.\cite{kalogera1996}, which are
\begin{equation}
\begin{array}{r}
    a_{\rm{f}}=G\left(M_{\rm{comp}}+M_{\rm NS}\right)\bigg[\displaystyle\frac{2 G\left(M_{\rm{comp}}+M_{\rm{NS}}\right)}{a_{\rm{i}}}-v_{\text{kick}}^{2}-v_{\text{orb}}^{2} -2 v_{\text{kick},y} v_{\text{orb}}\bigg]^{-1}
\end{array}
\end{equation}
and
\begin{equation}
    1-e^{2}=\frac{\left(v_{\text {kick},y}^{2}+v_{\text  {kick},z}^{2}+v_{\text {orb}}^{2}+2 v_{\text {kick},y} v_{\text {orb}}\right)a_{\rm{i}}^{2}}{G\left(M_{\rm comp}+M_{\rm NS}\right)a_{\rm{f}}}
\end{equation}
where $G$ is the Gravitational constant, $a_{\rm{i}}$ and $a_{\rm{f}}$ are the pre- and post-SN semi-major axis, and $v_{\mathrm{kick},i}$ represents the $i$-th component of the kick velocity $\mathbf{v}_{\text{kick}}$, respectively. Here, the pre-SN orbital velocity of the NS progenitor relative to its companion $v_{\text{orb}}$ is given by\cite{kalogera1996}
\begin{equation}
    v_{\text {orb}}^2=G\left(M_{\rm comp}+M_{\rm CO}\right)\left(\frac{2}{r}-\frac{1}{a_{\rm{i}}}\right) ,
\end{equation}
where $r$ represents the orbital separation between the NS progenitor.

The binary after the SN explosion will be disrupted if the natal kick is too strong, when (e.g., ref.\cite{callister2021})
\begin{equation}
    \beta<\frac{1}{2}+\frac{v_{\text {kick }}^{2}}{2 v_{\text {orb }}^{2}}+\frac{\mathbf{v}_{\text {kick }} \cdot \mathbf{v}_{\text {orb }}}{v_{\text {orb }}^{2}} ,
\end{equation}
where $\beta$, defined in Equation (7) of ref.\cite{kalogera1996}, represents the pre-SN-to-post-SN ratio of the binary's total mass. For binaries surviving after the SN kicks, the merger time via gravitational-wave emission can be calculated as\cite{mandel2020},
\begin{equation}
\label{equ:merger_time1}
    T_{\rm merger}=\frac{5}{256} \frac{c^{5} a_{\rm f}^{4}}{G^{3} (M_{\rm comp} + M_{\rm NS} )^{2} m_{\rm r}} T(e) , 
\end{equation}
where $c$ is the speed of light and $m_{\rm{r}}$ is the binary’s reduced mass, and the function $T(e)$ is given by
\begin{equation}
\label{equ:merger_time2}
    T(e) = \left(1+0.27 e^{10}+0.33 e^{20}+0.2 e^{1000}\right)\left(1-e^{2}\right)^{7 / 2}.
\end{equation}

\subsection{Disrupted and Merger Fractions of Binary Systems with SN Kicks Considered.} With the post-SN binary properties updated, we repeat $10^5$ times to obtain the average value of the kick and also its associated angle for each initial system by considering two different companions, including a $1.4M_\odot$ NS and an $8\,M_\odot$ BH. Our simulated fractions of the post-SN NS binaries (including BNS and NS--BH systems) that can survive and merge within Hubble time are displayed in \ref{fig:Disruptfraction_Supplement} and \ref{fig:Mergerfraction_Supplement}, respectively.

For the binary systems composed of a low-mass helium star with an initial mass from $M_{\rm{He,i}}\sim2.4\,M_\odot$ to $\sim3.2\,M_\odot$, the SN kicks are usually incapable of separating the systems because of a small number of materials ejected during the explosions of the helium stars. These post-SN NS binaries usually have low-eccentricity orbits\cite{tauris2015}. However, only those helium stars in very close binaries $P_{\rm{orb,i}}\lesssim0.4-0.7\,{\rm{d}}$, can generate close-orbit NS binaries which can finally merge within Hubble Time.

The core-collapses of high-mass helium stars ($M_{\rm{He,i}}\gtrsim4-5\,M_\odot$) with a compact-object companion can have much larger SN kicks to disrupt a large fraction of systems. If $M_{\rm{He,i}}\lesssim10\,M_\odot$, the survived NS binaries might have some extreme characteristics, e.g., large proper motions and large orbital eccentricities. Following Equations (\ref{equ:merger_time1}) and (\ref{equ:merger_time2}), these large-eccentricity NS binaries can spend less time inspiraling and merge earlier than those systems with low-eccentricity orbits. Thus, as shown in \ref{fig:Mergerfraction_Supplement}, although most NS mergers are derived from close systems with an initial orbital period $P_{\rm{orb,i}}\lesssim1\,{\rm{d}}$, we find that a part of merger NS binaries could be also originated from wide binary systems with $P_{\rm{orb,i}}\sim1-10\,{\rm{d}}$. If $M_{\rm{He,i}}\gtrsim10\,M_\odot$, almost all binary systems would be separated due to extremely strong SN kicks.

\section{Inferred Properties of the NS Companion with the Hulse-Taylor Pulsar} \label{sup_sec:hulse_taylor_pulsar}

\subsection{Progenitor Mass.}

We use Equation (\ref{equ:mu_kick}) to estimate the natal kick of newborn NSs for different helium stars in binary systems. As shown in \ref{fig:Mukick_Supplement}, in order to generate an NS kick velocity of $\gtrsim200\,\mathrm{km\,s}^{-1}$ inferred from the observations of the Hulse-Taylor pulsar\cite{tauris2015}, the initial helium star mass is required to be $M_{\rm{He,i}}\gtrsim6.3\,M_\odot$.

\subsection{Period and Period Derivative.}

By integrating the pulsar spin deceleration equation, i.e., $\dot{\Omega}_{\rm{NS}}=-K\Omega^{n}_{\rm{NS}}$ with the braking index $n$ and $\Omega_{\rm{NS}}=2\pi/P_{\rm{NS}}${, the} period $P_{\rm{NS}}$ and the period derivative $\dot{P}_{\rm{NS}}$ for an NS at time $t$ are\cite{tauris2017}
\begin{equation}
\label{equ:Period}
    P_{\rm{NS}}=P_{\rm{NS},0}\left[1+(n-1)\frac{\dot{P}_{\rm{NS},0}}{P_{\rm{NS},0}}t\right]^{1/(n-1)},
\end{equation}
\begin{equation}
\label{equ:Period_dot}
    \dot{P}_{\rm{NS}}=\dot{P}_{\rm{NS,0}}\left(\frac{P_{\rm{NS}}}{P_{\rm{NS},0}}\right)^{2-n},
\end{equation}
where $P_{\rm{NS,0}}$ and $\dot{P}_{\rm{NS,0}}$ represent the present-day values of the period and its derivative, respectively. Present Hulse-Taylor pulsar\cite{hulse1975}, as the first-born NS in the binary system, has a spin period of $P_{\rm{NS,0}}=59.0\,{\rm{ms}}$ and a period derivative of $\dot{P}_{\rm{NS,0}}=8.63\times10^{-18}$. By defining $n=3$, a maximum age of the Hulse-Taylor pulsar can be solved by Equation (\ref{equ:Period}) for $P_{\rm{NS}}\approx0$ as $t=\tau_0=108.7\,{\rm{Myr}}$. This yields a pre-SN orbital period of $P_{\rm{orb,f}}=0.03-0.8\,{\rm{d}}$ (see detailed simulations in ref.\cite{tauris2017}), which is approximate to the initial orbital period. 

For an initial helium star mass $M_{\rm{He,i}}\gtrsim6.3\,M_\odot$ in a close binary with an initial orbital period $P_{\rm{orb,i}}\approx0.03-0.8\,{\rm{d}}$, the initial rotational energy of the second-born NS could be from a few $10^{51}\,{\rm{erg}}$ to $\sim2\times10^{52}\,{\rm{erg}}$, roughly corresponding to an initial rotation period of $P_{\rm{mag,i}}\approx1-3\,{\rm{ms}}$. Because the second-born NS was likely to be a millisecond magnetar, we estimate its initial period derivative $\dot{P}_{\rm{mag,i}}$ based on Equation (\ref{equ:Period_dot}) by adopting the averaged period and period derivative of galactic magnetar (i.e., $P_{\rm{mag,0}}\sim8\,{\rm s}$ and $\dot{P}_{\rm{mag,0}}\sim10^{-11}$)\cite{olausen2014} as their present-day values. Thus, by setting $P_{\rm{mag,i}}=2\,{\rm{ms}}$, we can get its derivative as $\dot{P}_{\rm{mag,i}}=4\times10^{-8}$. After $\sim108.7\,{\rm{Myr}}$ of spin-down, at present, the magnetar orbiting the Hulse-Taylor pulsar may have a rotation period of $\sim700\,{\rm{s}}$ and a period derivative of $\sim10^{-13}$ currently located under the death line.

\clearpage

\captionsetup[table]{name=Supplementary Information Table}

\begin{table*}
\centering
\caption{\textbf{Distributions of $M_{\rm{ej}}$ and $E_{\rm{rot,i}}$ for Our Collected SLSNe-I\cite{yu2017}, GRB-SNe\cite{lv2018} and SNe Ic-BL\cite{lyman2016,taddia2019}.} }
\label{table:SLSN}
{\footnotesize
\begin{tabular}{lcclcc}
\hline\hline
\textbf{SLSN} & $M_{\rm{ej}}/M_\odot$ & \multicolumn{1}{c|}{$\log_{10}(E_{\rm{rot,i}}/{\rm{erg}})$} & \textbf{SN Ic-BL} & $M_{\rm{ej}}/M_\odot$ & $\log_{10}(E_{\rm{rot,i}}/{\rm{erg}})$ \\
\hline
CSS121015 & $2.51\pm0.27$ & \multicolumn{1}{c|}{$51.53\pm0.34$} & SN2002ap & $2.0^{+0.8}_{-0.7}$ & $51.30^{+0.22}_{-0.26}$ \\
DES13S2cmm & $0.90\pm0.08$ & \multicolumn{1}{c|}{$50.90\pm0.16$} & SN2003jd & $2.5^{+0.9}_{-0.5}$ & $51.43^{+0.15}_{-0.13}$ \\
DES14X2taz & $6.4\pm2.2$ & \multicolumn{1}{c|}{$51.70\pm0.70$} & SN2005kz & $8.1^{+3.7}_{-2.6}$ & $52.25^{+0.21}_{-0.26}$ \\
Gaia16apd & $3.65\pm0.53$ & \multicolumn{1}{c|}{$51.67\pm0.46$} & SN2007ru & $2.2^{+1.1}_{-1.1}$ & $51.67^{+0.18}_{-0.33}$ \\
iPTF13ajg & $4.0\pm1.7$ & \multicolumn{1}{c|}{$52.0\pm1.7$} & SN2009bb & $1.9^{+0.6}_{-0.5}$ & $51.52^{+0.22}_{-0.16}$ \\
iPTF13ehe & $6.8\pm4.3$ & \multicolumn{1}{c|}{$51.5\pm1.9$} & PTF10bzf & $2.8\pm0.7$ & $51.78\pm0.12$ \\
LSQ12dlf & $2.02\pm0.22$ & \multicolumn{1}{c|}{$51.03\pm0.35$} & PTF10ciw & $5.7\pm8.4$ & $52.30\pm0.50$ \\
LSQ14bdq & $14.98\pm0.90$ & \multicolumn{1}{c|}{$51.97\pm0.17$} & PTF10gvb & $3.3\pm2.2$ & $51.48\pm0.30$ \\
LSQ14mo & $1.35\pm0.08$ & \multicolumn{1}{c|}{$50.98\pm0.48$} & PTF10qts & $2.9\pm0.5$ & $51.90\pm0.09$ \\
PS1-10bzj & $2.3\pm0.38$ & \multicolumn{1}{c|}{$51.35\pm0.71$} & PTF10tqv & $1.6\pm0.8$ & $51.60\pm0.23$ \\
PS1-10ky & $2.5\pm0.47$ & \multicolumn{1}{c|}{$51.49\pm0.60$} & PTF10vgv & $0.6\pm0.6$ & $50.48\pm0.33$ \\
PS1-11ap & $2.41\pm0.15$ & \multicolumn{1}{c|}{$51.05\pm0.17$} & PTF10xem & $11.1\pm8.9$ & $51.90\pm0.38$ \\
PS1-14bj & $17.9\pm2.4$ & \multicolumn{1}{c|}{$51.39\pm0.43$} & PTF10ysd & $10.4\pm12.7$ & $51.85\pm0.71$ \\
PS15br & $0.85\pm0.04$ & \multicolumn{1}{c|}{$50.71\pm0.08$} & PTF10aavz & $5.0\pm3.5$ & $51.60\pm0.50$ \\
PTF10hgi & $2.33\pm0.39$ & \multicolumn{1}{c|}{$50.78\pm0.55$} & PTF11cmh & $1.7\pm0.8$ & $51.30\pm0.30$ \\
PTF11rks & $0.83\pm0.25$ & \multicolumn{1}{c|}{$50.4\pm1.1$} & PTF11img & $5.2\pm2.2$ & $52.00\pm0.20$ \\
PTF12dam & $7.3\pm1.0$ & \multicolumn{1}{c|}{$51.48\pm0.44$} & PTF11lbm & $0.6\pm0.4$ & $50.60\pm0.25$ \\
SCP06F6 & $2.9\pm2.5$ & \multicolumn{1}{c|}{$51.5\pm2.4$} & PTF12as & $1.2\pm0.9$ & $50.90\pm0.38$ \\
SN2005ap & $0.81\pm0.14$ & \multicolumn{1}{c|}{$51.40\pm0.41$} & PTF12eci & $1.9\pm1.3$ & $51.00\pm0.40$ \\
SN2007bi & $5.99\pm0.77$ & \multicolumn{1}{c|}{$51.39\pm0.36$} & iPTF13u & $1.8\pm0.7$ & $51.70\pm0.18$ \\
SN2008es & $3.01\pm0.69$ & \multicolumn{1}{c|}{$51.54\pm0.68$} & iPTF13alq & $2.4\pm0.7$ & $51.70\pm0.14$ \\
SN2010gx & $2.8\pm0.12$ & \multicolumn{1}{c|}{$51.50\pm0.25$} & iPTF13dnt & $7.0\pm2.6$ & $52.30\pm0.15$ \\
SN2010kd & $7.60\pm0.69$ & \multicolumn{1}{c|}{$51.45\pm0.37$} & iPTF13ebw & $5.3\pm1.0$ & $52.30\pm0.10$ \\
SN2011ke & $2.23\pm0.13$ & \multicolumn{1}{c|}{$51.40\pm0.29$} & iPTF14dby & $5.1\pm4.0$ & $51.60\pm0.25$ \\
SN2011kf & $2.10\pm0.38$ & \multicolumn{1}{c|}{$51.94\pm0.93$} & iPTF14gaq & $3.0\pm0.9$ & $52.00\pm0.20$ \\
SN2011kl & $0.51\pm0.06$ & \multicolumn{1}{c|}{$50.14\pm0.36$} & iPTF15dqg & $1.7\pm0.8$ & $51.30\pm0.20$ \\
SN2012il & $1.60\pm0.42$ & \multicolumn{1}{c|}{$50.9\pm1.0$} & iPTF15eov & $6.1\pm0.5$ & $52.00\pm0.05$ \\
SN2013dg & $1.76\pm0.12$ & \multicolumn{1}{c|}{$50.95\pm0.30$} & iPTF16asu & $0.9\pm0.1$ & $51.70\pm0.06$ \\
SN2013hx & $3.47\pm0.33$ & \multicolumn{1}{c|}{$51.66\pm0.30$} & iPTF16gox & $2.4\pm1.2$ & $51.90\pm0.25$ \\
SN2015bn & $5.28\pm0.28$ & \multicolumn{1}{c|}{$51.59\pm0.12$} & iPTF16ilj & $6.6\pm2.5$ & $52.00\pm0.30$ \\
SSS120810 & $5.59\pm0.94$ & \multicolumn{1}{c|}{$51.8\pm1.0$} & iPTF17cw & $4.5\pm1.8$ & $52.00\pm0.20$ \\
\hline\hline
\textbf{GRB-SN} & $M_{\rm{ej}}/M_\odot$ & $\log_{10}(E_{\rm{rot,i}}/{\rm{erg}})$ & \textbf{GRB-SN} & $M_{\rm{ej}}/M_\odot$ & $\log_{10}(E_{\rm{rot,i}}/{\rm{erg}})$ \\
\hline
SN1998bw & $6.80\pm0.57$ & $52.12\pm0.08$ & SN2008hw & $2.3\pm1.0$ & $51.95\pm0.56$\\
SN2001ke & $4.44\pm0.82$ & $52.25\pm0.50$ & SN2009nz & $4.69\pm0.13$ & $51.91\pm0.02$ \\
SN2002lt & $7.2\pm6.0$ & $52.45\pm0.46$ & SN2010bh & $2.47\pm0.23$ & $51.96\pm0.09$ \\
SN2003dh & $5.1\pm1.7$ & $52.08\pm0.32$ & SN2010ma & $1.3\pm0.4$ & $52.00\pm0.60$ \\
SN2003lw & $8.22\pm0.76$ & $52.20\pm0.09$ & SN2012bz & $6.10\pm0.49$ & $52.18\pm0.08$ \\
SN2005nc & $4.8\pm1.1$ & $52.28\pm0.40$ & SN2013dx & $3.0\pm0.1$ & $51.91\pm0.05$ \\
SN2006aj & $2.58\pm0.55$ & $51.79\pm0.02$ & SN2013fu & $6.71\pm0.20$ & $52.27\pm0.43$ \\
SN2008d & $5.3\pm1.0$ & $51.78\pm0.50$ & iPTF14bfu & $5.0\pm2.0$ & $52.30\pm0.50$ \\
\hline
\end{tabular}
}
\end{table*}

\clearpage

\begin{table*}
\centering
\caption{\textbf{The Fitting Results of the Derived Parameters for the Collected FBOTs.} The columns from left to right are (1) the FBOT event, (2) the ejecta mass in units of $M_\odot$, (3) the initial magnetar rotational energy in units of ${\rm{erg}}$, (4) the polar magnetic field strength in units of ${\rm{G}}$. Median values with 90\% credible intervals are listed. }
\label{table:FBOT}
{\small
\begin{tabular}{lccc}
\hline\hline
FBOT & $\log_{10}(M_{\rm{ej}}/M_\odot)$ & $\log_{10}(E_{\rm{rot,i}}/{\rm{erg}})$ & $\log_{10}(B_{\rm{p}}/{\rm{G}})$\\ \hline
AT2018cow & $-0.97^{+0.02}_{-0.02}$ & $51.38^{+0.02}_{-0.02}$ & $15.50^{+0.00}_{-0.00}$ \\
DES13C3bcok & $-0.59^{+0.08}_{-0.08}$ & $50.02^{+0.14}_{-0.08}$ & $15.19^{+0.09}_{-0.05}$ \\
DES13C3uig & $-0.95^{+0.22}_{-0.28}$ & $49.76^{+0.14}_{-0.12}$ & $14.55^{+0.19}_{-0.49}$ \\
DES13X1hav & $-1.53^{+0.37}_{-0.32}$ & $50.22^{+0.34}_{-0.14}$ & $14.55^{+0.19}_{-0.49}$ \\
DES13X3gms & $-0.39^{+0.10}_{-0.10}$ & $50.36^{+0.08}_{-0.06}$ & $14.88^{+0.19}_{-0.49}$ \\
DES13X3npb & $-0.43^{+0.13}_{-0.11}$ & $50.16^{+0.20}_{-0.16}$ & $15.17^{+0.07}_{-0.12}$ \\
DES13X3nyg & $-1.00^{+0.20}_{-0.25}$ & $50.44^{+0.16}_{-0.14}$ & $14.95^{+0.04}_{-0.05}$ \\
DES14C3tvw & $-0.62^{+0.19}_{-0.20}$ & $50.12^{+0.10}_{-0.08}$ & $14.93^{+0.05}_{-0.08}$ \\
DES14X1bnh & $-0.87^{+0.26}_{-0.35}$ & $50.62^{+0.22}_{-0.12}$ & $14.56^{+0.16}_{-0.44}$ \\
DES15C3lpq & $-0.66^{+0.09}_{-0.10}$ & $50.18^{+0.06}_{-0.06}$ & $14.96^{+0.02}_{-0.02}$ \\
DES15C3nat & $-1.00^{+0.11}_{-0.14}$ & $49.90^{+0.06}_{-0.04}$ & $14.89^{+0.07}_{-0.08}$ \\
DES15C3opk & $-0.64^{+0.11}_{-0.12}$ & $50.66^{+0.08}_{-0.08}$ & $14.77^{+0.06}_{-0.08}$ \\
DES15C3opp & $-0.55^{+0.21}_{-0.22}$ & $49.66^{+0.46}_{-0.32}$ & $15.56^{+0.09}_{-0.10}$ \\
DES15E2nqh & $-0.72^{+0.18}_{-0.23}$ & $50.30^{+0.14}_{-0.12}$ & $14.95^{+0.04}_{-0.04}$ \\
DES15S1fli & $-1.03^{+0.22}_{-0.29}$ & $50.22^{+0.14}_{-0.10}$ & $14.77^{+0.10}_{-0.14}$ \\
DES15S1fll & $-0.52^{+0.15}_{-0.17}$ & $49.84^{+0.18}_{-0.16}$ & $15.31^{+0.05}_{-0.06}$ \\ 
DES15X3mxf & $-0.51^{+0.07}_{-0.07}$ & $50.76^{+0.06}_{-0.06}$ & $15.24^{+0.02}_{-0.03}$ \\ 
DES16C1cbd & $-0.81^{+0.17}_{-0.19}$ & $50.24^{+0.12}_{-0.14}$ & $14.86^{+0.06}_{-0.08}$ \\
DES16C3gin & $-0.73^{+0.05}_{-0.05}$ & $50.04^{+0.06}_{-0.04}$ & $15.15^{+0.01}_{-0.01}$ \\
DES16E1bir & $0.16^{+0.12}_{-0.35}$ & $52.14^{+0.28}_{-0.52}$ & $14.77^{+0.11}_{-0.25}$ \\
DES16E2pv & $-1.24^{+0.30}_{-0.42}$ & $50.28^{+0.28}_{-0.14}$ & $14.75^{+0.23}_{-0.41}$ \\
DES16X3cxn & $-0.87^{+0.18}_{-0.17}$ & $50.20^{+0.14}_{-0.14}$ & $14.96^{+0.03}_{-0.03}$ \\
DES16X3ega & $-0.47^{+0.02}_{-0.02}$ & $50.20^{+0.02}_{-0.02}$ & $15.04^{+0.01}_{-0.01}$ \\
HSC17bhyl & $-1.17^{+0.35}_{-0.25}$ & $50.06^{+0.34}_{-0.26}$ & $15.32^{+0.03}_{-0.03}$ \\
HSC17btum & $-0.16^{+0.12}_{-0.13}$ & $50.54^{+0.26}_{-0.28}$ & $15.44^{+0.08}_{-0.09}$ \\
Koala & $-0.98^{+0.17}_{-0.18}$ & $51.18^{+0.10}_{-0.10}$ & $15.10^{+0.06}_{-0.07}$ \\
PS1-10bjp & $-0.97^{+0.02}_{-0.02}$ & $50.12^{+0.04}_{-0.04}$ & $15.48^{+0.01}_{-0.01}$ \\
PS1-11qr & $-0.68^{+0.15}_{-0.21}$ & $50.70^{+0.14}_{-0.14}$ & $14.31^{+0.26}_{-0.18}$ \\
PS1-12brf & $-0.94^{+0.17}_{-0.21}$ & $49.90^{+0.20}_{-0.20}$ & $15.36^{+0.04}_{-0.05}$ \\
PS1-12bv & $-0.98^{+0.29}_{-0.24}$ & $50.50^{+0.24}_{-0.16}$ & $14.42^{+0.19}_{-0.30}$ \\
PS1-13duy & $-0.86^{+0.38}_{-0.39}$ & $50.62^{+0.52}_{-0.30}$ & $14.95^{+0.38}_{-0.31}$ \\
PS1-13dwm & $-0.85^{+0.21}_{-0.26}$ & $49.72^{+0.48}_{-0.40}$ & $15.79^{+0.12}_{-0.11}$ \\
PTF10iam & $-1.50^{+0.15}_{-0.08}$ & $50.64^{+0.06}_{-0.06}$ & $14.10^{+0.03}_{-0.03}$ \\
SNLS04D4ec & $-1.02^{+0.07}_{-0.08}$ & $50.50^{+0.06}_{-0.04}$ & $14.41^{+0.08}_{-0.09}$ \\
SNLS05D2bk & $-1.43^{+0.02}_{-0.02}$ & $50.60^{+0.02}_{-0.02}$ & $14.12^{+0.02}_{-0.02}$ \\
SNLS06D1hc & $0.14^{+0.03}_{-0.04}$ & $51.58^{+0.04}_{-0.06}$ & $13.15^{+0.06}_{-0.04}$ \\
\hline
\end{tabular}
}
\end{table*}

\clearpage

\begin{figure*}
    \centering
    \includegraphics[width = 0.9\linewidth]{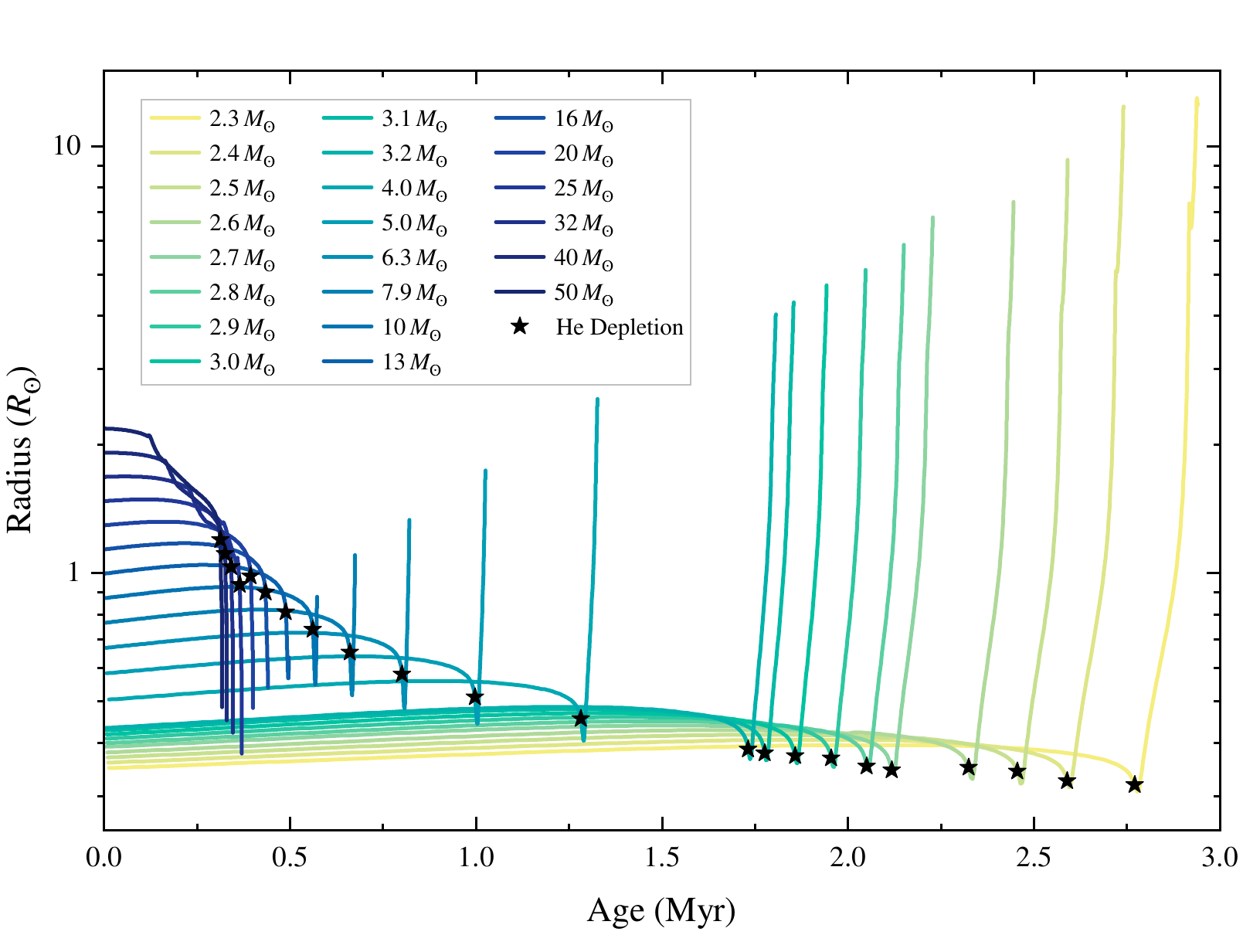}
    \caption{{\textbf{Radius of helium star as a function of the helium star age.} 
    The evolution of the radius for helium stars with identical initial orbital periods but different initial masses (see label). Black star points represent the depletion of central helium of the star.}}
    \label{fig:RadiusAge_Supplement}
\end{figure*}

\clearpage

\begin{figure*}
    \centering
    \includegraphics[width = 0.80\linewidth]{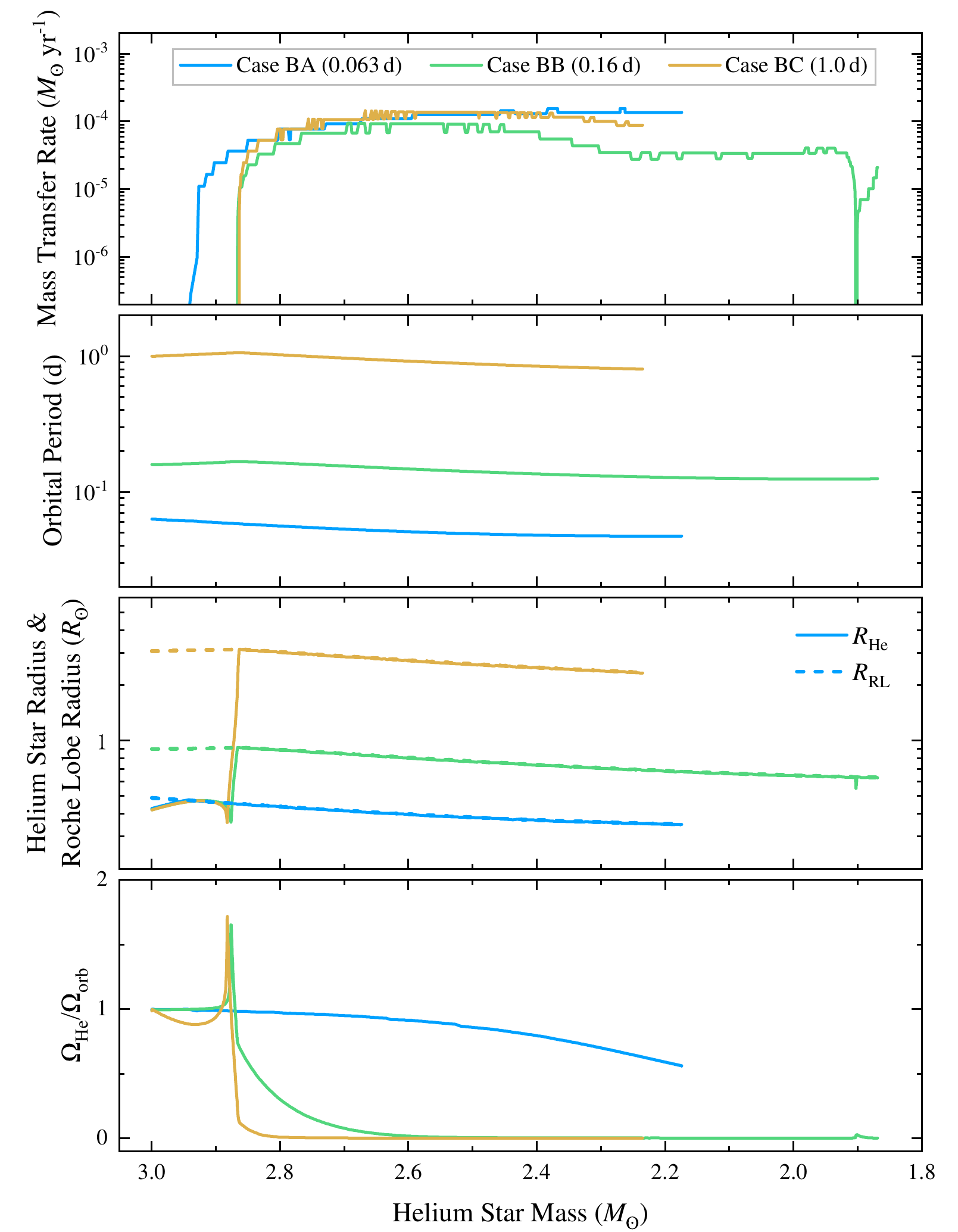}
    \caption{{\textbf{Evolution Parameters of Close-orbit Helium Stars.} (I) Mass transfer rate, (II) orbital period, (III) helium star radius (solid lines), Roche lobe radius (dashed lines), and (IV) rotational frequency ratio of helium star to the orbit as a function of helium star mass for three binary evolutionary sequences starting with the same initial masses of two components (initial helium star mass of $3\,M_\odot$ and NS mass of $1.4\,M_\odot$) but different initial orbital periods ($0.063$, $0.16$ and $1.0\,{\rm{d}}$). The three binary systems with different initial orbital periods undergo case BA (blue), case BB (green), and case BC (yellow) mass transfer, respectively.}}
    \label{fig:MassTransferRate_Supplement}
\end{figure*}

\clearpage

\begin{figure*}
    \centering
    \includegraphics[width = 0.85\linewidth]{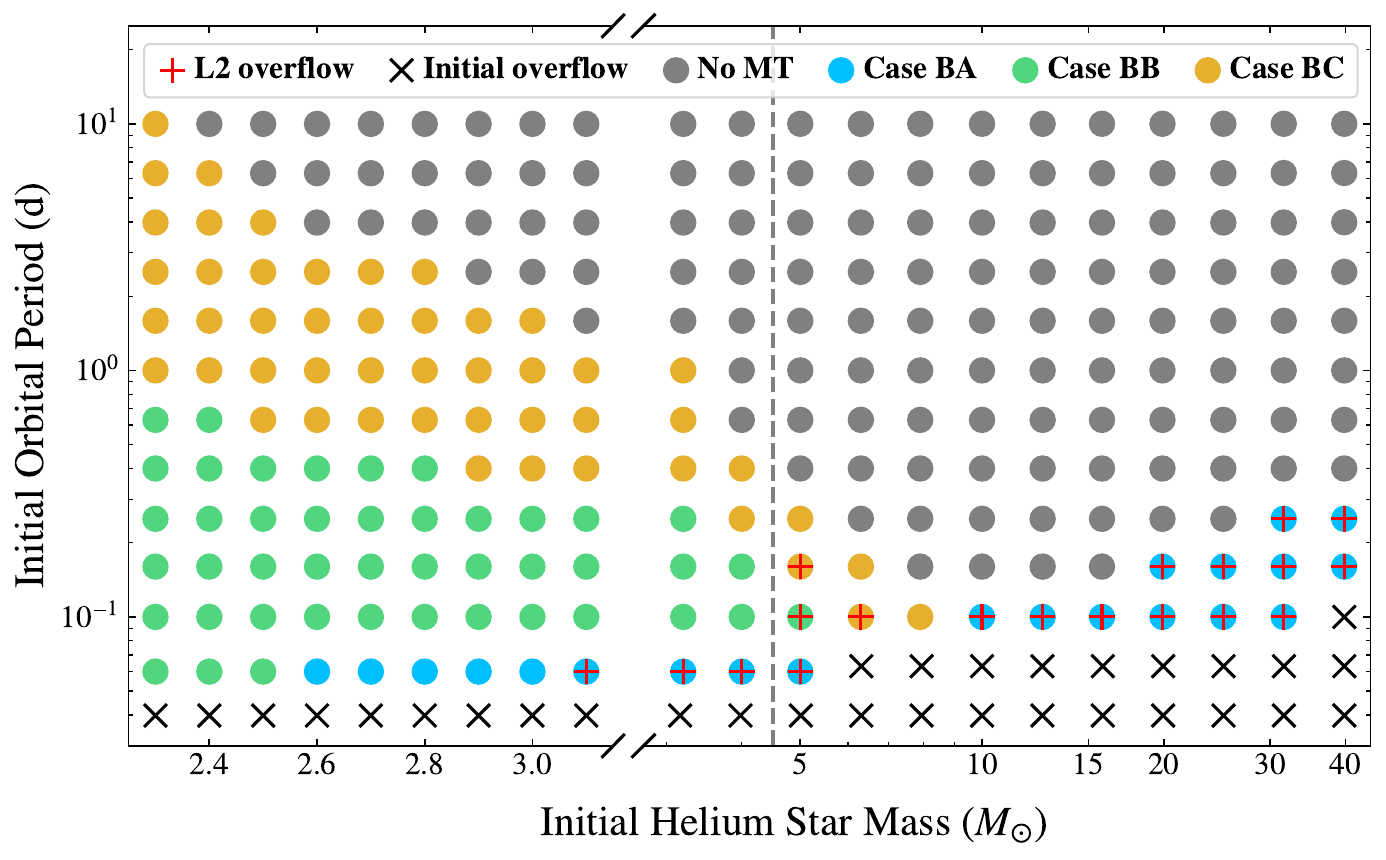}
    \caption{\textbf{Outcomes of binary systems as a function of the initial helium star mass and initial orbital period.} Black crosses: initial overflow through the first lagrangian point ($L_1$); red pluses: overflowing the second lagrangian point ($L_2$). Case BA (blue circles): mass transfer occurring during core helium burning phase; Case BB (green circles): mass transfer occurring during shell helium burning phase; Case BC (yellow circles): mass transfer occurring during core carbon burning phase and beyond; Gray triangle: no mass transfer. The boundary between FBOTs and other magnetar-driven SESNe is marked as the gray dashed line. }
    \label{fig:MassTransferCase_Supplement}
\end{figure*}

\clearpage

\begin{figure*}
\centering
\includegraphics[width=0.87\linewidth, trim = 6 0 8 0, clip]{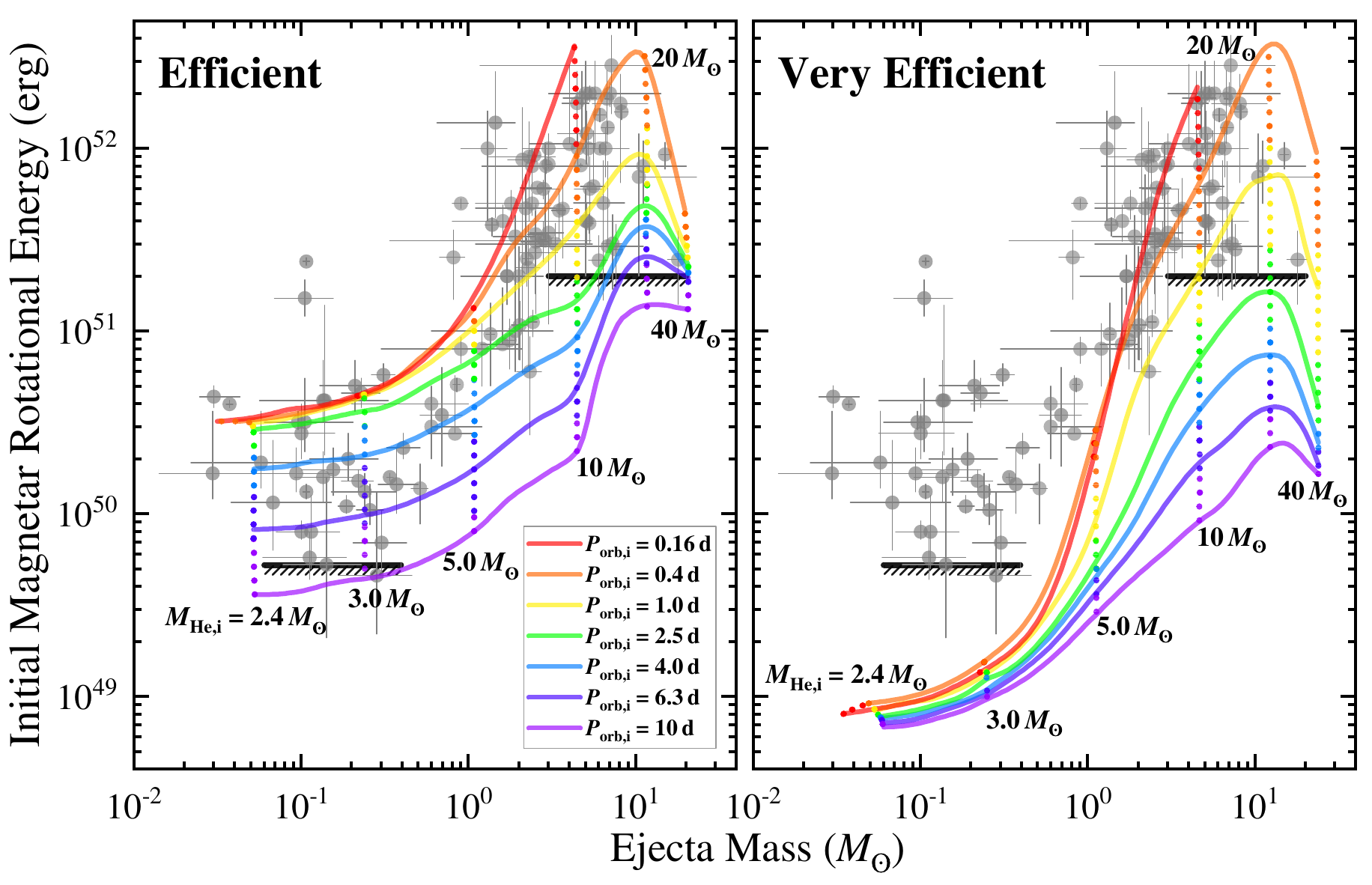}
\caption{{{\bf Dependence of simulated $E_{\rm{rot,i}}-M_{\rm{ej}}$ relationships on AM transport model.} Two different AM transport  mechanisms, including an efficient transport (left panel) and a very efficient transport (right panel), are adopted. The colored solid and dashed lines correspond to different initial orbital periods $P_{\rm{orb,i}}$ and initial helium star masses $M_{\rm{He,i}}$ as labeled.}}
\label{fig:Dependence_Model_Supplement}
\end{figure*}

\clearpage

\begin{figure*}
\centering
\includegraphics[width=0.70\linewidth, trim = 2 0 2 0, clip]{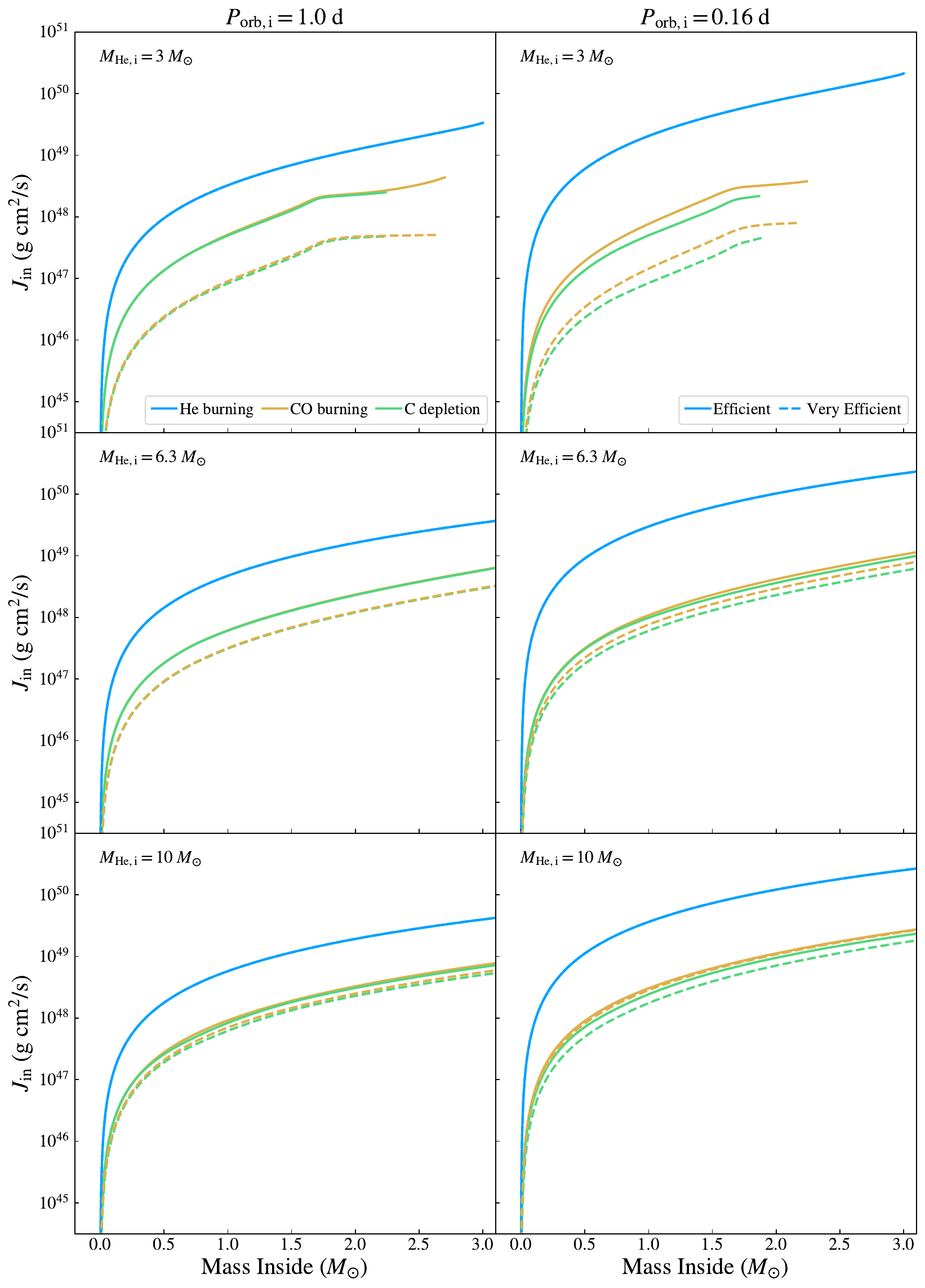}
\caption{{\bf AM distribution at different phases as a function of the mass within helium star.} {We present results by adopting three different initial helium star masses: $M_{\rm He,i}=3\,M_\odot$(top panels), $6.3\,M_\odot$ (middle panels), and $10\,M_\odot$ (bottom panels), as well as two initial orbital periods, i.e., $P_{\rm orb,i}=1.0\,{\rm d}$ (left panels) and $0.16\,{\rm d}$ (right panels). Two different AM transfer mechanisms, including an efficient transport (solid lines) and a very efficient transport (dashed lines), are considered. Different colors indicate the evolutionary phases of the helium star: blue for the helium burning phase, green for the carbon-oxygen burning phase, and red for the carbon depletion phase.}}
\label{fig:Dependence_Jin_Model_Supplement}
\end{figure*}

\clearpage

\begin{figure*}
\centering
\includegraphics[width=0.70\linewidth, trim = 0 0 0 0, clip]{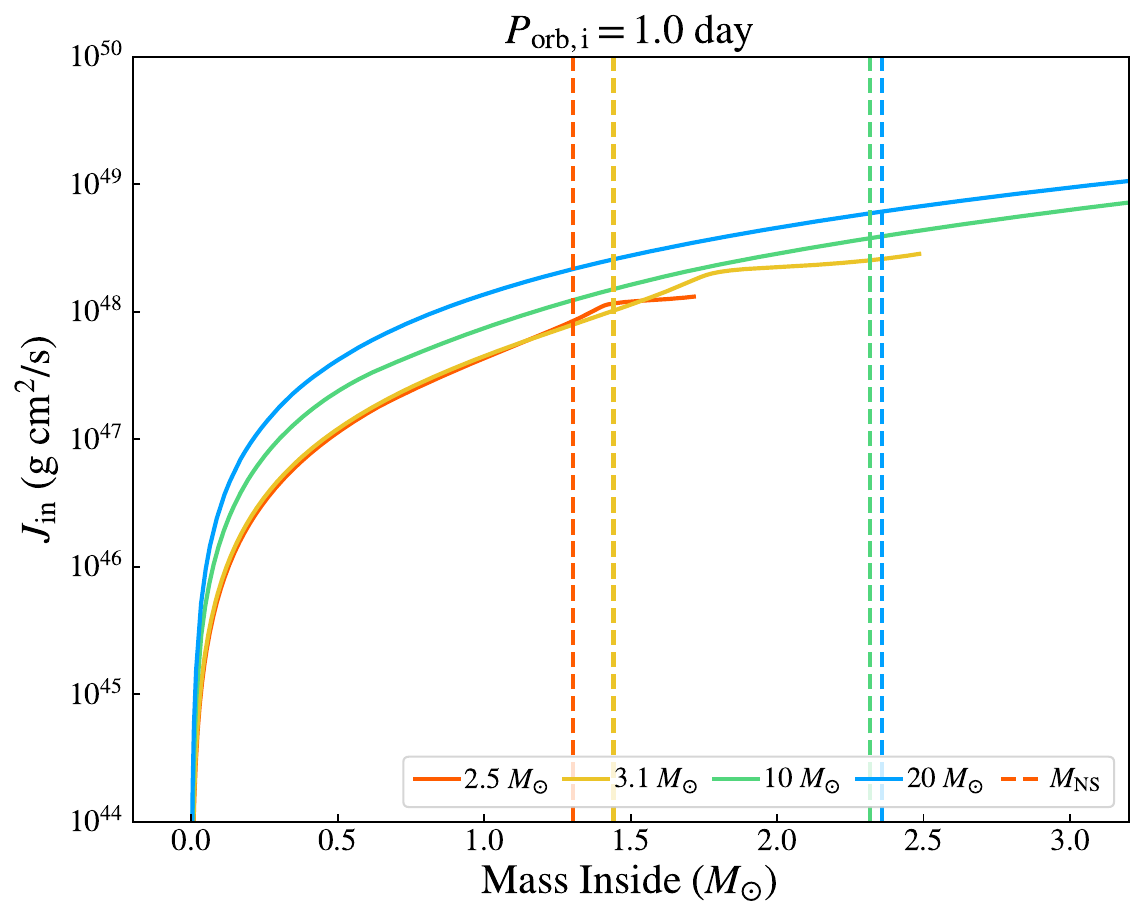}
\caption{{\bf AM distribution in the carbon depletion phases of different helium stars.} {Four different initial masses of helium stars are shown: 2.5 $M_{\odot}$ (orange line), 3.1 $M_{\odot}$ (golden line), 10 $M_{\odot}$ (light green line), and 20 $M_{\odot}$ (blue line). The dashed lines represent the corresponding baryonic NS mass.}}
\label{fig:Dependence_Jin_Core_Supplement}
\end{figure*}

\clearpage

\begin{figure*}
\centering
\includegraphics[width=01\linewidth, trim = 6 0 8 0, clip]{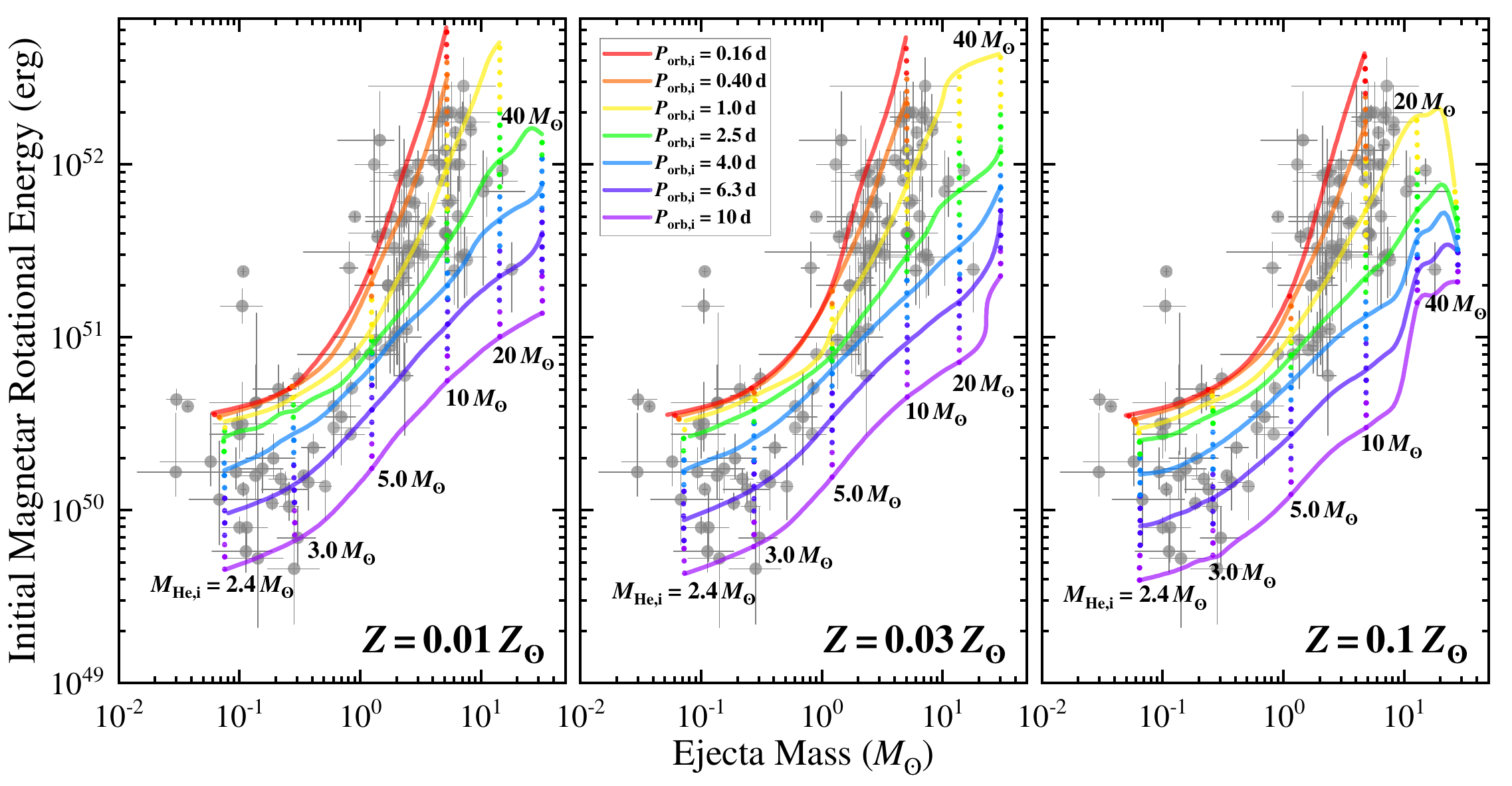}
\caption{{{\bf Dependence of simulated $E_{\rm{rot,i}}-M_{\rm{ej}}$ relationships on metallicity in environments with $Z<0.01\,Z_\odot$.} Three different metallicity environments, including $0.01\,Z_\odot$ (left panel), $0.03\,Z_\odot$ (left panel) and $0.1\,Z_\odot$, are adopted. The labels are consistent with those shown in \ref{fig:Dependence_Model_Supplement}.}}
\label{fig:Dependence_Metallicity_Supplement}
\end{figure*}

\clearpage

\begin{figure*}
\centering
\includegraphics[width=01\linewidth, trim = 6 0 8 0, clip]{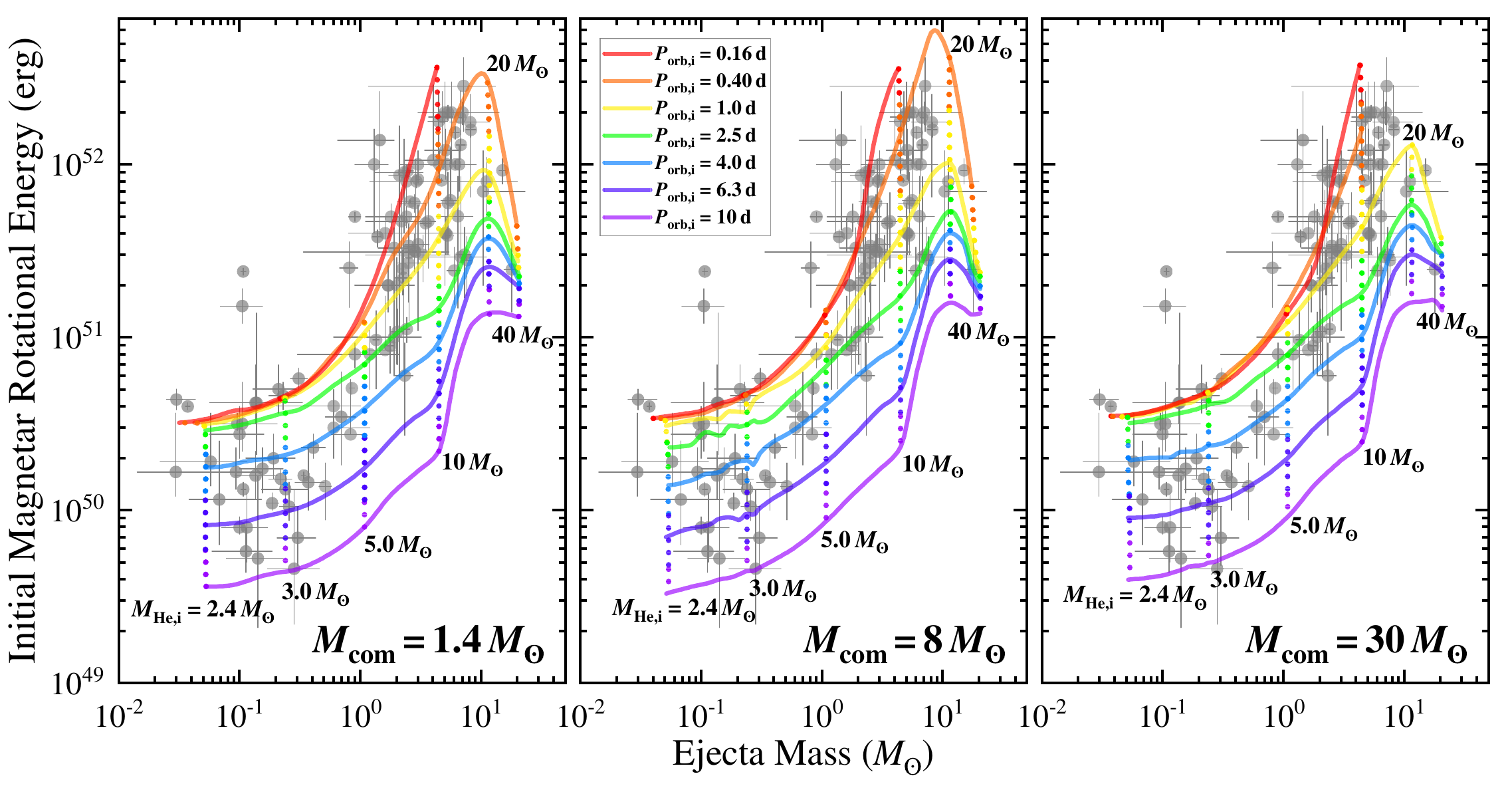}
\caption{{\bf Dependence of simulated $E_{\rm{rot,i}}-M_{\rm{ej}}$ relationships on companion mass.} {Three different companion masses, including $1.4\,M_\odot$ (left panel), $8\,M_\odot$ (middle panel) and $30\,M_\odot$ (right panel), are adopted. The labels are consistent with those shown in \ref{fig:Dependence_Model_Supplement}.}}
\label{fig:Dependence_Companion_Mass_Supplement}
\end{figure*}

\clearpage

\begin{figure*}
\centering
\includegraphics[width=0.70\linewidth, trim = 6 0 8 0, clip]{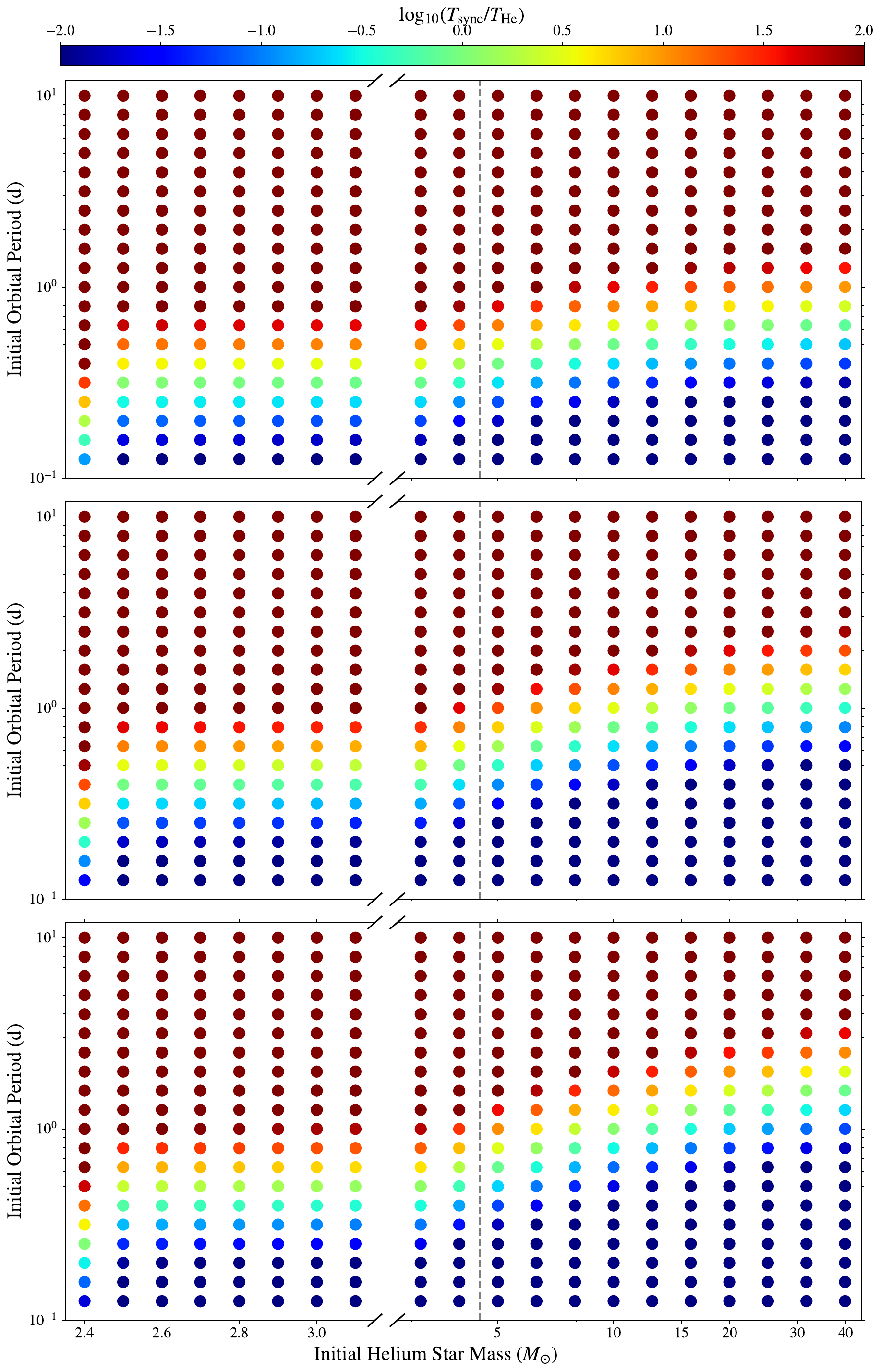}
\caption{{\bf The ratio between the synchronization timescale and lifetime of helium star as a function of the initial helium star mass and initial orbital period} {Three different companion masses, including $M_{\rm com}=1.4\,M_\odot$ (top panel), $8\,M_\odot$ (middle panel), and $30\,M_\odot$ (bottom panel), are considered. The boundary between FBOTs and other magnetar-driven SESNe is marked as the gray dashed line.}}
\label{fig:Dependence_Lifetime_Supplement}
\end{figure*}

\clearpage

\begin{figure*}
\centering
\includegraphics[width=0.77\linewidth, trim = 6 0 8 10, clip]{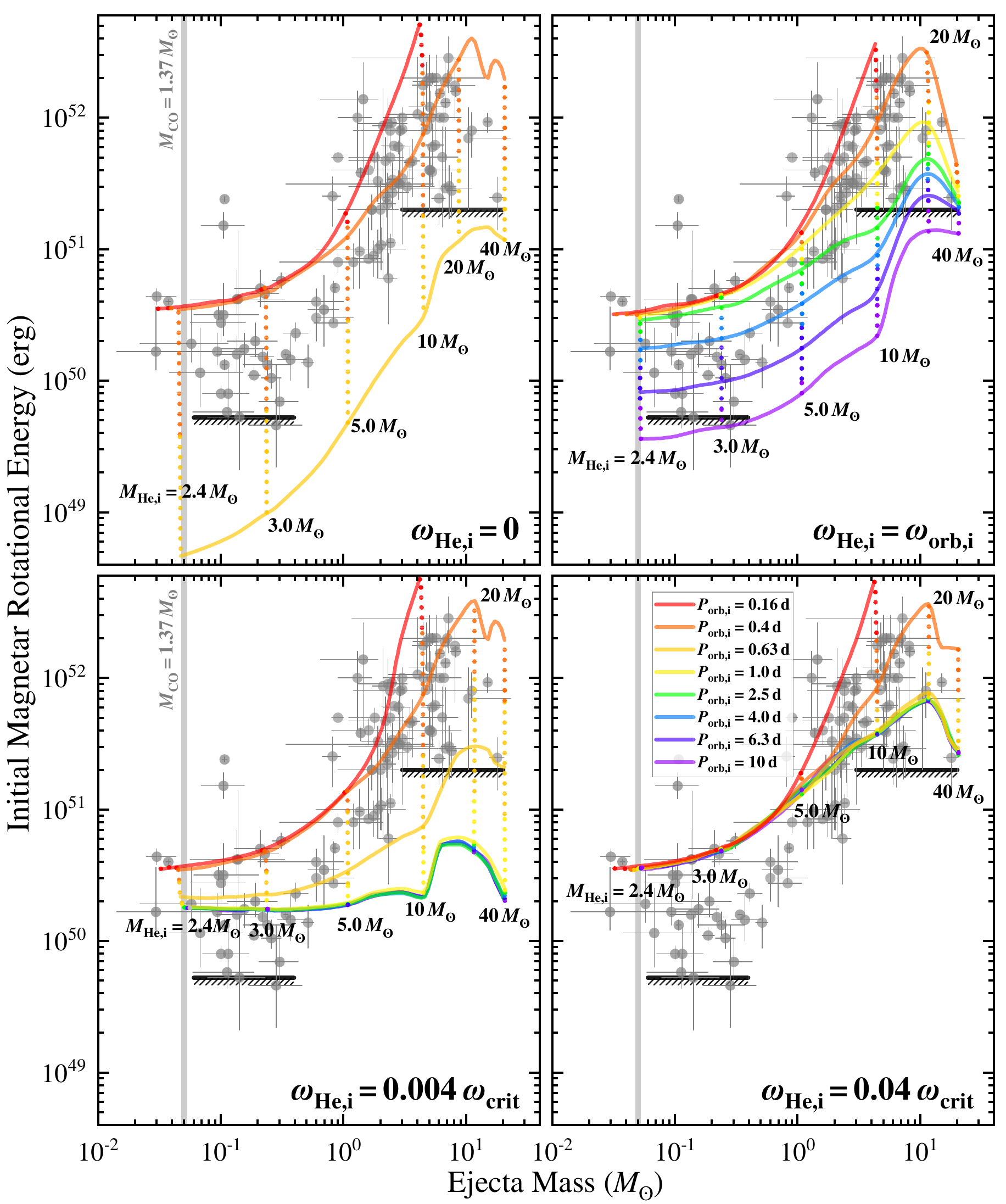}
\caption{{\bf Dependence of simulated $E_{\rm{rot,i}}-M_{\rm{ej}}$ relationships on initial rotation of helium stars.} { Four different initial rotation conditions, including $\omega_{\rm He,i}=0$ (left top panel), $\omega_{\rm He,i}=\omega_{\rm orb,i}$ (i.e., initial tidal synchronization; right top panel), $\omega_{\rm He,i}=0.004\,\omega_{\rm crit}$, and $\omega_{\rm He,i}=0.04\,\omega_{\rm crit}$ are considered. The labels are consistent with those shown in \ref{fig:Dependence_Model_Supplement}.} }
\label{fig:Dependence_Initial_Rotation_Supplement}
\end{figure*}

\clearpage

\begin{figure*}
\centering
\includegraphics[width=0.77\linewidth, trim = 6 0 8 0, clip]{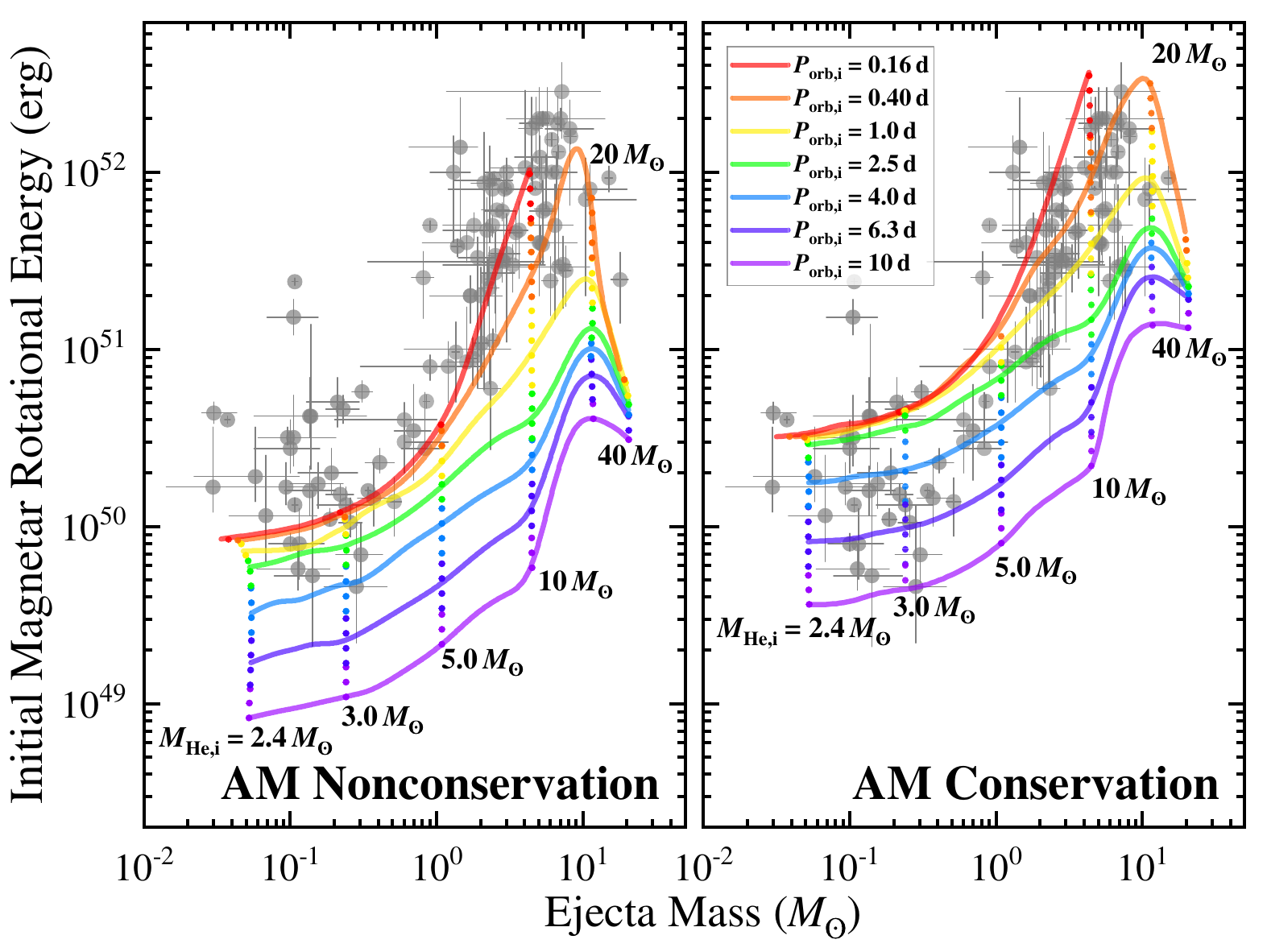}
\caption{{\bf Dependence of simulated $E_{\rm{rot,i}}-M_{\rm{ej}}$ relationships on AM conservation.} {Two different AM conservation scenarios are shown: 50\% conserved (left panel) and 100\% conserved (right panel). The labels are consistent with those shown in \ref{fig:Dependence_Model_Supplement}.}}
\label{fig:Dependence_Conservation_Supplement}
\end{figure*}

\clearpage

\begin{figure*}
\centering
\includegraphics[width=01\linewidth, trim = 6 0 8 0, clip]{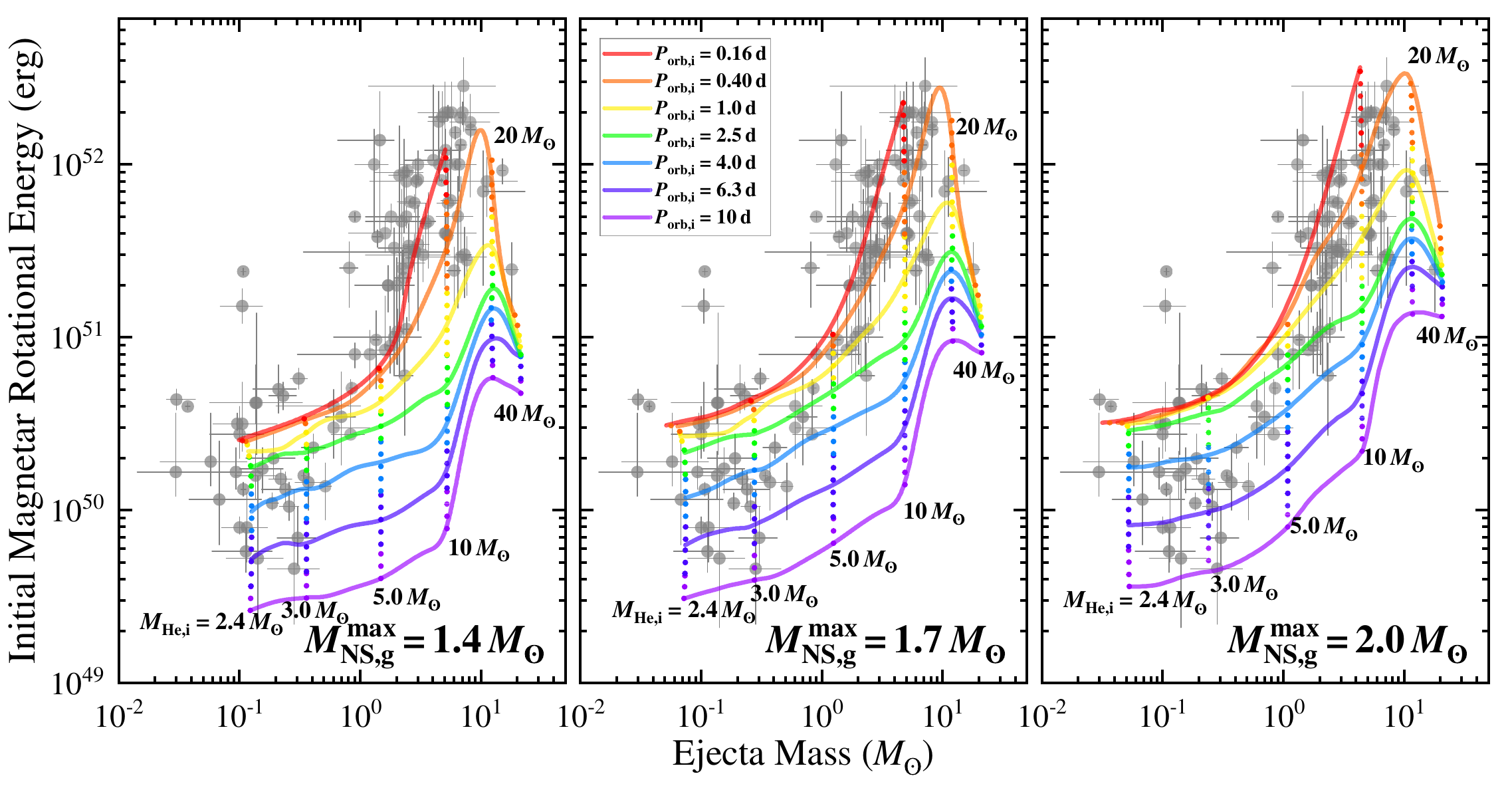}
\caption{{\bf Dependence of simulated $E_{\rm{rot,i}}-M_{\rm{ej}}$ relationships on the maximum NS gravitational mass.} {Three different maximum NS gravitational masses, including $1.4\,M_\odot$ (left panel), $1.7\,M_\odot$ (middle panel) and $2.0\,M_\odot$ (right panel), are adopted. The labels are consistent with those shown in \ref{fig:Dependence_Model_Supplement}.}}
\label{fig:Dependence_NS_Maximum_Mass_Supplement}
\end{figure*}

\clearpage

\begin{figure*}
\centering
\includegraphics[width=0.8\linewidth, trim = 0 0 0 0, clip]{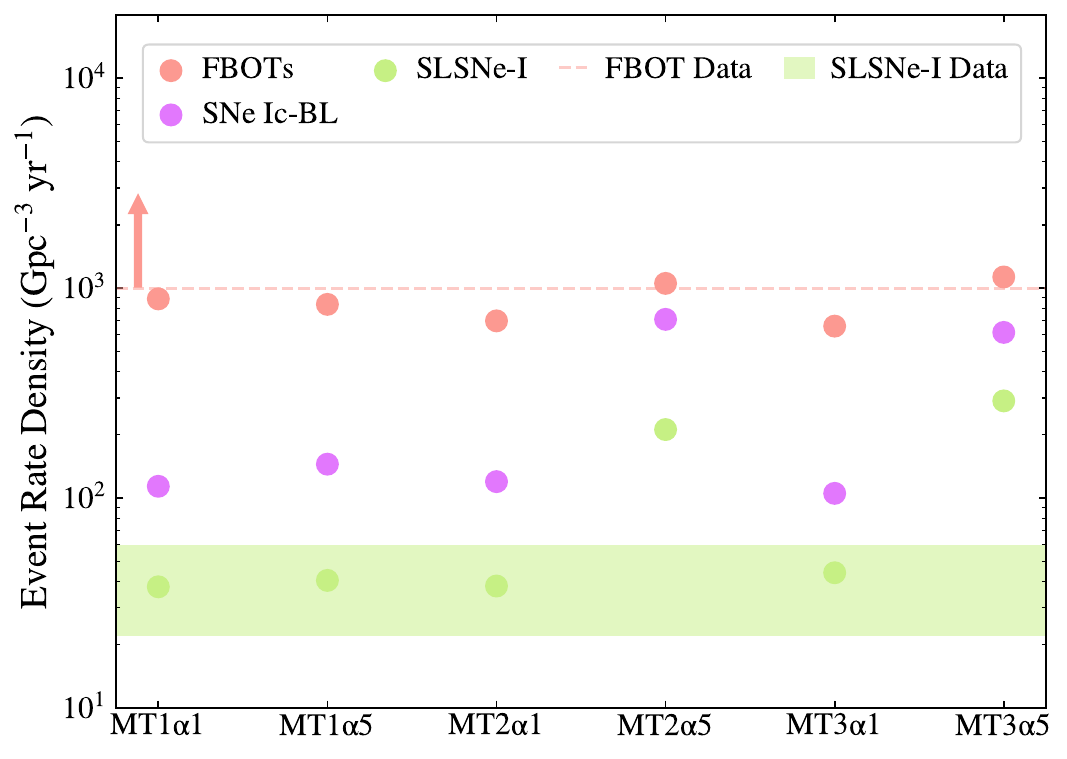}
\caption{{\bf Simulated Local Event rate densities of magnetar-driven SESNe according to different models.} {The labels are consistent with those listed in Figure \ref{fig:EventRate_MainText}. The red dashed line represents the lower limit of the observed FBOT event rate density at a redshift of $0.05<z<1.56$, while the green area indicates the observed range of local event rate density for SLSNe-I.}}
\label{fig:EventRate_Comp_Supplement}
\end{figure*}

\clearpage

\begin{figure*}
\centering
\includegraphics[width=0.95\linewidth, trim = 0 0 0 0, clip]{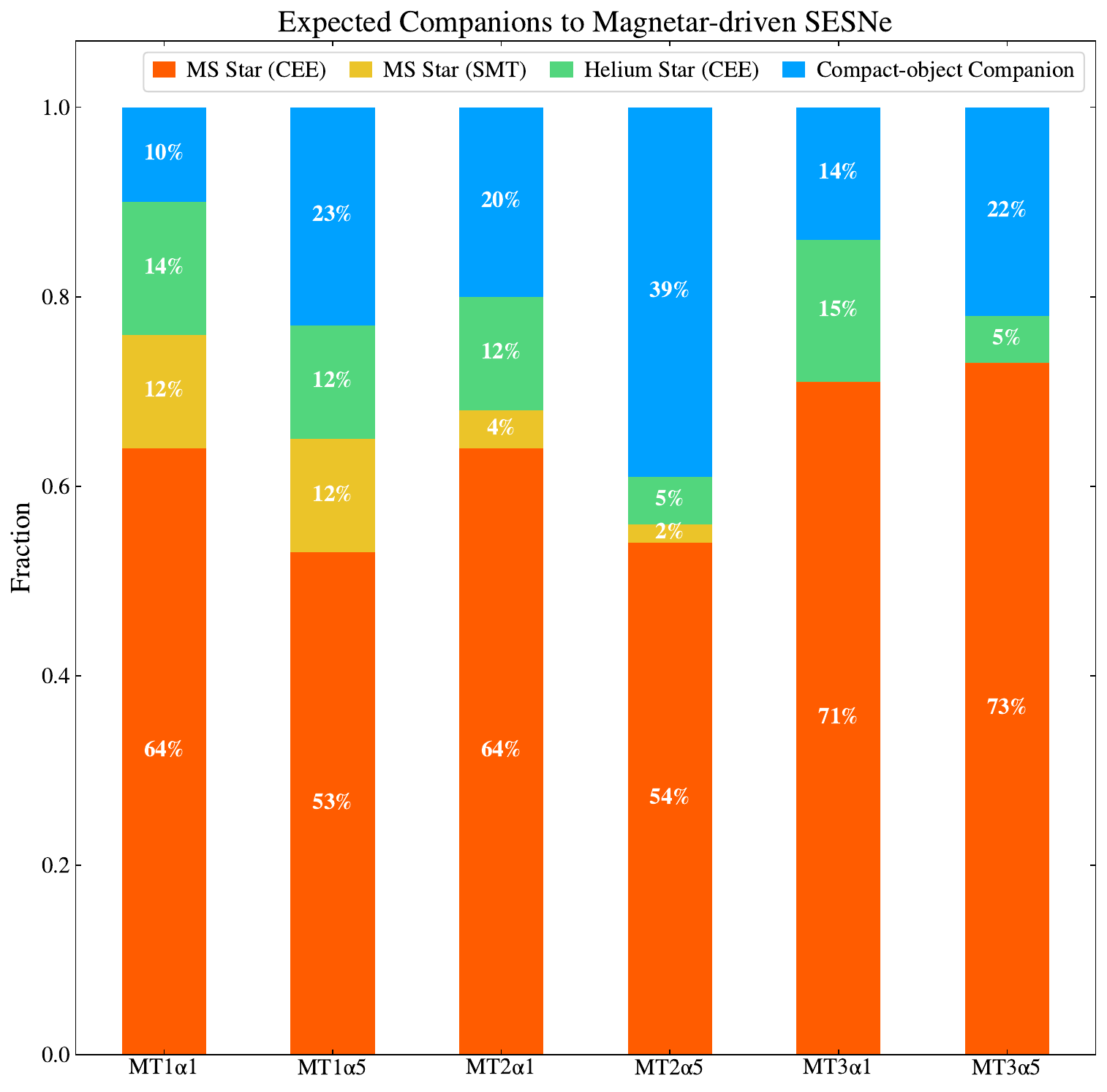}
\caption{{\bf Possible companions at the moment of magnetar-driven SESNe according to different models.} {The labels are consistent with those listed in Figure \ref{fig:Companions_MainText}. }}
\label{fig:Channel_Ratio_Supplement}
\end{figure*}

\clearpage

\begin{figure*}
\centering
\includegraphics[width=0.95\linewidth, trim = 0 0 0 0, clip]{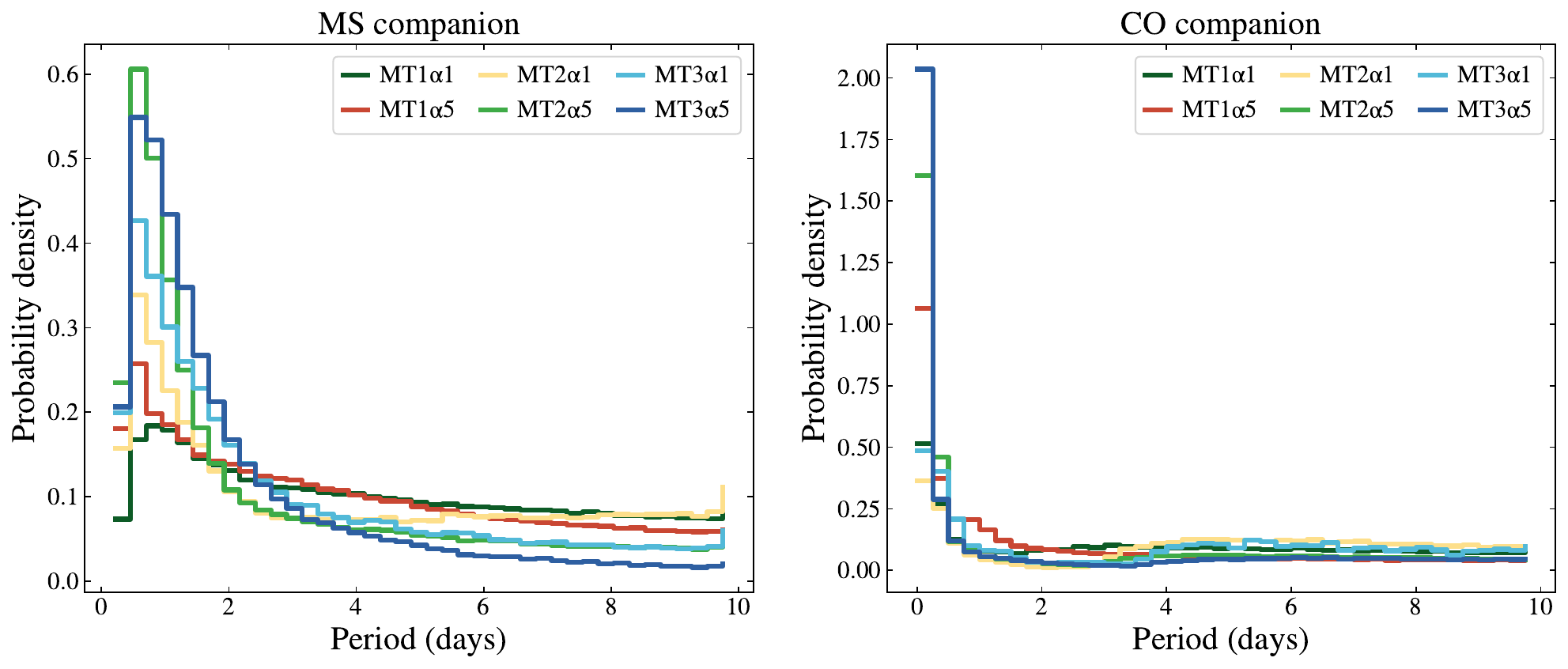}
\caption{{\bf Initial orbital period distributions of helium star binaries according to different models.} {The period distribution of helium star binaries with different companions (left panel: main sequence (MS) companion, right panel: compact objects (CO) companion) from different models (see labels).}}
\label{fig:Period_distribution_Supplement}
\end{figure*}

\clearpage

\begin{figure*}
\centering
\includegraphics[width=0.96\linewidth, trim = 0 0 0 0, clip]{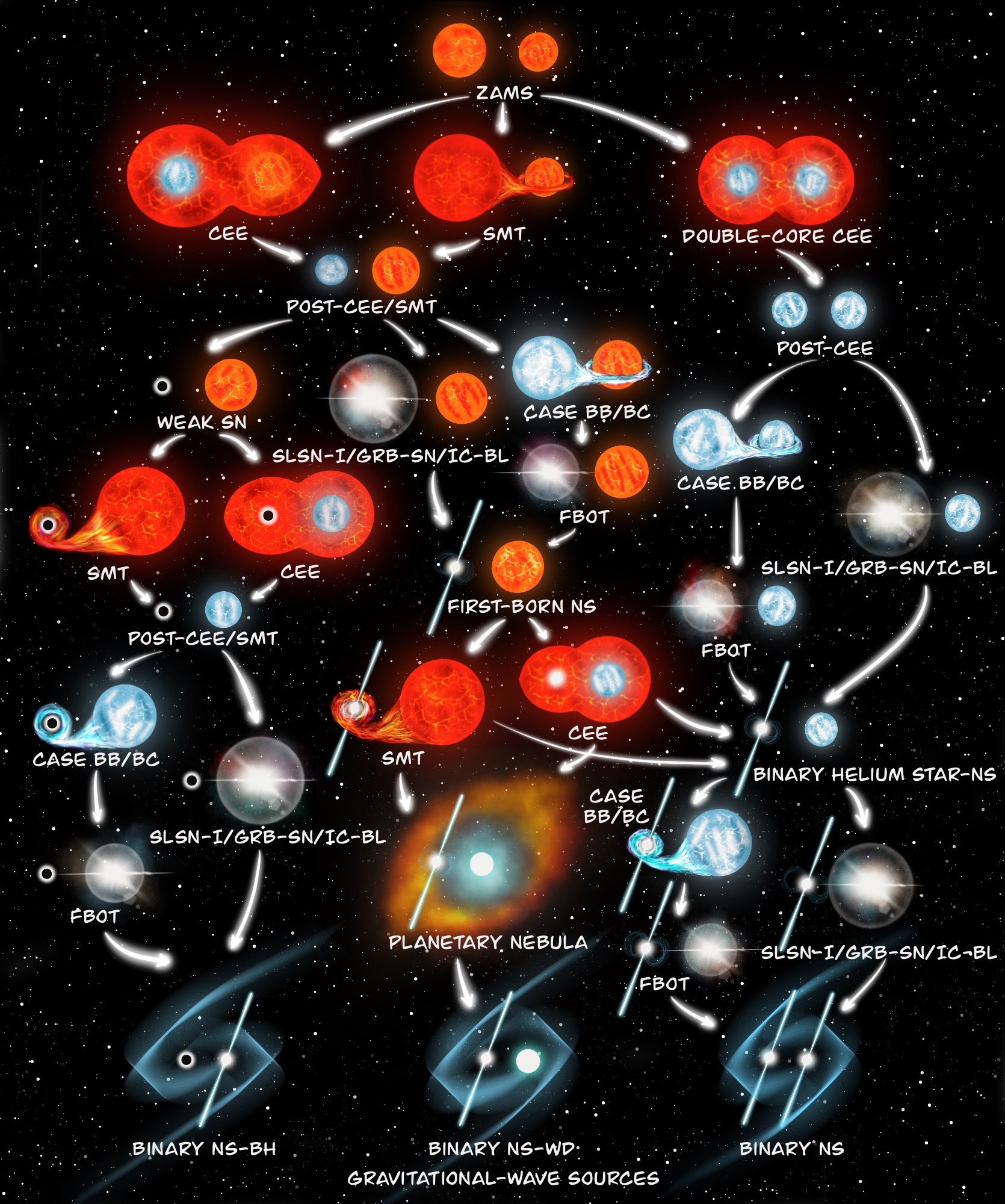}
\caption{{\bf Illustration of the formation channels of magnetar-driven SESNe and close-orbit NS binaries.} {Magnetar-driven SESNe could be the first and/or the second SNe in binary systems. These systems could finally leave behind close-orbit NS binaries if the systems are not disintegrated by SN kicks.  } }
\label{fig:illustration_Supplement}
\end{figure*}

\clearpage

\begin{figure*}
    \centering
    \includegraphics[width = 0.85\linewidth, trim = 0 40 0 0, clip]{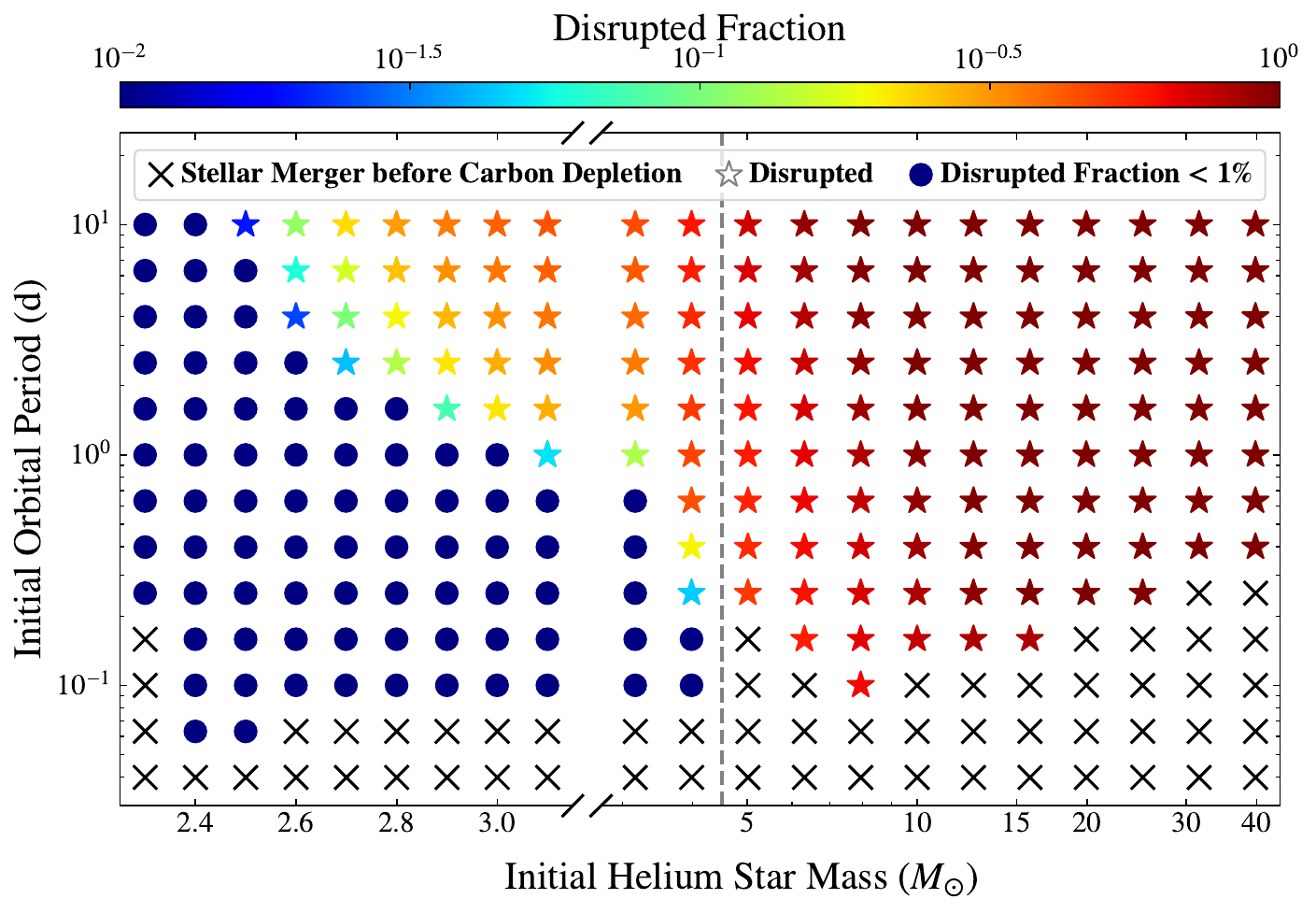}
    \includegraphics[width = 0.85\linewidth, trim = 0 0 0 57, clip]{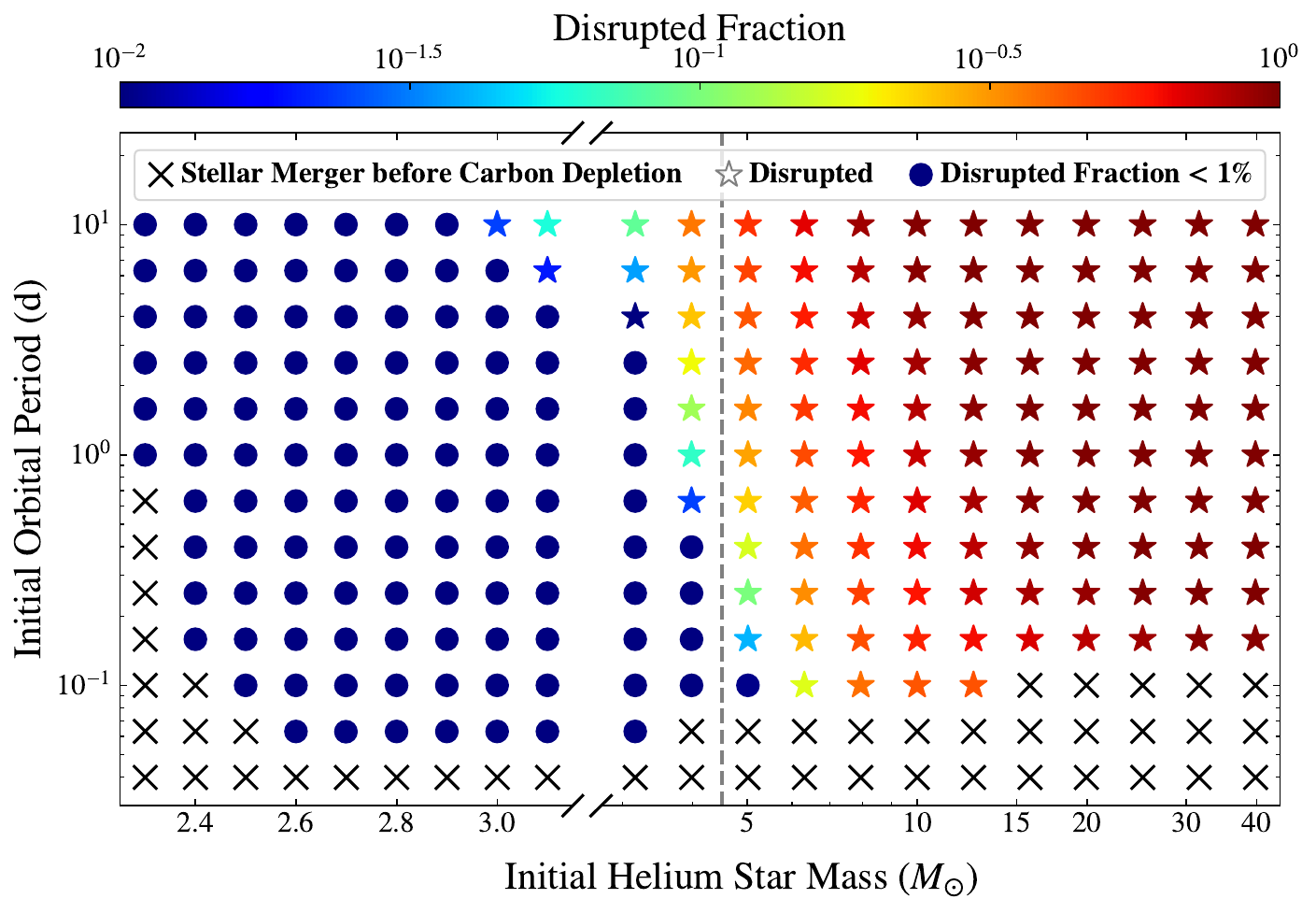}
    \caption{\textbf{Disrupted fraction of binary systems as a function of the initial helium star mass and initial orbital period.} Two companions are considered, i.e., a $1.4\,M_\odot$ NS (top panel) and a $8\,M_\odot$ BH (bottom panel). Star points show the disrupted systems with the color contour indicating the disrupted fraction. Black crosses and circle points represent the cases of stellar merger and binary systems with disrupted fraction $<1\%$, respectively.}
    \label{fig:Disruptfraction_Supplement}
\end{figure*}

\clearpage

\begin{figure*}
    \centering
    \includegraphics[width = 0.85\linewidth, trim = 0 40 0 0, clip]{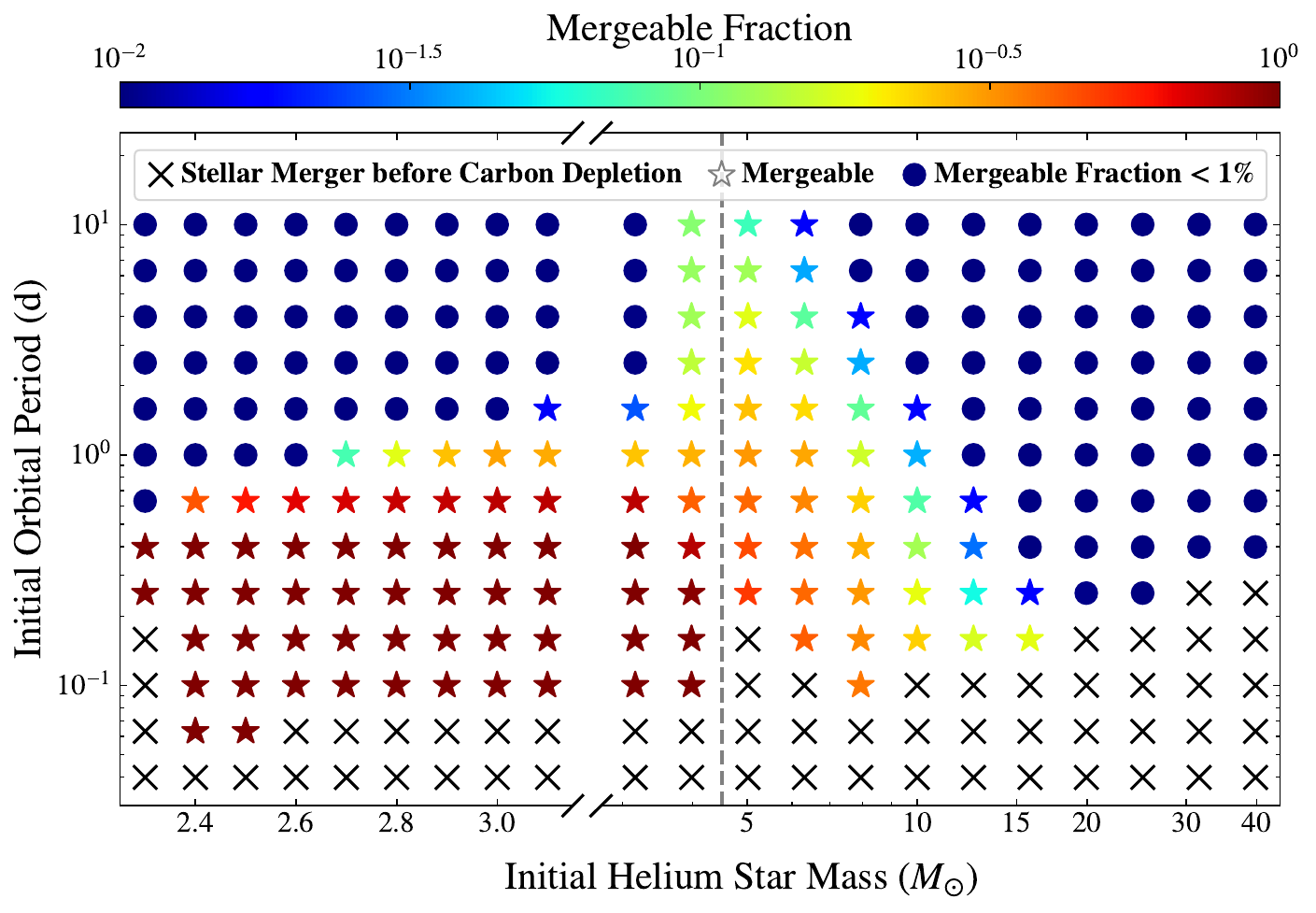}
    \includegraphics[width = 0.85\linewidth, trim = 0 0 0 57, clip]{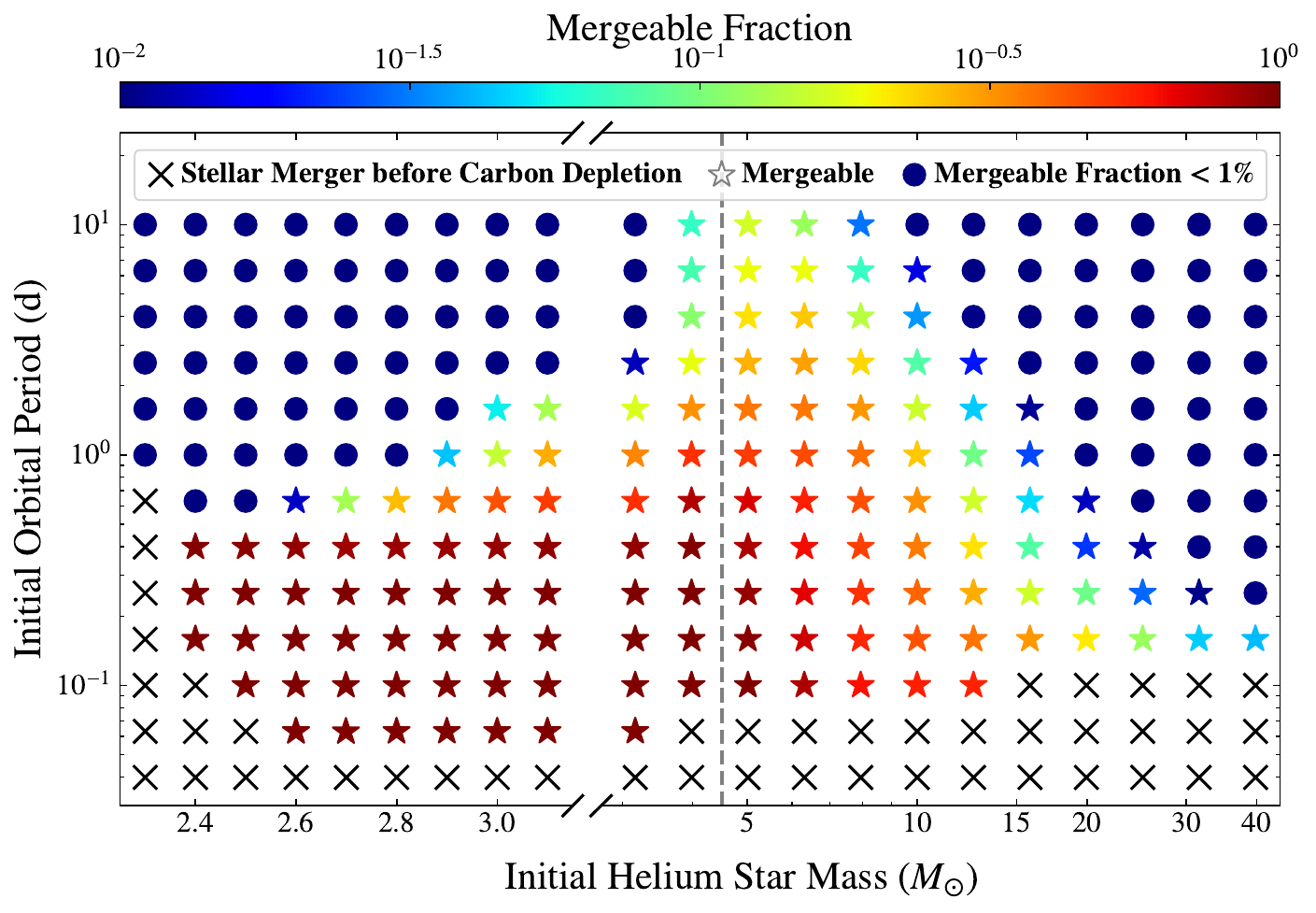}
    \caption{\textbf{Merger fraction of binary systems as a function of the initial helium star mass and initial orbital period.} Two companions are considered, i.e., a $1.4\,M_\odot$ NS (top panel) and a $8\,M_\odot$ BH (bottom panel). Star points show the merger systems within Hubble time with the color contour indicating the merger fraction. Black crosses and circle points represent the cases of stellar merger and binary systems with merger fraction $<1\%$, respectively.}
    \label{fig:Mergerfraction_Supplement}
\end{figure*}

\clearpage

\begin{figure*}
    \centering
    \includegraphics[width = 0.85\linewidth]{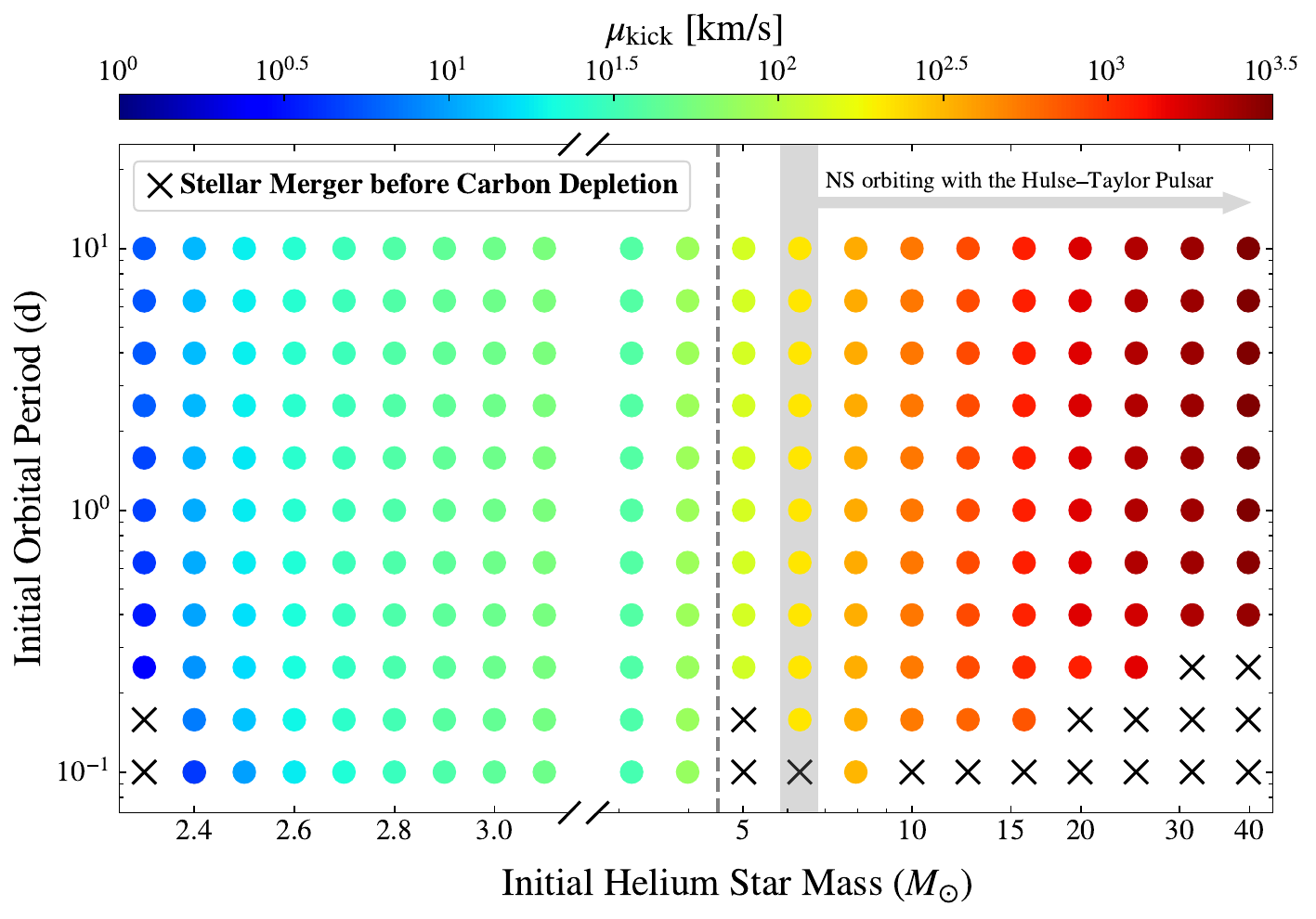}
    \caption{\textbf{$\mu_{\mathrm{kick}}$ of newborn NSs as a function of the initial helium star mass and initial orbital period.} Black crosses represent the cases of stellar merger. The color of circle points depicts the mean natal kick for newborn NSs. The gray vertical line and arrow show the initial helium star mass of NSs with $\mu_{\mathrm{kick}}$ larger than $200\ \mathrm{km\ s}^{-1}$.}
    \label{fig:Mukick_Supplement}
\end{figure*}

\clearpage

\end{supplinfo}

\newpage 
\newcommand{\araa}{Annu. Rev. Astron. Astr.}   
\newcommand{\aap}{Astron. Astrophys.}
\newcommand{\aj}{Astron. J.}         
\newcommand{\apj}{Astrophys. J.}
\newcommand{\apjl}{Astrophys. J. Lett.}      
\newcommand{\apjs}{Astrophys. J. Suppl. S.}
\newcommand{\mnras}{Mon. Not. R. Astron. Soc.}   
\newcommand{\nat}{Nature}
\newcommand{\pasp}{Publ. Astron. Soc. Pac.}  
\newcommand{\pasj}{Publ. Astron. Soc. Jpn.}
\newcommand{\aaps}{Astron. Astrophys. Supp.}
\newcommand{\prd}{Phys. Rev. D}
\newcommand{\prl}{Phys. Rev. Lett.}
\newcommand{\zap}{Zeitung f\"ur Astrophysik}
\newcommand{\prc}{Phys. Rev. C}

\bibliography{ms}

\begin{thebibliography}{100}
\expandafter\ifx\csname url\endcsname\relax
  \def\url#1{\texttt{#1}}\fi
\expandafter\ifx\csname urlprefix\endcsname\relax\def\urlprefix{URL }\fi
\providecommand{\bibinfo}[2]{#2}
\providecommand{\eprint}[2][]{\url{#2}}

\bibitem{filippenko1997}
\bibinfo{author}{{Filippenko}, A.~V.}
\newblock \bibinfo{title}{{Optical Spectra of Supernovae}}.
\newblock \emph{\bibinfo{journal}{\araa}} \textbf{\bibinfo{volume}{35}},
  \bibinfo{pages}{309--355} (\bibinfo{year}{1997}).

\bibitem{galyam2019}
\bibinfo{author}{{Gal-Yam}, A.}
\newblock \bibinfo{title}{{The Most Luminous Supernovae}}.
\newblock \emph{\bibinfo{journal}{\araa}} \textbf{\bibinfo{volume}{57}},
  \bibinfo{pages}{305--333} (\bibinfo{year}{2019}).

\bibitem{kasen2010}
\bibinfo{author}{{Kasen}, D.} \& \bibinfo{author}{{Bildsten}, L.}
\newblock \bibinfo{title}{{Supernova Light Curves Powered by Young Magnetars}}.
\newblock \emph{\bibinfo{journal}{\apj}} \textbf{\bibinfo{volume}{717}},
  \bibinfo{pages}{245--249} (\bibinfo{year}{2010}).

\bibitem{zhang2018}
\bibinfo{author}{{Zhang}, B.}
\newblock \emph{\bibinfo{title}{{The Physics of Gamma-Ray Bursts}}}
  (\bibinfo{publisher}{Cambridge University Press}, \bibinfo{year}{2018}).

\bibitem{woosley1993}
\bibinfo{author}{{Woosley}, S.~E.}
\newblock \bibinfo{title}{{Gamma-Ray Bursts from Stellar Mass Accretion Disks
  around Black Holes}}.
\newblock \emph{\bibinfo{journal}{\apj}} \textbf{\bibinfo{volume}{405}},
  \bibinfo{pages}{273} (\bibinfo{year}{1993}).

\bibitem{popham1999}
\bibinfo{author}{{Popham}, R.}, \bibinfo{author}{{Woosley}, S.~E.} \&
  \bibinfo{author}{{Fryer}, C.}
\newblock \bibinfo{title}{{Hyperaccreting Black Holes and Gamma-Ray Bursts}}.
\newblock \emph{\bibinfo{journal}{\apj}} \textbf{\bibinfo{volume}{518}},
  \bibinfo{pages}{356--374} (\bibinfo{year}{1999}).

\bibitem{usov1992}
\bibinfo{author}{{Usov}, V.~V.}
\newblock \bibinfo{title}{{Millisecond pulsars with extremely strong magnetic
  fields as a cosmological source of {\ensuremath{\gamma}}-ray bursts}}.
\newblock \emph{\bibinfo{journal}{\nat}} \textbf{\bibinfo{volume}{357}},
  \bibinfo{pages}{472--474} (\bibinfo{year}{1992}).

\bibitem{dai1998a}
\bibinfo{author}{{Dai}, Z.~G.} \& \bibinfo{author}{{Lu}, T.}
\newblock \bibinfo{title}{{Gamma-ray burst afterglows and evolution of
  postburst fireballs with energy injection from strongly magnetic millisecond
  pulsars}}.
\newblock \emph{\bibinfo{journal}{\aap}} \textbf{\bibinfo{volume}{333}},
  \bibinfo{pages}{L87--L90} (\bibinfo{year}{1998}).

\bibitem{wheeler2000}
\bibinfo{author}{{Wheeler}, J.~C.}, \bibinfo{author}{{Yi}, I.},
  \bibinfo{author}{{H{\"o}flich}, P.} \& \bibinfo{author}{{Wang}, L.}
\newblock \bibinfo{title}{{Asymmetric Supernovae, Pulsars, Magnetars, and
  Gamma-Ray Bursts}}.
\newblock \emph{\bibinfo{journal}{\apj}} \textbf{\bibinfo{volume}{537}},
  \bibinfo{pages}{810--823} (\bibinfo{year}{2000}).

\bibitem{zhang2001}
\bibinfo{author}{{Zhang}, B.} \& \bibinfo{author}{{M{\'e}sz{\'a}ros}, P.}
\newblock \bibinfo{title}{{Gamma-Ray Burst Afterglow with Continuous Energy
  Injection: Signature of a Highly Magnetized Millisecond Pulsar}}.
\newblock \emph{\bibinfo{journal}{\apjl}} \textbf{\bibinfo{volume}{552}},
  \bibinfo{pages}{L35--L38} (\bibinfo{year}{2001}).

\bibitem{metzger2011}
\bibinfo{author}{{Metzger}, B.~D.}, \bibinfo{author}{{Giannios}, D.},
  \bibinfo{author}{{Thompson}, T.~A.}, \bibinfo{author}{{Bucciantini}, N.} \&
  \bibinfo{author}{{Quataert}, E.}
\newblock \bibinfo{title}{{The protomagnetar model for gamma-ray bursts}}.
\newblock \emph{\bibinfo{journal}{\mnras}} \textbf{\bibinfo{volume}{413}},
  \bibinfo{pages}{2031--2056} (\bibinfo{year}{2011}).

\bibitem{lv2014}
\bibinfo{author}{{L{\"u}}, H.-J.} \& \bibinfo{author}{{Zhang}, B.}
\newblock \bibinfo{title}{{A Test of the Millisecond Magnetar Central Engine
  Model of Gamma-Ray Bursts with Swift Data}}.
\newblock \emph{\bibinfo{journal}{\apj}} \textbf{\bibinfo{volume}{785}},
  \bibinfo{pages}{74} (\bibinfo{year}{2014}).

\bibitem{zhang2022}
\bibinfo{author}{{Zhang}, Z.-D.}, \bibinfo{author}{{Yu}, Y.-W.} \&
  \bibinfo{author}{{Liu}, L.-D.}
\newblock \bibinfo{title}{{The Effects of a Magnetar Engine on the Gamma-Ray
  Burst-associated Supernovae: Application to Double-peaked SN 2006aj}}.
\newblock \emph{\bibinfo{journal}{\apj}} \textbf{\bibinfo{volume}{936}},
  \bibinfo{pages}{54} (\bibinfo{year}{2022}).

\bibitem{drout2014}
\bibinfo{author}{{Drout}, M.~R.} \emph{et~al.}
\newblock \bibinfo{title}{{Rapidly Evolving and Luminous Transients from
  Pan-STARRS1}}.
\newblock \emph{\bibinfo{journal}{\apj}} \textbf{\bibinfo{volume}{794}},
  \bibinfo{pages}{23} (\bibinfo{year}{2014}).

\bibitem{pursiainen2018}
\bibinfo{author}{{Pursiainen}, M.} \emph{et~al.}
\newblock \bibinfo{title}{{Rapidly evolving transients in the Dark Energy
  Survey}}.
\newblock \emph{\bibinfo{journal}{\mnras}} \textbf{\bibinfo{volume}{481}},
  \bibinfo{pages}{894--917} (\bibinfo{year}{2018}).

\bibitem{liu2022}
\bibinfo{author}{{Liu}, J.-F.}, \bibinfo{author}{{Zhu}, J.-P.},
  \bibinfo{author}{{Liu}, L.-D.}, \bibinfo{author}{{Yu}, Y.-W.} \&
  \bibinfo{author}{{Zhang}, B.}
\newblock \bibinfo{title}{{Magnetar Engines in Fast Blue Optical Transients and
  Their Connections with SLSNe, SNe Ic-BL, and lGRBs}}.
\newblock \emph{\bibinfo{journal}{\apjl}} \textbf{\bibinfo{volume}{935}},
  \bibinfo{pages}{L34} (\bibinfo{year}{2022}).

\bibitem{sawada2022}
\bibinfo{author}{{Sawada}, R.}, \bibinfo{author}{{Kashiyama}, K.} \&
  \bibinfo{author}{{Suwa}, Y.}
\newblock \bibinfo{title}{{On the Energy Source of Ultrastripped Supernovae}}.
\newblock \emph{\bibinfo{journal}{\apj}} \textbf{\bibinfo{volume}{927}},
  \bibinfo{pages}{223} (\bibinfo{year}{2022}).

\bibitem{wiseman2020}
\bibinfo{author}{{Wiseman}, P.} \emph{et~al.}
\newblock \bibinfo{title}{{The host galaxies of 106 rapidly evolving transients
  discovered by the Dark Energy Survey}}.
\newblock \emph{\bibinfo{journal}{\mnras}} \textbf{\bibinfo{volume}{498}},
  \bibinfo{pages}{2575--2593} (\bibinfo{year}{2020}).

\bibitem{fang2019}
\bibinfo{author}{{Fang}, Q.}, \bibinfo{author}{{Maeda}, K.},
  \bibinfo{author}{{Kuncarayakti}, H.}, \bibinfo{author}{{Sun}, F.} \&
  \bibinfo{author}{{Gal-Yam}, A.}
\newblock \bibinfo{title}{{A hybrid envelope-stripping mechanism for massive
  stars from supernova nebular spectroscopy}}.
\newblock \emph{\bibinfo{journal}{Nature Astronomy}}
  \textbf{\bibinfo{volume}{3}}, \bibinfo{pages}{434--439}
  (\bibinfo{year}{2019}).

\bibitem{Spruit2002}
\bibinfo{author}{{Spruit}, H.~C.}
\newblock \bibinfo{title}{{Dynamo action by differential rotation in a stably
  stratified stellar interior}}.
\newblock \emph{\bibinfo{journal}{\aap}} \textbf{\bibinfo{volume}{381}},
  \bibinfo{pages}{923--932} (\bibinfo{year}{2002}).

\bibitem{woosley2006GRB}
\bibinfo{author}{{Woosley}, S.~E.} \& \bibinfo{author}{{Heger}, A.}
\newblock \bibinfo{title}{{The Progenitor Stars of Gamma-Ray Bursts}}.
\newblock \emph{\bibinfo{journal}{\apj}} \textbf{\bibinfo{volume}{637}},
  \bibinfo{pages}{914--921} (\bibinfo{year}{2006}).

\bibitem{yoon2005}
\bibinfo{author}{{Yoon}, S.~C.} \& \bibinfo{author}{{Langer}, N.}
\newblock \bibinfo{title}{{Evolution of rapidly rotating metal-poor massive
  stars towards gamma-ray bursts}}.
\newblock \emph{\bibinfo{journal}{\aap}} \textbf{\bibinfo{volume}{443}},
  \bibinfo{pages}{643--648} (\bibinfo{year}{2005}).

\bibitem{aguileradena2018}
\bibinfo{author}{{Aguilera-Dena}, D.~R.}, \bibinfo{author}{{Langer}, N.},
  \bibinfo{author}{{Moriya}, T.~J.} \& \bibinfo{author}{{Schootemeijer}, A.}
\newblock \bibinfo{title}{{Related Progenitor Models for Long-duration
  Gamma-Ray Bursts and Type Ic Superluminous Supernovae}}.
\newblock \emph{\bibinfo{journal}{\apj}} \textbf{\bibinfo{volume}{858}},
  \bibinfo{pages}{115} (\bibinfo{year}{2018}).

\bibitem{Song2023}
\bibinfo{author}{{Song}, C.-Y.} \& \bibinfo{author}{{Liu}, T.}
\newblock \bibinfo{title}{{Long-duration Gamma-ray Burst Progenitors and
  Magnetar Formation}}.
\newblock \emph{\bibinfo{journal}{arXiv e-prints}}
  \bibinfo{pages}{arXiv:2301.05401} (\bibinfo{year}{2023}).

\bibitem{cantiello2007}
\bibinfo{author}{{Cantiello}, M.}, \bibinfo{author}{{Yoon}, S.~C.},
  \bibinfo{author}{{Langer}, N.} \& \bibinfo{author}{{Livio}, M.}
\newblock \bibinfo{title}{{Binary star progenitors of long gamma-ray bursts}}.
\newblock \emph{\bibinfo{journal}{\aap}} \textbf{\bibinfo{volume}{465}},
  \bibinfo{pages}{L29--L33} (\bibinfo{year}{2007}).

\bibitem{eldridge2011}
\bibinfo{author}{{Eldridge}, J.~J.}, \bibinfo{author}{{Langer}, N.} \&
  \bibinfo{author}{{Tout}, C.~A.}
\newblock \bibinfo{title}{{Runaway stars as progenitors of supernovae and
  gamma-ray bursts}}.
\newblock \emph{\bibinfo{journal}{\mnras}} \textbf{\bibinfo{volume}{414}},
  \bibinfo{pages}{3501--3520} (\bibinfo{year}{2011}).

\bibitem{demink2013}
\bibinfo{author}{{de Mink}, S.~E.}, \bibinfo{author}{{Langer}, N.},
  \bibinfo{author}{{Izzard}, R.~G.}, \bibinfo{author}{{Sana}, H.} \&
  \bibinfo{author}{{de Koter}, A.}
\newblock \bibinfo{title}{{The Rotation Rates of Massive Stars: The Role of
  Binary Interaction through Tides, Mass Transfer, and Mergers}}.
\newblock \emph{\bibinfo{journal}{\apj}} \textbf{\bibinfo{volume}{764}},
  \bibinfo{pages}{166} (\bibinfo{year}{2013}).

\bibitem{mandel2016}
\bibinfo{author}{{Mandel}, I.} \& \bibinfo{author}{{de Mink}, S.~E.}
\newblock \bibinfo{title}{{Merging binary black holes formed through chemically
  homogeneous evolution in short-period stellar binaries}}.
\newblock \emph{\bibinfo{journal}{\mnras}} \textbf{\bibinfo{volume}{458}},
  \bibinfo{pages}{2634--2647} (\bibinfo{year}{2016}).

\bibitem{eldridge2019}
\bibinfo{author}{{Eldridge}, J.~J.}, \bibinfo{author}{{Stanway}, E.~R.} \&
  \bibinfo{author}{{Tang}, P.~N.}
\newblock \bibinfo{title}{{A consistent estimate for gravitational wave and
  electromagnetic transient rates}}.
\newblock \emph{\bibinfo{journal}{\mnras}} \textbf{\bibinfo{volume}{482}},
  \bibinfo{pages}{870--880} (\bibinfo{year}{2019}).

\bibitem{ghodla2023}
\bibinfo{author}{{Ghodla}, S.}, \bibinfo{author}{{Eldridge}, J.~J.},
  \bibinfo{author}{{Stanway}, E.~R.} \& \bibinfo{author}{{Stevance}, H.~F.}
\newblock \bibinfo{title}{{Evaluating chemically homogeneous evolution in
  stellar binaries: electromagnetic implications - ionizing photons, SLSN-I,
  GRB, Ic-BL}}.
\newblock \emph{\bibinfo{journal}{\mnras}} \textbf{\bibinfo{volume}{518}},
  \bibinfo{pages}{860--877} (\bibinfo{year}{2023}).

\bibitem{chrimes2020}
\bibinfo{author}{{Chrimes}, A.~A.}, \bibinfo{author}{{Stanway}, E.~R.} \&
  \bibinfo{author}{{Eldridge}, J.~J.}
\newblock \bibinfo{title}{{Binary population synthesis models for core-collapse
  gamma-ray burst progenitors}}.
\newblock \emph{\bibinfo{journal}{\mnras}} \textbf{\bibinfo{volume}{491}},
  \bibinfo{pages}{3479--3495} (\bibinfo{year}{2020}).

\bibitem{lunnan2014}
\bibinfo{author}{{Lunnan}, R.} \emph{et~al.}
\newblock \bibinfo{title}{{Hydrogen-poor Superluminous Supernovae and
  Long-duration Gamma-Ray Bursts Have Similar Host Galaxies}}.
\newblock \emph{\bibinfo{journal}{\apj}} \textbf{\bibinfo{volume}{787}},
  \bibinfo{pages}{138} (\bibinfo{year}{2014}).

\bibitem{chen2017}
\bibinfo{author}{{Chen}, T.-W.} \emph{et~al.}
\newblock \bibinfo{title}{{Superluminous supernova progenitors have a
  half-solar metallicity threshold}}.
\newblock \emph{\bibinfo{journal}{\mnras}} \textbf{\bibinfo{volume}{470}},
  \bibinfo{pages}{3566--3573} (\bibinfo{year}{2017}).

\bibitem{japelj2018}
\bibinfo{author}{{Japelj}, J.} \emph{et~al.}
\newblock \bibinfo{title}{{Host galaxies of SNe Ic-BL with and without long
  gamma-ray bursts}}.
\newblock \emph{\bibinfo{journal}{\aap}} \textbf{\bibinfo{volume}{617}},
  \bibinfo{pages}{A105} (\bibinfo{year}{2018}).

\bibitem{modjaz2020}
\bibinfo{author}{{Modjaz}, M.} \emph{et~al.}
\newblock \bibinfo{title}{{Host Galaxies of Type Ic and Broad-lined Type Ic
  Supernovae from the Palomar Transient Factory: Implications for Jet
  Production}}.
\newblock \emph{\bibinfo{journal}{\apj}} \textbf{\bibinfo{volume}{892}},
  \bibinfo{pages}{153} (\bibinfo{year}{2020}).

\bibitem{izzard2004}
\bibinfo{author}{{Izzard}, R.~G.}, \bibinfo{author}{{Ramirez-Ruiz}, E.} \&
  \bibinfo{author}{{Tout}, C.~A.}
\newblock \bibinfo{title}{{Formation rates of core-collapse supernovae and
  gamma-ray bursts}}.
\newblock \emph{\bibinfo{journal}{\mnras}} \textbf{\bibinfo{volume}{348}},
  \bibinfo{pages}{1215--1228} (\bibinfo{year}{2004}).

\bibitem{detmers2008}
\bibinfo{author}{{Detmers}, R.~G.}, \bibinfo{author}{{Langer}, N.},
  \bibinfo{author}{{Podsiadlowski}, P.} \& \bibinfo{author}{{Izzard}, R.~G.}
\newblock \bibinfo{title}{{Gamma-ray bursts from tidally spun-up Wolf-Rayet
  stars?}}
\newblock \emph{\bibinfo{journal}{\aap}} \textbf{\bibinfo{volume}{484}},
  \bibinfo{pages}{831--839} (\bibinfo{year}{2008}).

\bibitem{Bogomazov2009}
\bibinfo{author}{{Bogomazov}, A.~I.} \& \bibinfo{author}{{Popov}, S.~B.}
\newblock \bibinfo{title}{{Magnetars, gamma-ray bursts, and very close
  binaries}}.
\newblock \emph{\bibinfo{journal}{Astronomy Reports}}
  \textbf{\bibinfo{volume}{53}}, \bibinfo{pages}{325--333}
  (\bibinfo{year}{2009}).

\bibitem{qin2018}
\bibinfo{author}{{Qin}, Y.} \emph{et~al.}
\newblock \bibinfo{title}{{The spin of the second-born black hole in coalescing
  binary black holes}}.
\newblock \emph{\bibinfo{journal}{\aap}} \textbf{\bibinfo{volume}{616}},
  \bibinfo{pages}{A28} (\bibinfo{year}{2018}).

\bibitem{bavera2022}
\bibinfo{author}{{Bavera}, S.~S.} \emph{et~al.}
\newblock \bibinfo{title}{{Probing the progenitors of spinning binary
  black-hole mergers with long gamma-ray bursts}}.
\newblock \emph{\bibinfo{journal}{\aap}} \textbf{\bibinfo{volume}{657}},
  \bibinfo{pages}{L8} (\bibinfo{year}{2022}).

\bibitem{fuller2022}
\bibinfo{author}{{Fuller}, J.} \& \bibinfo{author}{{Lu}, W.}
\newblock \bibinfo{title}{{The spins of compact objects born from helium stars
  in binary systems}}.
\newblock \emph{\bibinfo{journal}{\mnras}} \textbf{\bibinfo{volume}{511}},
  \bibinfo{pages}{3951--3964} (\bibinfo{year}{2022}).

\bibitem{bravo2011}
\bibinfo{author}{{Bravo}, E.} \& \bibinfo{author}{{Badenes}, C.}
\newblock \bibinfo{title}{{Is the metallicity of their host galaxies a good
  measure of the metallicity of Type Ia supernovae?}}
\newblock \emph{\bibinfo{journal}{\mnras}} \textbf{\bibinfo{volume}{414}},
  \bibinfo{pages}{1592--1606} (\bibinfo{year}{2011}).

\bibitem{metzger2015}
\bibinfo{author}{{Metzger}, B.~D.}, \bibinfo{author}{{Margalit}, B.},
  \bibinfo{author}{{Kasen}, D.} \& \bibinfo{author}{{Quataert}, E.}
\newblock \bibinfo{title}{{The diversity of transients from magnetar birth in
  core collapse supernovae}}.
\newblock \emph{\bibinfo{journal}{\mnras}} \textbf{\bibinfo{volume}{454}},
  \bibinfo{pages}{3311--3316} (\bibinfo{year}{2015}).

\bibitem{sukhbold2016}
\bibinfo{author}{{Sukhbold}, T.}, \bibinfo{author}{{Ertl}, T.},
  \bibinfo{author}{{Woosley}, S.~E.}, \bibinfo{author}{{Brown}, J.~M.} \&
  \bibinfo{author}{{Janka}, H.~T.}
\newblock \bibinfo{title}{{Core-collapse Supernovae from 9 to 120 Solar Masses
  Based on Neutrino-powered Explosions}}.
\newblock \emph{\bibinfo{journal}{\apj}} \textbf{\bibinfo{volume}{821}},
  \bibinfo{pages}{38} (\bibinfo{year}{2016}).

\bibitem{gomez2022}
\bibinfo{author}{{Gomez}, S.}, \bibinfo{author}{{Berger}, E.},
  \bibinfo{author}{{Nicholl}, M.}, \bibinfo{author}{{Blanchard}, P.~K.} \&
  \bibinfo{author}{{Hosseinzadeh}, G.}
\newblock \bibinfo{title}{{Luminous Supernovae: Unveiling a Population Between
  Superluminous and Normal Core-collapse Supernovae}}.
\newblock \emph{\bibinfo{journal}{arXiv e-prints}}
  \bibinfo{pages}{arXiv:2204.08486} (\bibinfo{year}{2022}).

\bibitem{sun2022}
\bibinfo{author}{{Sun}, N.-C.}, \bibinfo{author}{{Maund}, J.~R.} \&
  \bibinfo{author}{{Crowther}, P.~A.}
\newblock \bibinfo{title}{{A UV census of the environments of stripped-envelope
  supernovae}}.
\newblock \emph{\bibinfo{journal}{arXiv e-prints}}
  \bibinfo{pages}{arXiv:2209.05283} (\bibinfo{year}{2022}).

\bibitem{tauris2015}
\bibinfo{author}{{Tauris}, T.~M.}, \bibinfo{author}{{Langer}, N.} \&
  \bibinfo{author}{{Podsiadlowski}, P.}
\newblock \bibinfo{title}{{Ultra-stripped supernovae: progenitors and fate}}.
\newblock \emph{\bibinfo{journal}{\mnras}} \textbf{\bibinfo{volume}{451}},
  \bibinfo{pages}{2123--2144} (\bibinfo{year}{2015}).

\bibitem{suwa2015}
\bibinfo{author}{{Suwa}, Y.}, \bibinfo{author}{{Yoshida}, T.},
  \bibinfo{author}{{Shibata}, M.}, \bibinfo{author}{{Umeda}, H.} \&
  \bibinfo{author}{{Takahashi}, K.}
\newblock \bibinfo{title}{{Neutrino-driven explosions of ultra-stripped Type Ic
  supernovae generating binary neutron stars}}.
\newblock \emph{\bibinfo{journal}{\mnras}} \textbf{\bibinfo{volume}{454}},
  \bibinfo{pages}{3073--3081} (\bibinfo{year}{2015}).

\bibitem{frohmaier2021}
\bibinfo{author}{{Frohmaier}, C.} \emph{et~al.}
\newblock \bibinfo{title}{{From core collapse to superluminous: the rates of
  massive stellar explosions from the Palomar Transient Factory}}.
\newblock \emph{\bibinfo{journal}{\mnras}} \textbf{\bibinfo{volume}{500}},
  \bibinfo{pages}{5142--5158} (\bibinfo{year}{2021}).

\bibitem{prajs2017}
\bibinfo{author}{{Prajs}, S.} \emph{et~al.}
\newblock \bibinfo{title}{{The volumetric rate of superluminous supernovae at z
  {\ensuremath{\sim}} 1}}.
\newblock \emph{\bibinfo{journal}{\mnras}} \textbf{\bibinfo{volume}{464}},
  \bibinfo{pages}{3568--3579} (\bibinfo{year}{2017}).

\bibitem{cooke2012}
\bibinfo{author}{{Cooke}, J.} \emph{et~al.}
\newblock \bibinfo{title}{{Superluminous supernovae at redshifts of 2.05 and
  3.90}}.
\newblock \emph{\bibinfo{journal}{\nat}} \textbf{\bibinfo{volume}{491}},
  \bibinfo{pages}{228--231} (\bibinfo{year}{2012}).

\bibitem{shivvers2017}
\bibinfo{author}{{Shivvers}, I.} \emph{et~al.}
\newblock \bibinfo{title}{{Revisiting the Lick Observatory Supernova Search
  Volume-limited Sample: Updated Classifications and Revised Stripped-envelope
  Supernova Fractions}}.
\newblock \emph{\bibinfo{journal}{\pasp}} \textbf{\bibinfo{volume}{129}},
  \bibinfo{pages}{054201} (\bibinfo{year}{2017}).

\bibitem{AguileraDena2023}
\bibinfo{author}{{Aguilera-Dena}, D.~R.} \emph{et~al.}
\newblock \bibinfo{title}{{Stripped-envelope stars in different metallicity
  environments. II. Type I supernovae and compact remnants}}.
\newblock \emph{\bibinfo{journal}{\aap}} \textbf{\bibinfo{volume}{671}},
  \bibinfo{pages}{A134} (\bibinfo{year}{2023}).

\bibitem{ertl2020}
\bibinfo{author}{{Ertl}, T.}, \bibinfo{author}{{Woosley}, S.~E.},
  \bibinfo{author}{{Sukhbold}, T.} \& \bibinfo{author}{{Janka}, H.~T.}
\newblock \bibinfo{title}{{The Explosion of Helium Stars Evolved with Mass
  Loss}}.
\newblock \emph{\bibinfo{journal}{\apj}} \textbf{\bibinfo{volume}{890}},
  \bibinfo{pages}{51} (\bibinfo{year}{2020}).

\bibitem{schneider2021}
\bibinfo{author}{{Schneider}, F.~R.~N.}, \bibinfo{author}{{Podsiadlowski}, P.}
  \& \bibinfo{author}{{M{\"u}ller}, B.}
\newblock \bibinfo{title}{{Pre-supernova evolution, compact-object masses, and
  explosion properties of stripped binary stars}}.
\newblock \emph{\bibinfo{journal}{\aap}} \textbf{\bibinfo{volume}{645}},
  \bibinfo{pages}{A5} (\bibinfo{year}{2021}).

\bibitem{tauris2017}
\bibinfo{author}{{Tauris}, T.~M.} \emph{et~al.}
\newblock \bibinfo{title}{{Formation of Double Neutron Star Systems}}.
\newblock \emph{\bibinfo{journal}{\apj}} \textbf{\bibinfo{volume}{846}},
  \bibinfo{pages}{170} (\bibinfo{year}{2017}).

\bibitem{ViganaGomez2018}
\bibinfo{author}{{Vigna-G{\'o}mez}, A.} \emph{et~al.}
\newblock \bibinfo{title}{{On the formation history of Galactic double neutron
  stars}}.
\newblock \emph{\bibinfo{journal}{\mnras}} \textbf{\bibinfo{volume}{481}},
  \bibinfo{pages}{4009--4029} (\bibinfo{year}{2018}).

\bibitem{hulse1975}
\bibinfo{author}{{Hulse}, R.~A.} \& \bibinfo{author}{{Taylor}, J.~H.}
\newblock \bibinfo{title}{{Discovery of a pulsar in a binary system.}}
\newblock \emph{\bibinfo{journal}{\apjl}} \textbf{\bibinfo{volume}{195}},
  \bibinfo{pages}{L51--L53} (\bibinfo{year}{1975}).

\bibitem{yu2017}
\bibinfo{author}{{Yu}, Y.-W.}, \bibinfo{author}{{Zhu}, J.-P.},
  \bibinfo{author}{{Li}, S.-Z.}, \bibinfo{author}{{L{\"u}}, H.-J.} \&
  \bibinfo{author}{{Zou}, Y.-C.}
\newblock \bibinfo{title}{{A Statistical Study of Superluminous Supernovae
  Using the Magnetar Engine Model and Implications for Their Connection with
  Gamma-Ray Bursts and Hypernovae}}.
\newblock \emph{\bibinfo{journal}{\apj}} \textbf{\bibinfo{volume}{840}},
  \bibinfo{pages}{12} (\bibinfo{year}{2017}).

\bibitem{lv2018}
\bibinfo{author}{{L{\"u}}, H.-J.} \emph{et~al.}
\newblock \bibinfo{title}{{Gamma-Ray Burst/Supernova Associations: Energy
  Partition and the Case of a Magnetar Central Engine}}.
\newblock \emph{\bibinfo{journal}{\apj}} \textbf{\bibinfo{volume}{862}},
  \bibinfo{pages}{130} (\bibinfo{year}{2018}).

\bibitem{lyman2016}
\bibinfo{author}{{Lyman}, J.~D.} \emph{et~al.}
\newblock \bibinfo{title}{{Bolometric light curves and explosion parameters of
  38 stripped-envelope core-collapse supernovae}}.
\newblock \emph{\bibinfo{journal}{\mnras}} \textbf{\bibinfo{volume}{457}},
  \bibinfo{pages}{328--350} (\bibinfo{year}{2016}).

\bibitem{taddia2019}
\bibinfo{author}{{Taddia}, F.} \emph{et~al.}
\newblock \bibinfo{title}{{Analysis of broad-lined Type Ic supernovae from the
  (intermediate) Palomar Transient Factory}}.
\newblock \emph{\bibinfo{journal}{\aap}} \textbf{\bibinfo{volume}{621}},
  \bibinfo{pages}{A71} (\bibinfo{year}{2019}).

\bibitem{nicholl2017}
\bibinfo{author}{{Nicholl}, M.}, \bibinfo{author}{{Guillochon}, J.} \&
  \bibinfo{author}{{Berger}, E.}
\newblock \bibinfo{title}{{The Magnetar Model for Type I Superluminous
  Supernovae. I. Bayesian Analysis of the Full Multicolor Light-curve Sample
  with MOSFiT}}.
\newblock \emph{\bibinfo{journal}{\apj}} \textbf{\bibinfo{volume}{850}},
  \bibinfo{pages}{55} (\bibinfo{year}{2017}).

\bibitem{blanchard2020}
\bibinfo{author}{{Blanchard}, P.~K.}, \bibinfo{author}{{Berger}, E.},
  \bibinfo{author}{{Nicholl}, M.} \& \bibinfo{author}{{Villar}, V.~A.}
\newblock \bibinfo{title}{{The Pre-explosion Mass Distribution of Hydrogen-poor
  Superluminous Supernova Progenitors and New Evidence for a Mass-Spin
  Correlation}}.
\newblock \emph{\bibinfo{journal}{\apj}} \textbf{\bibinfo{volume}{897}},
  \bibinfo{pages}{114} (\bibinfo{year}{2020}).

\bibitem{chen2023}
\bibinfo{author}{{Chen}, Z.~H.} \emph{et~al.}
\newblock \bibinfo{title}{{The Hydrogen-poor Superluminous Supernovae from the
  Zwicky Transient Facility Phase I Survey. II. Light-curve Modeling and
  Characterization of Undulations}}.
\newblock \emph{\bibinfo{journal}{\apj}} \textbf{\bibinfo{volume}{943}},
  \bibinfo{pages}{42} (\bibinfo{year}{2023}).

\bibitem{foremanmackey2013}
\bibinfo{author}{{Foreman-Mackey}, D.}, \bibinfo{author}{{Hogg}, D.~W.},
  \bibinfo{author}{{Lang}, D.} \& \bibinfo{author}{{Goodman}, J.}
\newblock \bibinfo{title}{{emcee: The MCMC Hammer}}.
\newblock \emph{\bibinfo{journal}{\pasp}} \textbf{\bibinfo{volume}{125}},
  \bibinfo{pages}{306} (\bibinfo{year}{2013}).

\bibitem{piro2011}
\bibinfo{author}{{Piro}, A.~L.} \& \bibinfo{author}{{Ott}, C.~D.}
\newblock \bibinfo{title}{{Supernova Fallback onto Magnetars and
  Propeller-powered Supernovae}}.
\newblock \emph{\bibinfo{journal}{\apj}} \textbf{\bibinfo{volume}{736}},
  \bibinfo{pages}{108} (\bibinfo{year}{2011}).

\bibitem{arnett1982}
\bibinfo{author}{{Arnett}, W.~D.}
\newblock \bibinfo{title}{{Type I supernovae. I - Analytic solutions for the
  early part of the light curve}}.
\newblock \emph{\bibinfo{journal}{\apj}} \textbf{\bibinfo{volume}{253}},
  \bibinfo{pages}{785--797} (\bibinfo{year}{1982}).

\bibitem{quimby2013}
\bibinfo{author}{{Quimby}, R.~M.}, \bibinfo{author}{{Yuan}, F.},
  \bibinfo{author}{{Akerlof}, C.} \& \bibinfo{author}{{Wheeler}, J.~C.}
\newblock \bibinfo{title}{{Rates of superluminous supernovae at z
  {\ensuremath{\sim}} 0.2}}.
\newblock \emph{\bibinfo{journal}{\mnras}} \textbf{\bibinfo{volume}{431}},
  \bibinfo{pages}{912--922} (\bibinfo{year}{2013}).

\bibitem{Paxton2011}
\bibinfo{author}{{Paxton}, B.} \emph{et~al.}
\newblock \bibinfo{title}{{Modules for Experiments in Stellar Astrophysics
  (MESA)}}.
\newblock \emph{\bibinfo{journal}{\apjs}} \textbf{\bibinfo{volume}{192}},
  \bibinfo{pages}{3} (\bibinfo{year}{2011}).

\bibitem{Paxton2013}
\bibinfo{author}{{Paxton}, B.} \emph{et~al.}
\newblock \bibinfo{title}{{Modules for Experiments in Stellar Astrophysics
  (MESA): Planets, Oscillations, Rotation, and Massive Stars}}.
\newblock \emph{\bibinfo{journal}{\apjs}} \textbf{\bibinfo{volume}{208}},
  \bibinfo{pages}{4} (\bibinfo{year}{2013}).

\bibitem{Paxton2015}
\bibinfo{author}{{Paxton}, B.} \emph{et~al.}
\newblock \bibinfo{title}{{Modules for Experiments in Stellar Astrophysics
  (MESA): Binaries, Pulsations, and Explosions}}.
\newblock \emph{\bibinfo{journal}{\apjs}} \textbf{\bibinfo{volume}{220}},
  \bibinfo{pages}{15} (\bibinfo{year}{2015}).

\bibitem{hu2022}
\bibinfo{author}{{Hu}, R.-C.} \emph{et~al.}
\newblock \bibinfo{title}{{A Channel to Form Fast-spinning Black Hole-Neutron
  Star Binary Mergers as Multimessenger Sources}}.
\newblock \emph{\bibinfo{journal}{\apj}} \textbf{\bibinfo{volume}{928}},
  \bibinfo{pages}{163} (\bibinfo{year}{2022}).

\bibitem{MLT1958}
\bibinfo{author}{{B{\"o}hm-Vitense}, E.}
\newblock \bibinfo{title}{{{\"U}ber die Wasserstoffkonvektionszone in Sternen
  verschiedener Effektivtemperaturen und Leuchtkr{\"a}fte. Mit 5
  Textabbildungen}}.
\newblock \emph{\bibinfo{journal}{\zap}} \textbf{\bibinfo{volume}{46}},
  \bibinfo{pages}{108} (\bibinfo{year}{1958}).

\bibitem{Langer1983}
\bibinfo{author}{{Langer}, N.}, \bibinfo{author}{{Fricke}, K.~J.} \&
  \bibinfo{author}{{Sugimoto}, D.}
\newblock \bibinfo{title}{{Semiconvective diffusion and energy transport}}.
\newblock \emph{\bibinfo{journal}{\aap}} \textbf{\bibinfo{volume}{126}},
  \bibinfo{pages}{207} (\bibinfo{year}{1983}).

\bibitem{Nugis2000}
\bibinfo{author}{{Nugis}, T.} \& \bibinfo{author}{{Lamers}, H.~J.~G.~L.~M.}
\newblock \bibinfo{title}{{Mass-loss rates of Wolf-Rayet stars as a function of
  stellar parameters}}.
\newblock \emph{\bibinfo{journal}{\aap}} \textbf{\bibinfo{volume}{360}},
  \bibinfo{pages}{227--244} (\bibinfo{year}{2000}).

\bibitem{Glebbeek2009}
\bibinfo{author}{{Glebbeek}, E.}, \bibinfo{author}{{Gaburov}, E.},
  \bibinfo{author}{{de Mink}, S.~E.}, \bibinfo{author}{{Pols}, O.~R.} \&
  \bibinfo{author}{{Portegies Zwart}, S.~F.}
\newblock \bibinfo{title}{{The evolution of runaway stellar collision
  products}}.
\newblock \emph{\bibinfo{journal}{\aap}} \textbf{\bibinfo{volume}{497}},
  \bibinfo{pages}{255--264} (\bibinfo{year}{2009}).

\bibitem{Higgins2021}
\bibinfo{author}{{Higgins}, E.~R.}, \bibinfo{author}{{Sander}, A.~A.~C.},
  \bibinfo{author}{{Vink}, J.~S.} \& \bibinfo{author}{{Hirschi}, R.}
\newblock \bibinfo{title}{{Evolution of Wolf-Rayet stars as black hole
  progenitors}}.
\newblock \emph{\bibinfo{journal}{\mnras}} \textbf{\bibinfo{volume}{505}},
  \bibinfo{pages}{4874--4889} (\bibinfo{year}{2021}).

\bibitem{Heger2000}
\bibinfo{author}{{Heger}, A.} \& \bibinfo{author}{{Langer}, N.}
\newblock \bibinfo{title}{{Presupernova Evolution of Rotating Massive Stars.
  II. Evolution of the Surface Properties}}.
\newblock \emph{\bibinfo{journal}{\apj}} \textbf{\bibinfo{volume}{544}},
  \bibinfo{pages}{1016--1035} (\bibinfo{year}{2000}).

\bibitem{fuller2019}
\bibinfo{author}{{Fuller}, J.}, \bibinfo{author}{{Piro}, A.~L.} \&
  \bibinfo{author}{{Jermyn}, A.~S.}
\newblock \bibinfo{title}{{Slowing the spins of stellar cores}}.
\newblock \emph{\bibinfo{journal}{\mnras}} \textbf{\bibinfo{volume}{485}},
  \bibinfo{pages}{3661--3680} (\bibinfo{year}{2019}).

\bibitem{Heger2023}
\bibinfo{author}{{Heger}, A.}, \bibinfo{author}{{M{\"u}ller}, B.} \&
  \bibinfo{author}{{Mandel}, I.}
\newblock \bibinfo{title}{{Black holes as the end state of stellar evolution:
  Theory and simulations}}.
\newblock \emph{\bibinfo{journal}{arXiv e-prints}}
  \bibinfo{pages}{arXiv:2304.09350} (\bibinfo{year}{2023}).

\bibitem{worley2008}
\bibinfo{author}{{Worley}, A.}, \bibinfo{author}{{Krastev}, P.~G.} \&
  \bibinfo{author}{{Li}, B.-A.}
\newblock \bibinfo{title}{{Nuclear Constraints on the Moments of Inertia of
  Neutron Stars}}.
\newblock \emph{\bibinfo{journal}{\apj}} \textbf{\bibinfo{volume}{685}},
  \bibinfo{pages}{390--399} (\bibinfo{year}{2008}).

\bibitem{gao2020}
\bibinfo{author}{{Gao}, H.} \emph{et~al.}
\newblock \bibinfo{title}{{Relation between gravitational mass and baryonic
  mass for non-rotating and rapidly rotating neutron stars}}.
\newblock \emph{\bibinfo{journal}{Frontiers of Physics}}
  \textbf{\bibinfo{volume}{15}}, \bibinfo{pages}{24603} (\bibinfo{year}{2020}).

\bibitem{piro2014}
\bibinfo{author}{{Piro}, A.~L.} \& \bibinfo{author}{{Morozova}, V.~S.}
\newblock \bibinfo{title}{{Transparent Helium in Stripped Envelope
  Supernovae}}.
\newblock \emph{\bibinfo{journal}{\apjl}} \textbf{\bibinfo{volume}{792}},
  \bibinfo{pages}{L11} (\bibinfo{year}{2014}).

\bibitem{Hurley2002}
\bibinfo{author}{{Hurley}, J.~R.}, \bibinfo{author}{{Tout}, C.~A.} \&
  \bibinfo{author}{{Pols}, O.~R.}
\newblock \bibinfo{title}{{Evolution of binary stars and the effect of tides on
  binary populations}}.
\newblock \emph{\bibinfo{journal}{\mnras}} \textbf{\bibinfo{volume}{329}},
  \bibinfo{pages}{897--928} (\bibinfo{year}{2002}).

\bibitem{sl2014}
\bibinfo{author}{{Shao}, Y.} \& \bibinfo{author}{{Li}, X.-D.}
\newblock \bibinfo{title}{{On the Formation of Be Stars through Binary
  Interaction}}.
\newblock \emph{\bibinfo{journal}{\apj}} \textbf{\bibinfo{volume}{796}},
  \bibinfo{pages}{37} (\bibinfo{year}{2014}).

\bibitem{sl2021}
\bibinfo{author}{{Shao}, Y.} \& \bibinfo{author}{{Li}, X.-D.}
\newblock \bibinfo{title}{{Population Synthesis of Black Hole Binaries with
  Compact Star Companions}}.
\newblock \emph{\bibinfo{journal}{\apj}} \textbf{\bibinfo{volume}{920}},
  \bibinfo{pages}{81} (\bibinfo{year}{2021}).

\bibitem{Webbink1984}
\bibinfo{author}{{Webbink}, R.~F.}
\newblock \bibinfo{title}{{Double white dwarfs as progenitors of R Coronae
  Borealis stars and type I supernovae.}}
\newblock \emph{\bibinfo{journal}{\apj}} \textbf{\bibinfo{volume}{277}},
  \bibinfo{pages}{355--360} (\bibinfo{year}{1984}).

\bibitem{Fragos2019}
\bibinfo{author}{{Fragos}, T.} \emph{et~al.}
\newblock \bibinfo{title}{{The Complete Evolution of a Neutron-star Binary
  through a Common Envelope Phase Using 1D Hydrodynamic Simulations}}.
\newblock \emph{\bibinfo{journal}{\apjl}} \textbf{\bibinfo{volume}{883}},
  \bibinfo{pages}{L45} (\bibinfo{year}{2019}).

\bibitem{Kroupa1993}
\bibinfo{author}{{Kroupa}, P.}, \bibinfo{author}{{Tout}, C.~A.} \&
  \bibinfo{author}{{Gilmore}, G.}
\newblock \bibinfo{title}{{The Distribution of Low-Mass Stars in the Galactic
  Disc}}.
\newblock \emph{\bibinfo{journal}{\mnras}} \textbf{\bibinfo{volume}{262}},
  \bibinfo{pages}{545--587} (\bibinfo{year}{1993}).

\bibitem{Kobulnicky2007}
\bibinfo{author}{{Kobulnicky}, H.~A.} \& \bibinfo{author}{{Fryer}, C.~L.}
\newblock \bibinfo{title}{{A New Look at the Binary Characteristics of Massive
  Stars}}.
\newblock \emph{\bibinfo{journal}{\apj}} \textbf{\bibinfo{volume}{670}},
  \bibinfo{pages}{747--765} (\bibinfo{year}{2007}).

\bibitem{Abt1983}
\bibinfo{author}{{Abt}, H.~A.}
\newblock \bibinfo{title}{{Normal and abnormal binary frequencies.}}
\newblock \emph{\bibinfo{journal}{\araa}} \textbf{\bibinfo{volume}{21}},
  \bibinfo{pages}{343--372} (\bibinfo{year}{1983}).

\bibitem{Sana2012}
\bibinfo{author}{{Sana}, H.} \emph{et~al.}
\newblock \bibinfo{title}{{Binary Interaction Dominates the Evolution of
  Massive Stars}}.
\newblock \emph{\bibinfo{journal}{Science}} \textbf{\bibinfo{volume}{337}},
  \bibinfo{pages}{444} (\bibinfo{year}{2012}).

\bibitem{Moe2017}
\bibinfo{author}{{Moe}, M.} \& \bibinfo{author}{{Di Stefano}, R.}
\newblock \bibinfo{title}{{Mind Your Ps and Qs: The Interrelation between
  Period (P) and Mass-ratio (Q) Distributions of Binary Stars}}.
\newblock \emph{\bibinfo{journal}{\apjs}} \textbf{\bibinfo{volume}{230}},
  \bibinfo{pages}{15} (\bibinfo{year}{2017}).

\bibitem{Madau2017}
\bibinfo{author}{{Madau}, P.} \& \bibinfo{author}{{Fragos}, T.}
\newblock \bibinfo{title}{{Radiation Backgrounds at Cosmic Dawn: X-Rays from
  Compact Binaries}}.
\newblock \emph{\bibinfo{journal}{\apj}} \textbf{\bibinfo{volume}{840}},
  \bibinfo{pages}{39} (\bibinfo{year}{2017}).

\bibitem{Madau2014}
\bibinfo{author}{{Madau}, P.} \& \bibinfo{author}{{Dickinson}, M.}
\newblock \bibinfo{title}{{Cosmic Star-Formation History}}.
\newblock \emph{\bibinfo{journal}{\araa}} \textbf{\bibinfo{volume}{52}},
  \bibinfo{pages}{415--486} (\bibinfo{year}{2014}).

\bibitem{Belczynski2020}
\bibinfo{author}{{Belczynski}, K.} \emph{et~al.}
\newblock \bibinfo{title}{{Evolutionary roads leading to low effective spins,
  high black hole masses, and O1/O2 rates for LIGO/Virgo binary black holes}}.
\newblock \emph{\bibinfo{journal}{\aap}} \textbf{\bibinfo{volume}{636}},
  \bibinfo{pages}{A104} (\bibinfo{year}{2020}).

\bibitem{Ivanova2003}
\bibinfo{author}{{Ivanova}, N.}, \bibinfo{author}{{Belczynski}, K.},
  \bibinfo{author}{{Kalogera}, V.}, \bibinfo{author}{{Rasio}, F.~A.} \&
  \bibinfo{author}{{Taam}, R.~E.}
\newblock \bibinfo{title}{{The Role of Helium Stars in the Formation of Double
  Neutron Stars}}.
\newblock \emph{\bibinfo{journal}{\apj}} \textbf{\bibinfo{volume}{592}},
  \bibinfo{pages}{475--485} (\bibinfo{year}{2003}).

\bibitem{Ivanova2011}
\bibinfo{author}{{Ivanova}, N.}
\newblock \bibinfo{title}{{Common Envelope: On the Mass and the Fate of the
  Remnant}}.
\newblock \emph{\bibinfo{journal}{\apj}} \textbf{\bibinfo{volume}{730}},
  \bibinfo{pages}{76} (\bibinfo{year}{2011}).

\bibitem{Gotberg2020}
\bibinfo{author}{{G{\"o}tberg}, Y.} \emph{et~al.}
\newblock \bibinfo{title}{{Contribution from stars stripped in binaries to
  cosmic reionization of hydrogen and helium}}.
\newblock \emph{\bibinfo{journal}{\aap}} \textbf{\bibinfo{volume}{634}},
  \bibinfo{pages}{A134} (\bibinfo{year}{2020}).

\bibitem{Ivanova2020}
\bibinfo{author}{{Ivanova}, N.}, \bibinfo{author}{{Justham}, S.} \&
  \bibinfo{author}{{Ricker}, P.}
\newblock \emph{\bibinfo{title}{{Common Envelope Evolution}}}
  (\bibinfo{year}{2020}).

\bibitem{Klencki2022}
\bibinfo{author}{{Klencki}, J.}, \bibinfo{author}{{Istrate}, A.},
  \bibinfo{author}{{Nelemans}, G.} \& \bibinfo{author}{{Pols}, O.}
\newblock \bibinfo{title}{{Partial-envelope stripping and nuclear-timescale
  mass transfer from evolved supergiants at low metallicity}}.
\newblock \emph{\bibinfo{journal}{\aap}} \textbf{\bibinfo{volume}{662}},
  \bibinfo{pages}{A56} (\bibinfo{year}{2022}).

\bibitem{VignaGomez2022}
\bibinfo{author}{{Vigna-G{\'o}mez}, A.} \emph{et~al.}
\newblock \bibinfo{title}{{Stellar response after stripping as a model for
  common-envelope outcomes}}.
\newblock \emph{\bibinfo{journal}{\mnras}} \textbf{\bibinfo{volume}{511}},
  \bibinfo{pages}{2326--2338} (\bibinfo{year}{2022}).

\bibitem{laplace2020}
\bibinfo{author}{{Laplace}, E.}, \bibinfo{author}{{G{\"o}tberg}, Y.},
  \bibinfo{author}{{de Mink}, S.~E.}, \bibinfo{author}{{Justham}, S.} \&
  \bibinfo{author}{{Farmer}, R.}
\newblock \bibinfo{title}{{The expansion of stripped-envelope stars:
  Consequences for supernovae and gravitational-wave progenitors}}.
\newblock \emph{\bibinfo{journal}{\aap}} \textbf{\bibinfo{volume}{637}},
  \bibinfo{pages}{A6} (\bibinfo{year}{2020}).

\bibitem{Marchant2016}
\bibinfo{author}{{Marchant}, P.}, \bibinfo{author}{{Langer}, N.},
  \bibinfo{author}{{Podsiadlowski}, P.}, \bibinfo{author}{{Tauris}, T.~M.} \&
  \bibinfo{author}{{Moriya}, T.~J.}
\newblock \bibinfo{title}{{A new route towards merging massive black holes}}.
\newblock \emph{\bibinfo{journal}{\aap}} \textbf{\bibinfo{volume}{588}},
  \bibinfo{pages}{A50} (\bibinfo{year}{2016}).

\bibitem{Powell2023}
\bibinfo{author}{{Powell}, J.}, \bibinfo{author}{{M{\"u}ller}, B.},
  \bibinfo{author}{{Aguilera-Dena}, D.~R.} \& \bibinfo{author}{{Langer}, N.}
\newblock \bibinfo{title}{{Three dimensional magnetorotational core-collapse
  supernova explosions of a 39 solar mass progenitor star}}.
\newblock \emph{\bibinfo{journal}{\mnras}} \textbf{\bibinfo{volume}{522}},
  \bibinfo{pages}{6070--6086} (\bibinfo{year}{2023}).

\bibitem{de2018}
\bibinfo{author}{{De}, K.} \emph{et~al.}
\newblock \bibinfo{title}{{A hot and fast ultra-stripped supernova that likely
  formed a compact neutron star binary}}.
\newblock \emph{\bibinfo{journal}{Science}} \textbf{\bibinfo{volume}{362}},
  \bibinfo{pages}{201--206} (\bibinfo{year}{2018}).

\bibitem{Yao2020}
\bibinfo{author}{{Yao}, Y.} \emph{et~al.}
\newblock \bibinfo{title}{{SN2019dge: A Helium-rich Ultra-stripped Envelope
  Supernova}}.
\newblock \emph{\bibinfo{journal}{\apj}} \textbf{\bibinfo{volume}{900}},
  \bibinfo{pages}{46} (\bibinfo{year}{2020}).

\bibitem{Moore2024}
\bibinfo{author}{{Moore}, T.} \emph{et~al.}
\newblock \bibinfo{title}{{SN 2023zaw: the low-energy explosion of an
  ultra-stripped star, with non-radioactive heating}}.
\newblock \emph{\bibinfo{journal}{arXiv e-prints}}
  \bibinfo{pages}{arXiv:2405.13596} (\bibinfo{year}{2024}).

\bibitem{macfadyen1999}
\bibinfo{author}{{MacFadyen}, A.~I.} \& \bibinfo{author}{{Woosley}, S.~E.}
\newblock \bibinfo{title}{{Collapsars: Gamma-Ray Bursts and Explosions in
  ``Failed Supernovae''}}.
\newblock \emph{\bibinfo{journal}{\apj}} \textbf{\bibinfo{volume}{524}},
  \bibinfo{pages}{262--289} (\bibinfo{year}{1999}).

\bibitem{zenati2020}
\bibinfo{author}{{Zenati}, Y.}, \bibinfo{author}{{Siegel}, D.~M.},
  \bibinfo{author}{{Metzger}, B.~D.} \& \bibinfo{author}{{Perets}, H.~B.}
\newblock \bibinfo{title}{{Nuclear burning in collapsar accretion discs}}.
\newblock \emph{\bibinfo{journal}{\mnras}} \textbf{\bibinfo{volume}{499}},
  \bibinfo{pages}{4097--4113} (\bibinfo{year}{2020}).

\bibitem{Gottlieb2024}
\bibinfo{author}{{Gottlieb}, O.}, \bibinfo{author}{{Renzo}, M.},
  \bibinfo{author}{{Metzger}, B.~D.}, \bibinfo{author}{{Goldberg}, J.~A.} \&
  \bibinfo{author}{{Cantiello}, M.}
\newblock \bibinfo{title}{{She's Got Her Mother's Hair: End-to-End Collapsar
  Simulations Unveil the Origin of Black Holes' Magnetic Field}}.
\newblock \emph{\bibinfo{journal}{arXiv e-prints}}
  \bibinfo{pages}{arXiv:2407.16745} (\bibinfo{year}{2024}).

\bibitem{mandel2020}
\bibinfo{author}{{Mandel}, I.} \& \bibinfo{author}{{M{\"u}ller}, B.}
\newblock \bibinfo{title}{{Simple recipes for compact remnant masses and natal
  kicks}}.
\newblock \emph{\bibinfo{journal}{\mnras}} \textbf{\bibinfo{volume}{499}},
  \bibinfo{pages}{3214--3221} (\bibinfo{year}{2020}).

\bibitem{kalogera1996}
\bibinfo{author}{{Kalogera}, V.}
\newblock \bibinfo{title}{{Orbital Characteristics of Binary Systems after
  Asymmetric Supernova Explosions}}.
\newblock \emph{\bibinfo{journal}{\apj}} \textbf{\bibinfo{volume}{471}},
  \bibinfo{pages}{352} (\bibinfo{year}{1996}).

\bibitem{callister2021}
\bibinfo{author}{{Callister}, T.~A.}, \bibinfo{author}{{Farr}, W.~M.} \&
  \bibinfo{author}{{Renzo}, M.}
\newblock \bibinfo{title}{{State of the Field: Binary Black Hole Natal Kicks
  and Prospects for Isolated Field Formation after GWTC-2}}.
\newblock \emph{\bibinfo{journal}{\apj}} \textbf{\bibinfo{volume}{920}},
  \bibinfo{pages}{157} (\bibinfo{year}{2021}).

\bibitem{olausen2014}
\bibinfo{author}{{Olausen}, S.~A.} \& \bibinfo{author}{{Kaspi}, V.~M.}
\newblock \bibinfo{title}{{The McGill Magnetar Catalog}}.
\newblock \emph{\bibinfo{journal}{\apjs}} \textbf{\bibinfo{volume}{212}},
  \bibinfo{pages}{6} (\bibinfo{year}{2014}).

\end{thebibliography}

\end{document}